\documentclass[12pt,letterpaper]{lsuetd}
\usepackage{setspace,dsfont,verbatim,indentfirst}
\usepackage{graphicx}
\usepackage[hidelinks=false]{hyperref}
\usepackage{subfigure}
\usepackage{amsmath}
\usepackage{amssymb}
\usepackage{epstopdf}

\newcommand{\avg}[1]{\langle #1 \rangle}
\makeatletter

\begin{document}
\renewcommand\@pnumwidth{1.55em}
\renewcommand\@tocrmarg{9.55em}
\renewcommand*\l@chapter{\@dottedtocline{0}{1.5em}{2.3em}}
\renewcommand*\l@figure{\@dottedtocline{1}{0em}{3.1em}}
\let\l@table\l@figure

\pagenumbering{roman}
\thispagestyle{empty}
\begin{center}
ADVANCES IN QUANTUM METROLOGY: CONTINUOUS VARIABLES IN PHASE SPACE

\vfill
\doublespacing
A Dissertation \\
\singlespacing
Submitted to the Graduate Faculty of the \\
Louisiana State University and \\
Agricultural and Mechanical College \\
in partial fulfillment of the \\
requirements for the degree of \\
Doctor of Philosophy\\
\doublespacing
in \\
                                       
The Department of Physics and Astronomy \\
\singlespacing
\vfill

by \\
Bryan Tomas Gard \\
B.S, Louisiana State University, 2012 \\
May 2016
\end{center}
\pagebreak



\chapter*{Acknowledgments}
\doublespacing
\vspace{0.55ex}
I must first thank Jonathan P. Dowling, for the proper balance of the ``carrot and stick" principle, which likely has some quantum uncertainty relation. Jon provides the ideal level of encouragement or threats, as the situation requires and keeps progress moving, while managing not to be unreasonable. If Jon does not have the answer, he can always point you in the right direction all while not making you feel inferior, for asking possibly ``dumb questions".

I also thank Hwang Lee for always having the insight to see the clever solution, when my attempt at an explanation confuses, more than explains. I would also like to comment that Hwang's presence at meetings always has a calming effect (perhaps to offset Jon's).

I thank the other members of my committee, Mark Wilde, Thomas Corbitt, and dean's representative Evgueni Nesterov, for offering their time and input to my work.

I appreciate the guidance and many many helpful discussions with Emanuel Knill at NIST-Boulder, where I spent two summers working on quantum research. The vastness of his knowledge in many sub disciplines of physics is truly humbling.

I would also like to acknowledge support from the National Physical Science Consortium (NPSC) and the National Institute of Standards and Technology (NIST) for supporting my graduate career through fellowship and summer internships. This support allowed me to focus on my studies, while taking classes, and focus on research, instead of juggling these duties along with teaching or grading duties. In large part, I believe this is why I am able to graduate in a somewhat brief time as a graduate student.

I would also like to thank the many enriching collaborators I have worked with. Researchers at the Boeing Corporation, Barbara Capron, Claudio Parazzoli, and Ben Koltenbah, have provided much assistance in the form of providing many discussions on an industry project in quantum metrology which served as invaluable tool to further my understanding of quantum optics and metrology. Also included in this project were many helpful discussions with Christopher Gerry.

I also must thank my wife, Lynn Gard, who somehow put up with my travels, long nights, never-ending typing, no money, complaining, headaches, and too many other issues to list. Always a source of comfort, I know I couldn't have accomplished this much without her.

Last but certainly not least, my parents, James and Janet Gard, who were always encouraging and at least pretended to listen to my many ``exciting" realizations in the fun world of quantum mechanics. I would also like to thank my brothers, Darin and Chris Gard for acknowledging that there can be only one true nerd in the family and accepting defeat.

\addcontentsline{toc}{chapter}{\hspace{-1.5em} {Acknowledgments} \vspace{12pt}}
\pagebreak


\singlespacing
\tableofcontents
\pagebreak



\renewenvironment{abstract}{{\hspace{-2.2em} \huge \textbf{\abstractname}} \par}{\pagebreak}
\addcontentsline{toc}{chapter}{\hspace{-1.5em} {Abstract}}
\begin{abstract}
\vspace{0.55ex}
\doublespacing
This dissertation serves as a general introduction to Wigner functions, phase space, and quantum metrology but also strives to be useful as a how-to guide for those who wish to delve into the realm of using continuous variables, to describe quantum states of light and optical interferometry. We include many of the introductory elements one needs to appreciate the advantages of this treatment as well as show many examples in an effort to make this dissertation more friendly.

In the initial segment of this dissertation, we focus on the advantages of Wigner functions and their use to describe many quantum states of light. We focus on coherent states and squeezed vacuum with a Mach Zehnder Interferometer for many of our examples, also used by experiments such as advanced LIGO. Later, we will also analyze this setup in more detail with a full example including the effects of many noise sources such as phase drift, photon loss, inefficient detectors, and thermal noise. In this setup, we also show the optimal measurement scheme, which is currently not employed in experiment. Throughout our metrology discussions, we will also discuss various quantum limits and use quantum Fisher information to show optimal bounds. When applicable, we also discuss the use of quantum Gaussian information and how it relates to our Wigner function treatment.

The remainder of our discussion focuses on investigating the effects of photon addition and subtraction to various states of light and analyze the nondeterministic nature of this process. We use examples of $m$ photon additions to a coherent state as well as discuss the properties of an $m$ photon subtracted thermal state. We also provide an argument that this process must \textit{always} be a nondeterministic one, or the ability to violate quantum limits becomes apparent. We show that using phase measurement as one's metric is much more restrictive, which limits the usefulness of photon addition and subtraction. When we consider SNR however, we show improved SNR statistics, at the cost of increased measurement time. In this case of SNR, we also quantify the efficiency of the photon addition and subtraction process.

\end{abstract}

\pagenumbering{arabic}
\addtocontents{toc}{\vspace{12pt} \hspace{-1.8em} Chapter \vspace{-.4em}}
\singlespacing
\setlength{\textfloatsep}{12pt plus 2pt minus 2pt}
\setlength{\intextsep}{6pt plus 2pt minus 2pt}
\chapter{Introduction}
\doublespacing
In order to analyze various quantum metrology configurations, we require a quantum mechanical description of light and the effects of common optical elements. There are many mathematical models that accomplish this, through the use of wave vectors \cite{Born1927,Dirac1939,LL1977}, density matrices \cite{bib:GerryKnight05,Sakurai,Nori,bib:NielsenChuang00,Fano1957}, and Wigner functions \cite{Wig1932,Isidro2008,Ford2007,Jafarov2007}, to name a few. In this dissertation, we will discuss the use of continuous variables in phase space \cite{Curtright2012,Nolte2010}, their advantages and potential issues. We will also use this treatment to describe common interferometer setups that involve parameter estimation \cite{Ji2008,Six2015,Paris2008}, quantify their photon statistics in terms of signal to noise ratio, and discuss the effects of many quantum optics techniques. Specifically, we will investigate an interferometric setup like LIGO \cite{GEO6001,dd1,GEO6002,italyVirgo,JapanTAMA} and also describe the effects of relatively exotic operations like photon addition and subtraction \cite{Agar1,Agar2,Zavatta2004a,Zavatta2005a,Braun,Gerry2012a,Gerry2014a,Zavatta2011,Josse2006}.

With the recent, first ever, direct detection of gravitational waves \cite{Abbott2016}, many large interferometers around the world continually attempt to measure further gravitational wave events \cite{GEO6001,GEO6002,LIGO1,LIGO2,JapanTAMA,italyVirgo}. The Laser Interferometer Gravitational-Wave Observatory (LIGO) in Livingston, Louisiana and Hanford, Washington, are two examples of such interferometers. The initial configuration of LIGO was comprised of a coherent state and vacuum coupled in a Michelson interferometer \cite{LIGO1} (henceforth, we refer to this as the classical setup). This scheme is a classical strategy and is limited to a classical bound on the phase variance measurement, the Shot Noise Limit (SNL) \cite{lloyd2004,dd2015}. The objective is to measure a relative phase shift induced in one arm of the interferometer by a passing gravitational wave. Recently, the first direct measurement of gravitational waves has been shown \cite{Abbott2016}. With this amazing accomplishment, comes the need for further measurements. Despite the remarkable precision obtained by this method, improvements are still possible. One such improvement for Advanced LIGO consists of input states of a coherent state and squeezed vacuum state, a configuration first proposed by Caves \cite{Caves1981} and shown to achieve a superior phase variance measurement as compared to the previous classical input states \cite{LIGO_SNL,LIGO_Squeezed,GEO_bound,Squeezed_interferometers}.

While there are many technical challenges in using a true quantum setup such as this, we show here that some of the measurement techniques previously used in the classical setup, are no longer optimal and even may exhibit problems with effects such as phase drift and thermal noise (another source of noise typically found in configurations like LIGO, radiation pressure noise, is not considered here). In order to achieve an optimal measurement scheme and investigate their resistance to phase drift, we turn to quantum measurements such as homodyne and parity measurements and compare them to a standard intensity measurement. We show that, under ideal conditions, the parity measurement achieves the smallest phase variance, but under noisy conditions, the parity measurement suffers greatly, while the homodyne measurement continues to achieve superior phase measurement. In general, we divide our results into two regimes, the low power regime ($|\alpha|^2<500$), in which different detection schemes can lead to significantly different phase variances, and the high power regime ($|\alpha|^2>10^{5}$), which applies to Advanced LIGO, and where all detection schemes approach the optimal bound.

The use of photon addition or subtraction is an implementation of noiseless amplification. First proposed by Agarwal and Tara \cite{Agar1,Agar2}, noiseless amplification can be used to enhance a general signal with no added noise, but with the requirement that it does so nondeterministically.  If one desires an amplification of signal, it must either come with additional noise (e.g. a deterministic squeezer \cite{Walls1983,bib:GerryKnight05}) or it must be probabilistic. Either of these cases ensures consistency with fundamental conditions such as no super-luminal communication.

Here we discuss the use of photon addition and subtraction as a probabilistic amplifier and its effects on various sources, including thermal and coherent light \cite{bib:GerryKnight05,Bachor,Leo2010,Tsang2011,Wang2008}. Unlike many past discussions of this implementation, we consider the case of photon addition and subtraction at the \emph{output} of a Mach-Zehnder Interferometer (MZI) \cite{MZI1}.  Since we are using an MZI model, we are then in the realm of metrology and can therefore use many previously developed techniques from this field. The reasoning behind using the probabilistic amplification operation at the output is simply a model of the limit of control over a specified system. In the case of an externally measured source, meaning a source one has no direct control over, deterministic amplification proves useless for a phase estimation problem, as the added noise always kills any benefit of the amplification. This limit extends to the metric of signal to noise ratio (SNR), where a deterministic amplifier always amplifies signal along with its noise, leaving SNR invariant at best. More concretely, this restriction means any modification to a standard MZI must be done after the phase shifter $\phi$. With this restriction in mind, the question then remains, since a deterministic amplifier doesn't provide any benefit, is there any hope for a probabilistic amplification process?

Recent discussions by Caves \cite{Caves2014} show the use of post selection schemes and their place in quantum metrology protocols. As we discuss later, we also show that post selection schemes alone do not allow for increased phase information and also discuss some of the pitfalls when using post selection schemes that can lead to deceivingly positive results. This result however, does not invalidate the usefulness of post selection schemes in metrology, when other metrics are considered.

\pagebreak
\singlespacing
\chapter{Wigner Functions in Phase Space}
\doublespacing
\section{Phase Space}
The use of continuous variables in phase space serves many purposes. While the choice of mathematical treatment is ultimately a choice of preference, here we will discuss the advantages of using continuous variables in phase space. A perhaps more standard approach to quantum metrology is with the use of wave vectors or density matrices. Mathematically, there are many choices available when considering which method to work in, all of which offer a full description of quantum mechanics, but a specific choice may offer computational simplicity or, as we argue here in the case of Wigner functions, offer a visual aspect as well as some connections to known measurements, such as Parity measurement.

\begin{figure}[!htb]
\centering
\includegraphics[width=0.75\columnwidth]{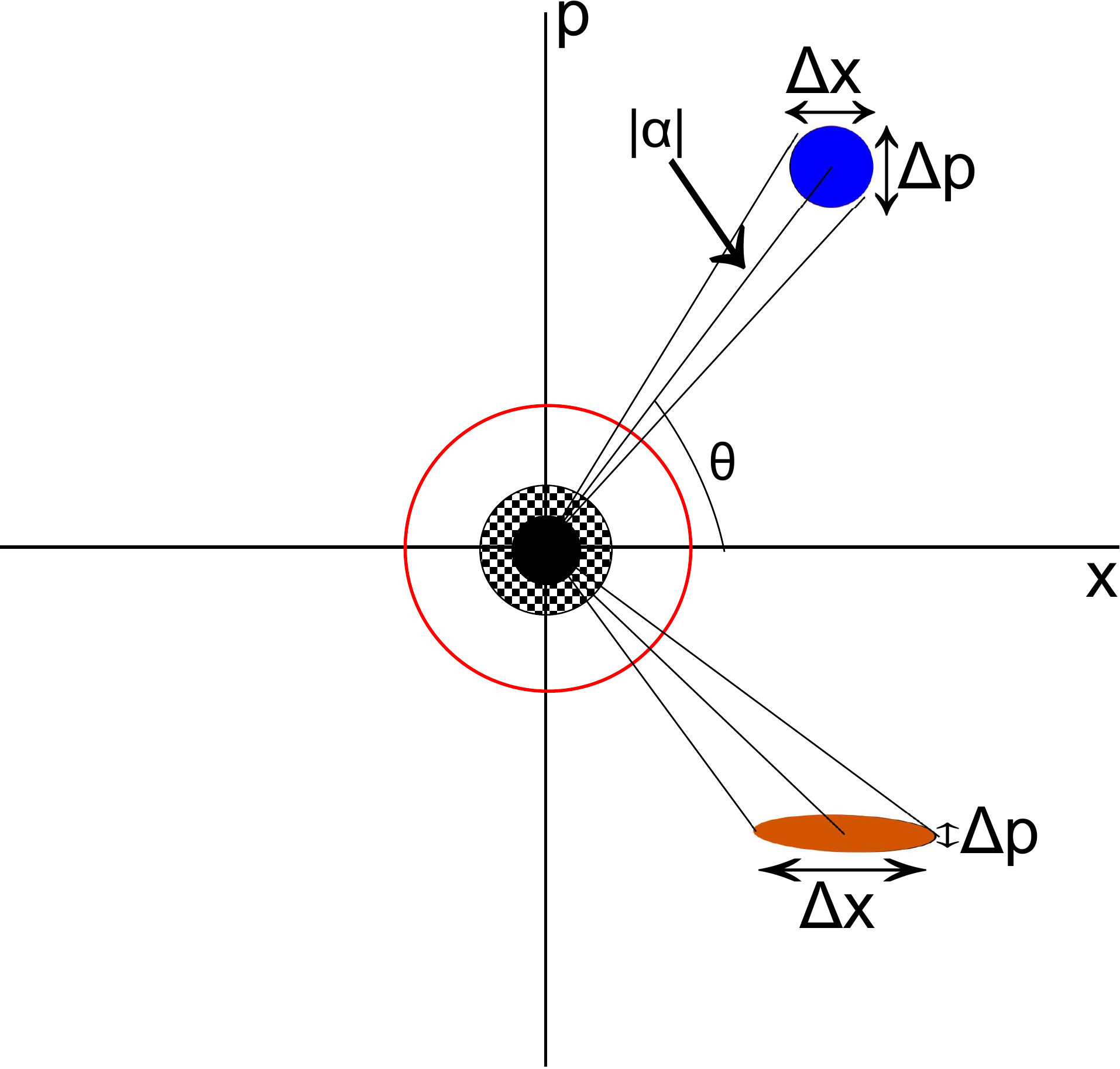}
\caption{Various states of light shown in phase space in terms of position, momentum space, (x,p). Fock state (red) shown as a ring. The radius of this ring depends on the photon number chosen. Vacuum state (black circle) shown at the center. Thermal state (checkerboard) shares partial overlap with the vacuum but  is always strictly larger. Coherent state (blue circle) shown in top right quadrant with phase angle $\theta$ and uncertainties in each quadrature also shown. This state is displaced from the center by an amount given by $|\alpha|$. A squeezed coherent state (orange) is shown in the bottom right quadrant and is squeezed to reduce the uncertainty in the $p$ quadrature.}
\label{fig:phasespace}
\end{figure}

We first begin with a visual description of phase space, in terms of the conjugate variables, $(x,p)$. This brief introduction is used as a simple illustration of various common states of light, in phase space, and we discuss a more rigorous mathematical approach in the following section. Shown in Figure~\ref{fig:phasespace}, we see many different states of light depicted in phase space. We note that the absence of photons, the vacuum state (black circle), and thermal state (checkerboard) partially overlap at the center of phase space, while a Fock state (red ring), a purely quantum mechanical state which contains exactly $n$ photons, is represented by a thin ring whose radius is determined by the chosen number state. A displaced vacuum state, or coherent state (blue circle), quantum mechanical description of laser light, is also shown along with a general uncertainty in its quadrature values due to the limits of quantum mechanics. The amount of displacement in this state, given by $|\alpha|$ is related to its average photon number by $\bar{n}=|\alpha|^2$. In general, all of these states can be squeezed, which trades uncertainty between its quadratures. The angle of this squeezing process determines which quadrature is enhanced, while the other suffers. A squeezed coherent state (orange) is shown and has been squeezed along the axis which enhances the $p$ quadrature while increasing the uncertainty in the $x$ quadrature, in compliance with the uncertainty principle relating these quadratures, $\Delta x \Delta p \geq \hbar/2$. Unless otherwise stated, for the remainder of this document, we use the natural units convention of $\hbar=c=1$. A review of these states of light can be found in \cite{bib:GerryKnight05}.

These displayed states of light compose a typical set of the most commonly described forms of light. More exotic forms of quantum light, such as photon added or subtracted states, two mode squeezed vacuum, Schr\"{o}dinger cat states, etc. can be visually depicted with various combinations of the states shown in Figure~\ref{fig:phasespace}. For example, Schr\"{o}dinger cat states can be shown by a superposition of two coherent states, while photon subtracted thermal states have the vacuum portion of a thermal state removed. These pictures in phase space of various forms of light can be very instructive when we consider things such as the various statistics of these states. We can see that the Fock state reduces to the vacuum state for $n=0$, but is not allowed to reduce to a single point, as this would violate the uncertainty principle. A similar comparison can be made for the thermal state. A thermal state of zero average photon number, also reduces to the vacuum state, as does a coherent state with no displacement. In this way, one may say that the vacuum state is the principal state of light, which other states are modifications of, through various optical processes.

\section{Wigner Functions}
Wigner functions, first introduced by Eugene Wigner in 1932 \cite{Wig1932}, is a quasi-probability distribution for a given state of light, in phase space. The term ``quasi" is used since these distributions may take on negative values, which also means they are not typical (classical) probabilities. One can show that any state whose Wigner function obtains a negative value, is a quantum state, but this statement is not an if and only if statement, meaning all negative Wigner functions correspond to quantum states, but not all quantum states have Wigner functions which attain negative values.  This treatment of light is a full mathematical description and is connected to that of density matrices by \cite{Adesso2014},
\begin{equation}
W_\rho (\textbf{x},\textbf{p})=\frac{1}{\pi^N}\int_{\mathbb{R}^N}\langle \textbf{x}+\textbf{y}| \rho | \textbf{x}-\textbf{y}\rangle e^{2i y \cdot p} d^Ny,
\end{equation}
where $\textbf{y}$ are eigenvectors of the quadrature $\{ \hat{x}\}$ operators satisfying $\hat{x}|\textbf{y}\rangle=x |\textbf{y}\rangle$ and can also be connected to the so called characteristic function with the following $s-$ ordered relations,
\begin{equation}
\begin{split}
\chi_\rho^s(\xi)&=\textrm{Tr}[\rho \hat{D}(\xi)] e^{s ||\xi||^2/2}\\
\hat{D}_k(\alpha)&=e^{\alpha\hat{a}_k^\dagger-\alpha^*\hat{a}_k}\\
W_\rho^s(\xi)&=\frac{1}{\pi^2}\int_{\mathbb{R}^{2N}}\chi_\rho^s(\kappa)e^{i \kappa^\top \Omega \xi}d^{2N}\kappa\\
\Omega&=\underset{k=1}{\overset{N}{\oplus}}\omega, 
\quad \omega=\left(
\begin{array}{cc}
0&1\\
-1&0\\
\end{array}\right),
\end{split}
\end{equation}
with $\xi \in \mathbb{R}^{2N}, ||\cdot||$ standing for the Euclidean norm on $\mathbb{R}^{2N}$ and
for Wigner functions, $s=0$, Husimi Q-functions, $s=-1$ and P-functions, $s=1$. For our purposes, we will solely focus on Wigner functions for the remainder of this document. As mentioned earlier, the coherent state is also known as displaced vacuum by virtue of, $|\alpha\rangle_k=\hat{D}_k(\alpha)|0\rangle_k$. Note that the Wigner functions can be defined for any two conjugate variables, not always position, momentum space $(x,p)$. Another typical representation is in complex phase space $(\alpha, \alpha^*)$. These two bases are connected by the relations,
\begin{equation}
\hat{x}_k=\frac{1}{\sqrt{2}}(\hat{a}_k+\hat{a}_k^\dagger), 
\quad
\hat{p}_k= \frac{1}{i\sqrt{2}}(\hat{a}_k-\hat{a}_k^\dagger),
\quad
\hat{a}|\alpha\rangle=\alpha|\alpha\rangle.
\end{equation}
The position and momentum operators obey the bosonic commutation relations $[\hat{x}_k,\hat{p}_l]=i \delta_{kl}$, while the creation and annihilation operators obey 
\begin{equation}
[\hat{a}_k,\hat{a}_l^\dagger]=\delta_{kl},
\label{eq:comm1}
\end{equation}
where the commutator is defined by $[\hat{O},\hat{Q}]\equiv\hat{O}\hat{Q}-\hat{Q}\hat{O}$.
Now that the mathematical construction of Wigner functions had been covered, we turn to the practicality of using Wigner functions in quantum optical metrology.

Many familiar properties in terms of density matrices convert to Wigner functions, but the advantageous aspects are apparent that we transition from an infinite dimensional discrete sum to a continuous variable integral of size $2N$, where $N$ is the number of spatial modes. Specifically some of these properties are,
\[
\textrm{Tr}[\rho]=1=\int_{\mathbb{R}^{2N}}W_\rho(\kappa)d^{2N}\kappa=\chi_\rho(0),
\]
where this property is seen as the normalization requirement of any quantum state (which enforces that probabilities sum to one). From this simple constraint, we see that, instead of performing the trace of an \textit{infinite} sum, we instead integrate our Wigner function over a \textit{finite} set of $2N$ variables. While both of these techniques are typically straightforward, computationally we find that integrals are typically much more easily managed, without the need of typical ``truncation" tricks as used with sums. In general this comment applies to all measurements with Wigner functions, meaning we are typically able to obtain \textit{analyctical} results, while working with density matrices frequently (though, not always) results in \textit{numerical} answers. The purity of a quantum state is also commonly used to classify states. In terms of Wigner functions this is simply,
\[
\mu_\rho=\textrm{Tr}[\rho^2]=(2\pi)^N\int_{\mathbb{R}^{2N}}[W_\rho(\kappa)]^2d^{2N}\kappa=\int_{\mathbb{R}^{2N}}|\chi_\rho(\xi)|^2d^{2N}\xi,
\]
where the state is pure if $\mu_\rho=1$ and is mixed if $\mu_\rho<1$. This condition also has a pleasing visualization in terms of the Bloch sphere, where pure states lie on the surface of the unit sphere, while mixed states lie inside the volume of the unit sphere. In general, we can see that a trace  over a density matrix corresponds to an integral of our Wigner function. This idea extends to that of partial traces. For example, consider a two mode density matrix, $\rho_{AB}$. This state has the property, $\textrm{Tr}_B[\rho_{AB}]=\rho_{A}$, where we have traced over the ``B" mode. Similarly, our Wigner function has the property $\int W_{AB}  dB=W_A$ and we have integrated out the ``B" mode.

\section{Gaussian States}
\begin{figure}[!htb]
\centering
\includegraphics[width=0.75\columnwidth]{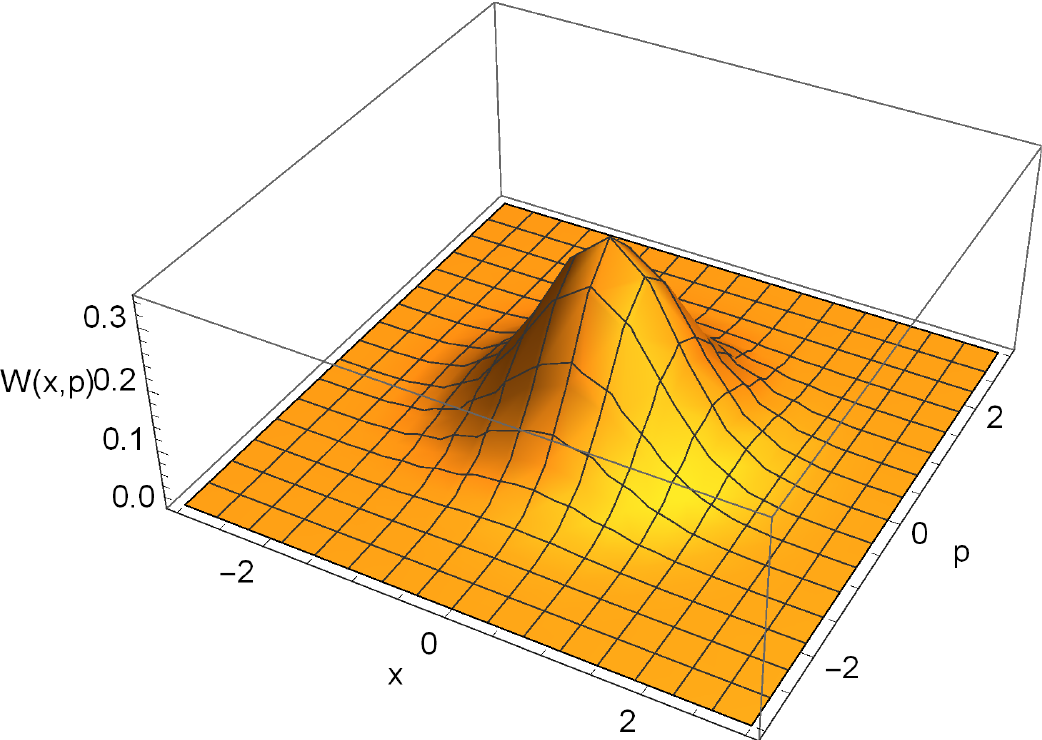}
\caption{Wigner function for the vacuum state as a function of the phase space quadratures, $x,p$. We can easily observe this state's Gaussian form.}
\label{fig:wigvac}
\end{figure}

We can now discuss another particular strength of working in phase space and the use of Wigner functions, that of Gaussian form, which classify many typical states of light. Any Wigner function that is Gaussian in form, has many simplifications that can be made. In this section we will review many of the properties of such states. A general Gaussian function can be written as,
\[
f(\textbf{x})=C \textrm{exp}(-\frac{1}{2}\textbf{x}^\top\textbf{Ax}+\textbf{b}^\top \textbf{x}),
\]
where, $\textbf{x}=(x_1,x_2,\dotsc,x_N)^\top,\textbf{b}=(b_1,b_2,\dotsc,b_N)^\top$ and $A$ is an $N \times N$ positive definite matrix \cite{Adesso2014} and $C$ ensures normalization, such that $\int f(\textbf{x}) d\textbf{x}=1$. In terms of Wigner functions then, the simplest example of a Gaussian form is the Wigner function of the vacuum state, given by,
\begin{equation}
W_{|0\rangle}(x,p)=\frac{1}{\pi} e^{-x^2-p^2},
\label{eq:vac1}
\end{equation}
which one can notice has the promised Gaussian form. Shown in Figure~\ref{fig:wigvac}, we see that we also may visualize these various states of light easily from the use of Wigner functions. Compared to our phase space picture shown in Figure~\ref{fig:phasespace}, we can notice that the phase space view is simply a projection (or slice) of the full Wigner function. 

Another example of a typical state of light, in terms of Wigner functions is the coherent state,

\begin{equation}
W_{|\alpha\rangle}(x,p)=\frac{1}{\pi}e^{-(x-\sqrt{2}|\alpha|\cos\theta)^2-(p-\sqrt{2}|\alpha|\sin\theta)^2},
\label{eq:coh1}
\end{equation}
where $|\alpha|$ is amplitude of the coherent state and $\theta$ is the phase angle. In this form, it is instructive to notice that the form is similar to that of Eq.~(\ref{eq:vac1}) but is displaced in both the $x$ and $p$ directions. The amount of displacement is controlled by the size of $\alpha$ and the direction of displacement is controlled by $\theta$. We can also show the form of a thermal state is given by,
\begin{equation}
W_{\textrm{th}}(x,p)=\frac{1}{\pi(2n_{th}+1)}e^{\frac{-x^2-p^2}{2n_{th}+1}},
\label{eq:th1}
\end{equation}
where $\bar{n}=2n_{\textrm{th}}$ is the average photon number in the thermal state. Again, one can connect this to the vacuum state for $\bar{n}=0$.

\begin{figure}[!htb]
\centering
\includegraphics[width=0.75\columnwidth]{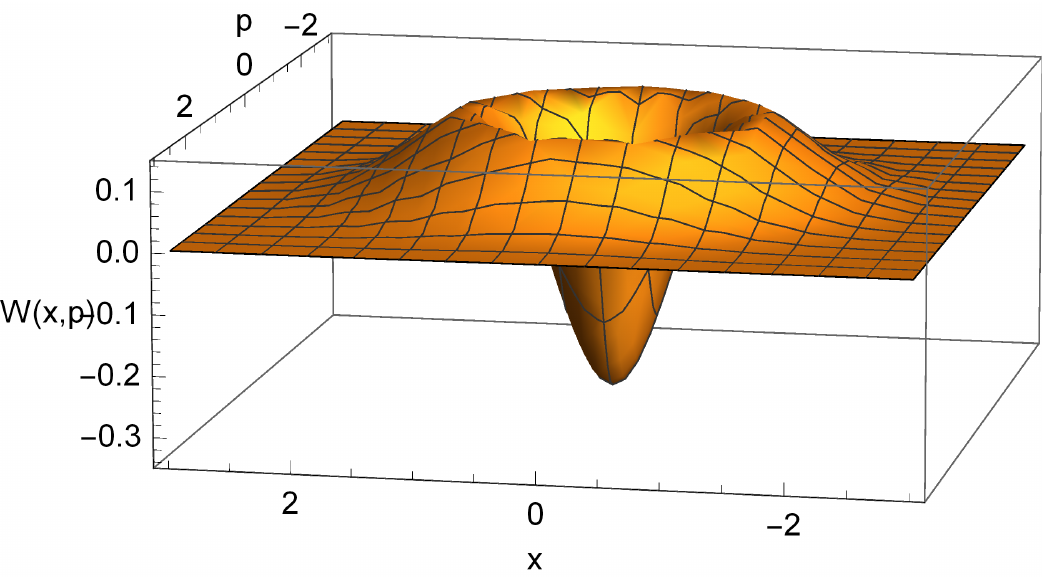}
\caption{Wigner function for the single-photon Fock state as a function of the phase space quadratures, $x,p$. We can observe that not only is this a non-Gaussian form but also attains negative values near the origin.}
\label{fig:wigfock1}
\end{figure}

All the previous Wigner functions adhere to the Gaussian form, but as an example of a non-Gaussian form, we turn to the simplest example of a Fock state, the single-photon state. All Wigner function Fock states can be described by 
\begin{equation}
W_{|n\rangle}(x,p)=\frac{1}{\pi}(-1)^nL_n[2(x^2+p^2)]e^{-x^2-p^2},
\label{eq:fockwig}
\end{equation}
where $L_n$ is the Laguerre polynomial. Note that for $n=0$ this reduces to Eq.~(\ref{eq:vac1}). For $n=1$, we then have the single-photon Fock state, which is necessarily non-Gaussian and shown in Figure~\ref{fig:wigfock1}. 
It is important to note that this Gaussian discussion is not specific to Wigner functions and is, in no way, a limitation on these usefulness of this treatment, but merely a choice of simplification. While, in general there is no problem with representing non-Gaussian forms with Wigner functions, the remainder of this section will discuss strategies that one can utilize when restricting to Gaussian only states. The advantage in using this restriction is in properties of the Gaussian form itself. Any Gaussian function can be fully described by its first and second moments. With our choice of basis, $(x,p)$, this amounts to only needing to specify $\langle \hat{x} \rangle, \langle \hat{p} \rangle$ and  $\langle \hat{x}^2 \rangle, \langle \hat{p}^2 \rangle$. In practice, the quantities we are really interested in are the mean and covariance. The mean and covariances given by
\begin{equation}
\begin{split}
d_i&=\langle \hat{R}_i\rangle\\
\sigma_{ij}&=\langle \hat{R}_i \hat{R}_j + \hat{R}_j \hat{R}_i \rangle -2\langle \hat{R}_i\rangle \langle \hat{R}_j\rangle,
\end{split}
\label{eq:moments}
\end{equation}
where $i,j\in [1,2]$, corresponding to the two conjugate phase space variables $(\hat{R}_1\equiv\hat{x},\hat{R}_2\equiv\hat{p})$. One can notice that for $i=j=1, \sigma_{11}=2(\langle\hat{x}^2\rangle - \langle\hat{x}\rangle^2)$, twice the variance of the $x$ quadrature. This factor of two is merely a convention and due to the definitions discussed earlier, but should be noted when comparing to references with other definitions. Now that we have established definitions of our mean and covariance, we can connect them back to Wigner functions with the relation,
\begin{equation}
W(\textbf{X})=\frac{1}{\pi^N \sqrt{\textrm{det}(\pmb{\sigma})}}e^{-(\textbf{X}-\textbf{d})^\top\pmb{\sigma}^{-1}(\textbf{X}-\textbf{d})},
\label{eq:wiggauss}
\end{equation}
where $N$ is the number of spatial modes, $\textbf{X,d}$ are vectors of phase space variables and means, respectively, and $\pmb{\sigma}$ is the full covariance matrix for the desired spatial modes. One may also notice that this construction lends itself to one measurement choice in particular, homodyne measurement, as we work specifically in the first and second moments of the phase space quadratures, which is exactly what a homodyne process attempts to measure. We will discuss this aspect more thoroughly, in later sections. This, along with a treatment of Quantum Fisher Information (QFI) are the main benefits from using a Gaussian-only treatment, but this discussion is also left for later sections.

\pagebreak
\singlespacing
\chapter{Interferometer Model}
\doublespacing
\section{Mach Zehnder Interferometer Model}
In order to fully model the interferometric process, we must have a description of the effect of various optical elements on the various states of light presented earlier. There are many choices of how to model states of light, but there are also choices in how to describe the propagation of this light through optical elements. In general, one can describe the movement of the the entire state through various the optical elements, a Schr\"{o}dinger picture, or describe the effect of these elements on the mode operators, a Heisenberg picture. While both are mathematically complete descriptions, they are not necessarily computationally equivalent \cite{gard2014} and therefore we choose to describe this propagation in the Heisenberg picture.

We have seen some examples of common types of light used in theoretical quantum optics, in terms of Wigner functions; here we will discuss how we model the evolution of these states through various optical elements, in terms of general Wigner functions, as well as the Gaussian restriction discussed earlier. First we will show how the mode operators, specifically our basis choice of $\hat{x},\hat{p}$, evolve through various optical elements. In general, each optical element is represented by a symplectic ($f(\beta) \Omega f(\beta)^\top=\Omega$) matrix, $f(\beta)$ of dimension $2N \times 2N$ and acts on a vector of phase space variables of length $2N$, where $N$ is the number of spatial modes, by the following input-output relation.
\begin{eqnarray}
f(\beta)\cdot
\left(
\begin{array}{c}
x_{1_i}\\
p_{1_i}\\
\vdots\\
x_{N_i}\\
p_{N_i}
\end{array}
\right)= 
\left(
\begin{array}{c}
x_{1_f}\\
p_{1_f}\\
\vdots\\
x_{N_f}\\
p_{N_f}
\end{array}
\right)
\end{eqnarray}
Perhaps the simplest transform to begin with is that of the displacement operator, briefly mentioned in the previous section. This operator can take a vacuum state into a coherent state and can be described by the transformation,
\begin{eqnarray}
D(\alpha,\theta)=
\sqrt{2}\left(
\begin{array}{cc}
|\alpha| \cos\theta&0\\
0&|\alpha|\sin\theta
\end{array} \right),
\label{eq:disp}
\end{eqnarray}
where $|\alpha|$ determines the magnitude of displacement and $\theta$, the direction.
For another example, a typical optical element, a beam splitter, splits the amplitude of two incident electric fields, according to a transmission parameter intrinsic to the beam splitter. The action of this device can be described by the matrix,
\begin{eqnarray}
BS(T)=
\left(
\begin{array}{cccc}
\sqrt{T}&0&\sqrt{1-T}&0\\
0&\sqrt{T}&0&\sqrt{1-T}\\
\sqrt{1-T}&0&-\sqrt{T}&0\\
0&\sqrt{1-T}&0&-\sqrt{T}\\
\end{array} \right),
\label{eq:bs}
\end{eqnarray}
where $0\leq T \leq1$ is the transmissivity of the beam splitter. Another common optical element is a phase shifter. This single mode element can be described by,
\begin{eqnarray}
PS(\phi)=
\left(
\begin{array}{cc}
\cos\phi&-\sin\phi\\
\sin\phi&\cos\phi
\end{array} \right),
\label{eq:ps1}
\end{eqnarray}
where $\phi$ is the phase imparted to a single mode, usually treated as an unknown parameter to be estimated. While this is a typical way to model a phase shifter, it is also instructive to construct a so-called symmetric phase shifter, which is used to form a balanced phase between two modes. This symmetric phase shifter is a two mode transform of the form,
\begin{eqnarray}
PS_2(\phi)=
\left(
\begin{array}{cccc}
\cos(\frac{\phi}{2})&-\sin(\frac{\phi}{2})&0&0\\
\sin(\frac{\phi}{2})&\cos(\frac{\phi}{2})&0&0\\
0&0&\cos(\frac{\phi}{2})&\sin\frac{\phi}{2})\\
0&0&-\sin(\frac{\phi}{2})&\cos(\frac{\phi}{2})\\
\end{array} \right) .
\label{eq:symmph}
\end{eqnarray}

Squeezing is a quantum operation that is commonly used to outperform classical only treatments. A single mode squeezing operator can be described by,
\begin{eqnarray}
S(r,\theta)=
\left(
\begin{array}{cc}
\cosh r+\cos\theta \sinh r&\sin\theta \sinh r\\
\sin\theta \sinh r&\cosh r - \cos\theta \sinh r
\end{array} \right),
\label{eq:sqzr}
\end{eqnarray}
where $r\geq0$ is the squeezing parameter and $0\leq \theta \leq 2\pi$ is the squeezing angle. The squeezing strength can be related to the gain of the squeezer by the relation $G=\cosh^2r$, for $G\geq1$. In some circumstances, it may be possible to assume $\theta=0$ and change variables using $r \rightarrow \cosh^{-1}(\sqrt{G})$ to obtain,
\begin{eqnarray}
S(G)=
\left(
\begin{array}{cc}
\sqrt{G}+\sqrt{G-1}&0\\
0&\sqrt{G}-\sqrt{G-1}
\end{array} \right),
\label{eq:sqzg}
\end{eqnarray}
which, if applicable, can greatly improve computation time. A two mode version of this squeezer is described similarly as,
\begin{eqnarray}
S_2(r,\theta)=
\left(
\begin{array}{cccc}
\cosh r&0&\sinh r \cos\theta&\sinh r \sin\theta\\
0&\cosh r&\sinh r \sin\theta&-\sinh r \cos\theta\\
\sinh r \cos\theta&\sinh r \sin\theta&\cosh r&0\\
\sinh r \sin\theta&-\sinh r \cos\theta&0&\cosh r\\
\end{array} \right),
\label{eq:sqz2}
\end{eqnarray}
which can also be written in terms of the gain of this squeezing process, if so desired. These squeezing processes have many different implementations, including the use of non-linear crystals and atomic clouds. The specific differences and challenges with each of these implementations is not the focus of our discussion however \cite{Slusher1985,Schulte2015}. We have described several typical optical elements and how they transform phase space variables; therefore we can now model various states of light and the action of many optical elements on these states. Since we have given various single and dual mode transforms, a natural question is how to deal with the case of mixing various combinations of single and dual mode transforms, as these are of different dimension. In treatments of density matrices, this is given by the tensor product, $\otimes$, but in phase space it is given by the direct sum, $\oplus$. For example, if we wish to pass a two mode Wigner function through a 50/50 beam splitter and a single mode displacement on the ``upper mode", we would use a full transform of,
\begin{eqnarray}
BS(1/2)\cdot(D(\alpha,\theta)\oplus\mathbb{I}_2)=
BS(1/2)\cdot
\left(
\begin{array}{cccc}
\sqrt{2}|\alpha| \cos\theta&0&0&0\\
0& \sqrt{2}|\alpha|\sin\theta&0&0\\
0&0&1&0\\
0&0&0&1\\
\end{array} \right)
\end{eqnarray}
where it is clear that we have simply inserted the displacement operator into the top-left block and the identity matrix into the bottom right block, as the bottom mode does not incur a transformation at this point. We also use the convention that the first entry listed in a direct sum labels the first mode, etc. Some caution should be used at this point as one can note that the direct sum of two single mode squeezers, $S(r,\theta)\oplus S(r,\theta) \neq S_2(r,\theta)$.

\begin{figure}[!htb]
\centering
\includegraphics[width=0.75\columnwidth]{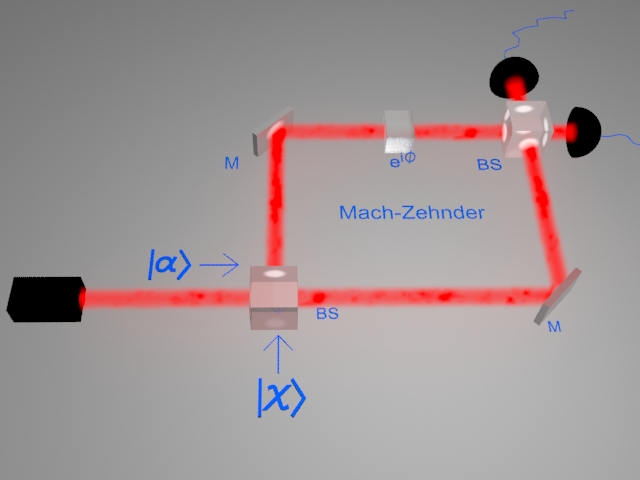}
\caption{A general Mach-Zehnder interferometer with coherent ($\alpha$) and squeezed vacuum ($\chi$) states as input. A phase shift, $\phi$, represents the phase difference between the two arms of the MZI, due to a path length difference between the two arms. Our goal is to estimate $\phi$, which in the case of LIGO, would be caused by a passing gravitational wave.}
\label{fig:MZI}
\end{figure}

Now as an illustrative example, let us consider a full model of a typical Mach-Zehnder Interferometer (MZI), which is mathematically equivalent to a Michelson Interferometer (MI), shown for a specific choice of input states, in Figure~\ref{fig:MZI}.
 For reasons discussed later, let us consider input states of a coherent state and squeezed vacuum. Since, in general, these two states are assumed to be initially independent, the two mode Wigner function for this choice is simply the product of the individual Wigner functions. The full Wigner function can then be written as,
\begin{equation}
W(\textbf{X})=\frac{1}{\pi^2}\textrm{Exp}(-2|\alpha|^2-p_1^2+2\sqrt{2}|\alpha|x_1-x_1^2-(\textrm{e}^{2r}p_2^2+\textrm{e}^{-2r}x_2^2))
\end{equation}
where $\textbf{X}$ is a list of the mode variables ${x_1,p_1,x_2,p_2}$, corresponding to the position and momentum components of each spatial mode. The average photon number in the coherent state is $N_{\textrm{coh}}=|\alpha|^2$ and in the squeezed vacuum state $N_{\textrm{sqz}}=\sinh^2{r}$. Both states have equal phases, as this gives rise to the optimal phase sensitivity (discussed later) and are taken to be $\theta_{\textrm{coh}}=\theta_{\textrm{sqz}}=0$.

The propagation of this Wigner function is accomplished by a simple transformation of the phase space variables through the MZI, dictated by its optical elements, discussed earlier. These transformations are described by
\begin{eqnarray}
BS(1/2)=\frac{1}{\sqrt{2}}
\left(
\begin{array}{cccc}
1&0&1&0\\
0&1&0&1\\
1&0&-1&0\\
0&1&0&-1\\
\end{array} \right)
\end{eqnarray}
where both beam splitters are fixed to be 50/50 and we have chosen to use a symmetric phase model, shown in Eq.~(\ref{eq:symmph}), in order to simplify calculations as well as agree with many other references \cite{Nori,dd2012}. Using these transforms, the total transform for phase space variables is given by,
\begin{eqnarray}
\left(
\begin{array}{cc}
x_{1f}\\
p_{1f}\\
x_{2f}\\
p_{2f}
\end{array}\right)
=BS(1/2) \cdot PS_2(\phi) \cdot BS(1/2) \cdot
\left(
\begin{array}{cc}
x_{1}\\
p_{1}\\
x_{2}\\
p_{2}
\end{array} \right)
\label{eq:MZI1}
\end{eqnarray}
From here, the final variables (denoted by $x_{1f}$ etc.) are inserted to the initial Wigner function to obtain the Wigner function at the output. This is a straightforward task but we do not show the result here as the state is fairly cumbersome. This process can be followed for the various choices of input states and showcases the basic method of Wigner function evolution, in the Heisenberg picture.

We can again discuss the use of a Gaussian restriction and show an alternative method to evolve a given Gaussian state of light through the MZI. This method utilizes Quantum Gaussian Information (QGI) \cite{Adesso2014} and lends itself to simplifying some measurement schemes, as well as allowing a particularly useful calculation of Quantum Fisher Information (QFI), discussed later. As we showed earlier, once a Gaussian form is assumed, one need only be concerned with the mean and covariance. It is these two quantities that we will evolve through our MZI. Initially, we require the mean and covariance of the various states of light, and luckily most have a very simple form, which showcases the usefulness of this method.  The mean of a given state of light is tied to its displacement value, which for any state centered at the origin, in phase space, is clearly zero. Therefore, states such as the thermal state and vacuum state (along with the squeezed versions of these) have a zero mean. For completeness, we show all the means in vector form as,
\begin{eqnarray}
\langle \hat{R}_{\textrm{vac}}\rangle=\langle \hat{R}_{\textrm{th}}\rangle=
\left(
\begin{array}{c}
0\\
0
\end{array}
\right) ,\quad
\langle \hat{R}_{\textrm{coh}}\rangle=
\sqrt{2}\left(
\begin{array}{c}
|\alpha|\cos\theta\\
|\alpha|\sin\theta
\end{array}
\right) ,
\end{eqnarray}
where it is understood that the first entry corresponds to $\langle \hat{x} \rangle$ and the second entry, $\langle \hat{p}\rangle$. The covariance matrix for each of these states is given by,
\begin{eqnarray}
\sigma_{\textrm{vac}}=\sigma_{\textrm{coh}}=
\left(
\begin{array}{cc}
1 &0\\
0 &1
\end{array}
\right) ,\quad
\sigma_{\textrm{th}}=
(2n_{\textrm{th}}+1)\left(
\begin{array}{cc}
1 &0\\
0 &1
\end{array}
\right).
\end{eqnarray}
Note that the coherent state carries its statistics \textit{only} in its mean, while the thermal state carries its statistics only in its covariance. Much like the process described earlier, we now need a way to evolve these two parameters (mean and covariance) through various optical elements. The transforms that act on these parameters are the same as we used earlier, with the only difference being in how the covariance evolves. Both mean and covariance evolve according to,
\begin{equation}
f(\beta)\cdot \avg{\hat{R}_i}=\avg{\hat{R}_f},\quad
f(\beta)\cdot\sigma_i\cdot f(\beta)^{\top}=\sigma_f,
\end{equation}
where $f(\beta)$ is given by the same optical element transforms shown in Eq.~(\ref{eq:disp})-(\ref{eq:sqz2}). At this point, our mean and covariance have evolved through various optical elements and we now have the statistics at the output. If so desired, we could now construct the Wigner function at the output with Eq.~(\ref{eq:wiggauss}) which would give the same Wigner function with the previous evolution method, described above. For completeness, we will again use the example of our MZI from Figure~\ref{fig:MZI} and list the mean and covariance, at the output, as our example input states, a coherent state and squeezed vacuum both maintain Gaussian form. This choice of input states and the fixed topology of our MZI evolve the mean and covariance to,
\begin{eqnarray}
\avg{\hat{R}_f}= \left(
\begin{array}{c}
\sqrt{2} |\alpha|\cos(\phi/2)\\
0\\
0\\
\sqrt{2} |\alpha|\sin(\phi/2)
\end{array} \right)
\end{eqnarray}
\begin{eqnarray}
\sigma_f= \left(
\begin{array}{cccc}
\gamma e^{-r} & 0 & 0 & e^{-r}\sin\phi\sinh r\\
0 & \lambda e^{r} & e^{r}\sin\phi\sinh r & 0 \\
0 & e^{r}\sin\phi\sinh r &\gamma e^{r} & 0 \\
e^{-r}\sin\phi\sinh r & 0 & 0 & \lambda e^{-r}
\end{array} \right),
\end{eqnarray}
where $\gamma=\cosh r +\cos\phi\sinh r$ and $\lambda=\cosh r -\cos\phi\sinh r$, where again we have fixed $\theta_{\textrm{coh}}=\theta_{\textrm{sqz}}=0$.

\pagebreak
\singlespacing
\chapter{Quantum Measurement}
\doublespacing
\section{Measurements using Wigner functions}
Now that we have shown ways in which one can evolve various states of light through an MZI, we can discuss the role of the detectors. While there are many different types of detectors, instead of restricting to specific physical processes of detection, we describe the detection process in terms of quantum operators, shown in  Eq.~(\ref{eq:meas}). In general, measurement operators are computed, in terms of Wigner functions, by utilizing \cite{Adesso2014},
\begin{equation}
\textrm{Tr}[\rho \hat{O}]=\int_{\mathbb{R}^{2N}}W_\rho^0(\kappa)\overset{\sim}{f}(\kappa) d^{2N}\kappa,
\label{eq:meas}
\end{equation}
where $\overset{\sim}{f}(\kappa)$ is a symmetrically ordered function of phase space operators, $\hat{O}=f(\hat{a}_k,\hat{a}_k^\dagger)$. A significant difference between the density matrix approach and Wigner functions is the requirement of symmetric ordering. Using Wigner functions, all measurements are assumed to be symmetrically ordered in their field operators and therefore specific measurements must take this symmetric ordering into account. For example, a typical measurement operator, the number operator (also defined as an intensity measurement), given by $\hat{n}=\hat{a}^\dagger\hat{a}=1/2(\hat{x}^2+\hat{p}^2)$ is not symmetrically ordered. In order to symmetrize this operator, we write it in symmetric form as,
\[\{\hat{a}^\dagger,\hat{a}\}_{\textrm{sym}}=\frac{1}{2}(\hat{a}^\dagger\hat{a}+\hat{a}\hat{a}^\dagger)\]
and utilize the commutation relations shown in Eq.~(\ref{eq:comm1}) to obtain,
\[\hat{n}=\{\hat{a}^\dagger,\hat{a}\}_{\textrm{sym}}-\frac{1}{2}
\]and therefore have that a measurement of intensity on a single mode Wigner function, in terms of symmetric ordering is given by
\begin{equation}
\langle \hat{n}\rangle= \frac{1}{2}\int(x^2+p^2)W(x,p)dxdp -\frac{1}{2},
\label{eq:num1}
\end{equation}
and its second moment, which requires similar, but more complicated symmetric ordering,
\begin{equation}
\langle \hat{n}^2\rangle= \frac{1}{4}\int(x^2+p^2)^2W(x,p)dxdp -\langle \hat{n}\rangle-\frac{1}{2}.
\label{eq:num2}
\end{equation}
While this symmetrization is always required when using Wigner functions, it is a straightforward process. Note that this requirement is needed for any products of the field operators, but a typical homodyne measurement, i.e. $\hat{x}=1/\sqrt{2}(\hat{a}+\hat{a}^\dagger)$, is already symmetrically ordered. 

Also of typical use as a state characterization tool is constructing the photon number distribution for a given state. This is physically done by running many trials of an experiment and performing number counting at either detector, each trial giving a certain number of photons. Over many trials then, we can construct the probabilities of the state having any number of photons. When using density matrices, if one works in the number basis, then this distribution is essentially calculated for free but luckily using Wigner functions, we can also obtain this distribution with a relatively simple calculation given by constructing a generating function and differentiating it according to,
\begin{equation}
\begin{split}
G(l)&=\frac{2}{1+l}\int \textrm{exp}\left[\frac{l-1}{l+1}(x_1^2+p_1^2)\right]W(x_1,p_1) dx_1 dp_1\\
P(n)&=\frac{1}{n!} \frac{\partial^nG(l)}{\partial l^n}\Big{|}_{l\rightarrow 0},
\end{split}
\label{eq:numdist}
\end{equation}
where $G(l)$ is our generating function, $P(n)$ our photon number distribution and it's clear that $n \in \mathbb{Z}$, since $P(n)$ is necessarily discrete. This construction allows us to calculate the photon number distribution for any state of light and use it to characterize the properties of this state, which we will utilize when we analyze our full examples later in this dissertation. This construction also demonstrates that we can recover the discrete nature of the quantized nature of photons, even when describing them in a continuous variable space.

\begin{figure}[!htb]
\centering
\includegraphics[width=0.6\columnwidth]{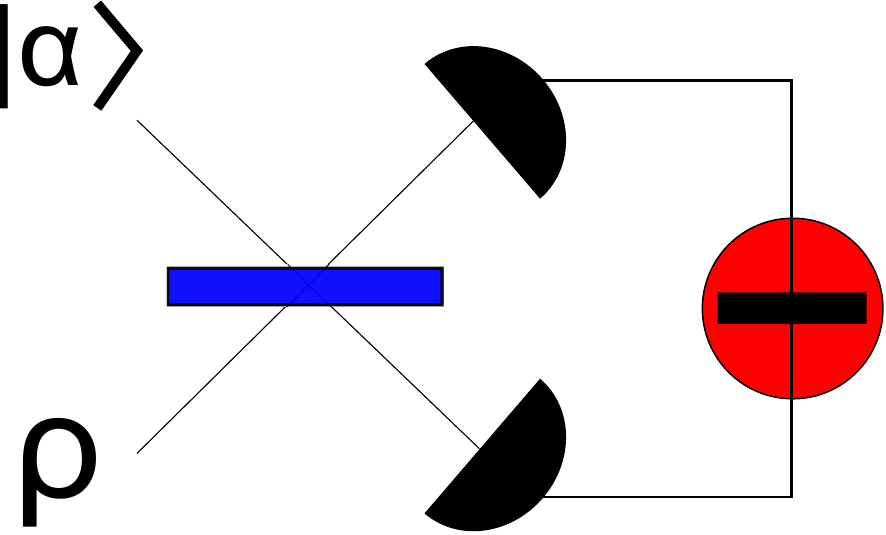}
\caption{A typical balanced homodyne measurement is performed with a target beam incident on a 50-50 beam splitter along with a coherent state. The two detectors then perform an intensity difference measurement.}
\label{fig:homod1}
\end{figure}

Investigating the so-called balanced homodyne measurements further, we see that it is typically discussed in terms of its implementation, that is, the target beam, is incident on a 50/50 beam splitter along with a coherent state (or commonly referred to as a local oscillator) of the same frequency as the target beam (typically this coherent state is derived from the same source as the input state). After the beam splitter, the two outputs are collected on detectors and an intensity difference measurement is performed. This process is shown in Figure~\ref{fig:homod1}.
In practice, the phase of the coherent state is adjusted, allowing a homodyne measurement along any arbitrary direction in phase space, $\hat{R}(\theta)=\frac{1}{\sqrt{2}}(\hat{a}e^{-i \theta}+\hat{a}^\dagger e^{i \theta})=(\hat{x}\cos\theta+\hat{p}\sin\theta)$. Therefore this process amounts to nothing but an attempt to measure the mean value of any arbitrary superposition of the state in phase space. This allows us to notice that we do not need to fully model this measurement scheme as in Figure~\ref{fig:homod1}, but can instead simply calculate any homodyne measurement as $\avg{\hat{R}(\theta)}=\avg{\hat{x}}\cos\theta+\avg{\hat{p}}\sin\theta$, on the state represented by $\rho$. For simplicity, if we consider one particular homodyne measurement, along  the $x$ direction then a specific homodyne measurement, along with its variance, can be calculated, using Wigner functions with,
\begin{equation}
\langle \hat{x}\rangle=\int x W(x,p)dxdp, \quad \avg{\hat{x}^2}=\int x^2 W(x,p)dxdp, \quad (\Delta x)^2= \avg{\hat{x}^2}-\avg{\hat{x}}^2
\label{eq:homod1}
\end{equation}
This technique also gives us some insight into the usefulness of a homodyne measurement, in terms of our Gaussian information techniques described earlier. Since this technique works directly in terms of mean and covariance, it's clear that homodyne measurements are essentially calculated for free with this treatment, as the output state's mean vector is exactly what a homodyne measurement attempts to capture. 

We can also notice that with a bit of work, we can use the Gaussian techniques to easily calculate an intensity measurement. This can be done by utilizing
\begin{equation}
\avg{\hat{n}}=\avg{\hat{a}^\dagger\hat{a}}=\frac{1}{2}(\avg{\hat{x}^2}+\avg{\hat{p}^2})=\frac{1}{2}((\Delta\hat{x})^2+\avg{\hat{x}}^2+(\Delta\hat{p})^2+\avg{\hat{p}}^2),
\label{eq:numgauss}
\end{equation}
where we have used the relation $(\Delta\hat{R})^2=\avg{\hat{R}^2}-\avg{\hat{R}}^2$. In the proper, symmetrical form, this can be written in terms of mean and covariance as,
\begin{equation}
\avg{\hat{n}}=\frac{1}{2}\left(\frac{\textrm{Tr}[\sigma]}{2}+\avg{\hat{x}}^2+\avg{\hat{p}}^2-1
\right).
\label{eq:numgauss2}
\end{equation}
Note that $\avg{\hat{R}}$ are quantities already calculated in this Gaussian treatment, and so we need only square them. Also of significance is that $\sigma$ is only a 2$\times$2 matrix (not infinite like for a density matrix) and its trace is trivial. Therefore, this detection scheme is also calculated very simply when using this Gaussian treatment. Calculating the second moment of this operator using Gaussian techniques, while possible, is not as simple as one would perhaps desire. If we assume $\avg{\hat{R}}=0$ for the chosen input state of light, this calculation becomes tractable, but its advantages over using the Wigner becomes less apparent. With this observation, we then can say, that Gaussian information techniques can greatly simply some calculations, when particular input states are assumed (Gaussian) and particular measurement schemes are considered (mainly homodyne). Contrasted with this Gaussian only treatment, we can then see that Wigner functions provide a completely general approach to handle any types of input state and any types of measurements.

Another typical measurement choice is one that we briefly mentioned earlier, an intensity difference measurement defined by $\avg{\hat{n}_1}-\avg{\hat{n}_2}$, where it is clear this is now a measurement on two spatial modes. However, one can easily see that this detection scheme is a direct adaptation of the single mode intensity measurement. Specifically this measurement can be calculated, in terms of Wigner functions, as,
\begin{equation}
\avg{\hat{n}_1}-\avg{\hat{n}_2}=\frac{1}{2}\left(
\int (x_1^2+p_1^2)W(x_1,p_1) dx_1 dp_1-\int  (x_2^2+p_2^2)W(x_2,p_2) dx_2 dp_2
\right)
\label{eq:numdiff}
\end{equation}

As a final example of another measurement choice, we consider the single mode parity measurement, defined as $\hat\Pi=(-1)^{\hat{a}^\dagger\hat{a}}$. Under this choice, we are able to utilize another benefit of describing our system in terms of Wigner functions, since it has been shown \cite{Plick2010} that the parity measurement satisfies $\avg{\hat\Pi}=\pi W(0,0)$, or in words, the expectation of the parity measurement is given by the value of the Wigner function, at the origin of phase space. From this property along with the property, $\avg{\hat{\Pi}^2}=1$, so $(\Delta\Pi)^2=1-\avg{\Pi}^2$, the parity measurement, is perhaps the simplest to calculate of the choices presented here, as an integral is not required.

\begin{figure}[!htb]
\centering
\includegraphics[width=0.95\columnwidth]{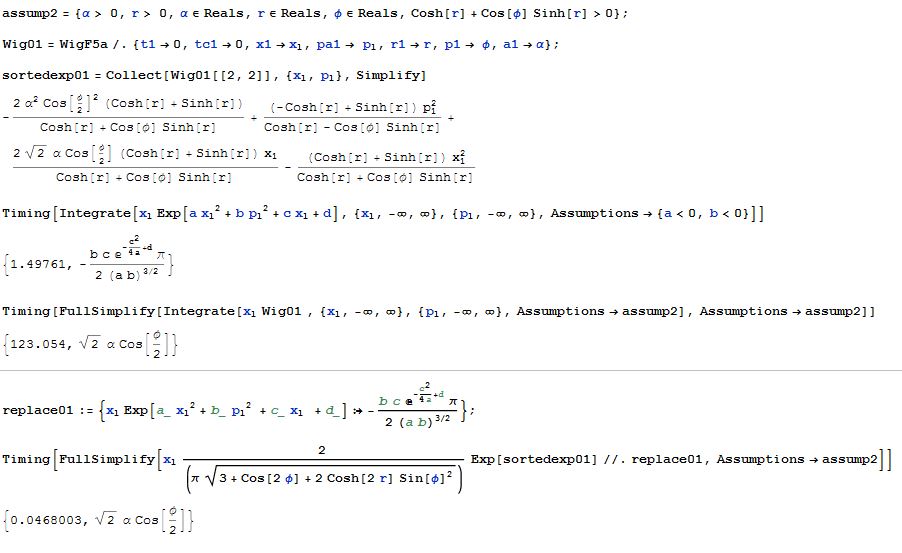}
\caption{Example of replacement method instead of directly integrating our Wigner function. Outputs show the time taken to calculate (in seconds) as well as the output. The fourth line from the bottom shows the direct integration time, along with its result, while the last line shows our replacement methods time and its result.}
\label{fig:code1}
\end{figure}

These cover some of the common types of measurements and while, in principle, there are many more, the methods discussed here showcase some of the benefits and properties of utilizing Wigner functions and Gaussian information. It seems only fair then to discuss some of the difficulties with using Wigner functions. Likely, the same difficulties exist for other, analytical methods, but these are computational issues and not problems with the construction itself. We have shown how a general quantum measurement can be calculated using Wigner functions, which involves integrating the quantum operator against the output Wigner function. In general we have seen that these Wigner functions are typically exponentials and can, depending on the input states, have very complicated exponents. We assume one would calculate these integrals using software such as Mathematica and the following uses formatting and language to follow this software, but many other software likely have a similar form. In the case of these complicated integrals, even with sufficient assumptions within the software, the computational time required to complete these integrals can be lengthy (greater than a day on a relatively powerful desktop PC). Instead of moving to a higher end PC (or access to a super computer), we note a technique that allows us to calculate such intractable integrals. This is only due to inefficiencies in how the software handles complicated integrals but is useful for our purposes and therefore relevant to include in our discussion. Instead of integrating something such as, $\int f(\beta) W(\textbf{X})d\textbf{X}$, a generic quantum measurement, where $W(\textbf{X})$ is given, for example, by $e^{f(|\alpha|, \theta_{\textrm{coh}}, r, \theta_{\textrm{sqz}}, \phi, x_1, p_1)}$, we instead form a list of replacement rules. First we group our Wigner functions exponent into terms dependent on the phase space operators $x_1, p_1$, with Mathematica's ``Collect" command. For example, if our Wigner function is named ``Wig01" then we would perform
\[
\textrm{Collect}[\textrm{Wig01}[[2,2]],\{x_1, p_1\},\textrm{Simplify}],
\]
where ``[[2,2]]" simply selects only the exponent of Wig01 (assuming the exponent is in the list position 2,2). With the form of this collection, we then see the form that our Wigner function takes and can integrate a general exponential of this form and create a replacement rule for this integration. In principle, integrating a general exponential of the same form as our Wigner function shouldn't provide any benefit, as the only difference is in what we consider constants, but the software seems to not understand this difference well. Once we have the form of the exponential, we integrate a general form and create a replacement rule for this form. Then, instead of integrating our specific Wigner function, we instead use the replacement rule with the formatted Wigner function. The collection of this process with a specific choice of a coherent state and squeezed vacuum, as input, and homodyne measurement is shown in Figure~\ref{fig:code1}. Here we show an example of this code as well as the times (in seconds) shown for the various calculations and while this calculation is easily manageable in either treatment, one can see the benefit of using our replacement method when the calculation times begin to become cumbersome. Note that the total time to calculate the general integral, perform the replacement, and simplify is $\tau_{\textrm{replace}}\simeq 1.53 s$, while a direct integration and simplification is $\tau_{\textrm{direct}}\simeq 123.05 s$, approximately 80 times longer. The main drawback of this method is the organization required for composing the specified Wigner function to a general form, but one can imagine forming a large database of these various forms which could alleviate some of the overhead of this process. In even more complicated cases, this replacement method still maintains its benefit over the direct integration method but typically requires more prep-work, depending on the chosen measurement scheme.

\section{Phase Measurement}
In the previous section we discussed how to utilize Wigner functions to perform quantum measurements.  In this section we will show how these measurements apply to a typical goal in metrology, performing a phase measurement \cite{Bela1995,Shapiro1991,Bajer1996}. As shown in Figure~\ref{fig:MZI}, we can use a MZI, along with an unknown phase $\phi$, to represent the unknown effect of a material (perhaps a material with an unknown index of refraction) interrogated inside the MZI. While the exact effect of this material can vary, depending on the physical situation, we can model the effects of this material as an unknown variable, $\phi$. If this is the model one uses, then there are some typical benchmarks we can discuss.

The so-called Shot Noise Limit (SNL) is typically defined for the MZI in terms of the total average photon number which enters it \cite{LIGO_SNL,Kurn2014}. First developed by Caves \cite{Caves1981} in a fixed setup of sending a coherent state and vacuum into an MZI, the SNL is defined by $(\Delta\phi)_{SNL}^2=1/\avg{\hat{n}}$. Using the example from previous sections, if we consider sending a coherent state and squeezed vacuum into an MZI, we then have, $(\Delta\phi)_{SNL}^2=(|\alpha|^2+\sinh^2r)^{-1}$. This sets a boundary on the variance of our unknown phase measurement distinguishing classical techniques from quantum techniques. A similar limit, the Heisenberg limit, while not a hard limit, is typically also used as a reference and defined by $(\Delta\phi)_{HL}^2=1/\avg{\hat{n}}^2$. This Heisenberg limit should be considered with caution as it is obtained from the uncertainty relation $(\Delta\phi)^2 (\Delta n)^2\geq 1$ and assuming that $(\Delta n) \leq \avg{\hat{n}}$, so that, $(\Delta n)^2 \leq \avg{\hat{n}}^2$. However, this assumption is violated for certain states of light, such as the two-mode squeezed vacuum, and it has been shown that this state indeed can achieve a phase estimate below the HL \cite{TMSV1}. Spefically, the HL is technically only a hard limit for states of \textit{definite} photon number (such as Fock states \cite{bib:GerryKnight05,Fried1953}) but not necessarily a hard limit for states of \textit{average} photon number (all Gaussian states \cite{Adesso2014} as well as some non-Gaussian). So while we may use the HL as a reference, it should be considered carefully.

There are two principal ways in which we can calculate the variance of our phase measurement. The first way one can show the phase variance of a chosen measurement is to use error propagation \cite{Goodman1960,Ku1966,Clifford1973} in the form of,
\begin{equation}
(\Delta\phi)^2=\frac{(\Delta\hat{O})^2}{|\partial\avg{\hat{O}}/\partial\phi|^2},
\label{eq:errorprop}
\end{equation}
where $\hat{O}$ is a chosen quantum operator. The use of this formula connects the variance of our unknown phase, to any quantum operator, with the utilization of the Taylor expansion according to,
\begin{equation}
\begin{split}
f &\approx  f^{0}+\frac{\partial f}{\partial \phi} \phi+\ldots\\
(\Delta f)^2&\approx \left|\frac{\partial f}{\partial \phi}\right|^2(\Delta \phi)^2,
\end{split}
\end{equation}
where $f$ is a non-linear function, and we have assumed that $\phi$ is the only parameter to be estimated and is uncorrelated with any other variables. This gives us a variance of $\phi$, a kind of quality of our measurement. In general, one particular quantum operator can outperform, giving a smaller phase variance, another quantum operator, and therefore the process of searching for the optimal measurement scheme can be very challenging. Note that in order to utilize Eq.~(\ref{eq:errorprop}), we require the first and second moments of the chosen operator, since $(\Delta\hat{O})^2= \avg{\hat{O}^2}-\avg{\hat{O}}^2$. This requirement may be trivial, as in the case of the parity operator or may be fairly complicated, as in the case of the intensity difference operator.

An alternative to using this error propagation treatment is with the use of Classical Fisher Information (CFI) \cite{Prok2011,Casella2003}. Instead of considering a specific quantum measurement, we instead consider probabilities of events (though, certain quantum measurements lead to probabilistic events). For example, the intensity measurement gives a measurement of the average number of photons entering a detector. A particular question we could ask in this case is,``What is the probability of the detector receiving exactly one photon?" The probability distribution for this case can then be used with the relation,
\begin{equation}
(\Delta\phi^2)^{-1}=\textrm{CFI}=\sum_{i}\frac{1}{P_{i}}\left(\frac{dP_i}{d\phi}\right)^2,
\label{eq:cfi1}
\end{equation}
with the condition that $\sum_{i} P_i=1$, that is the $P_i$'s represent  complete probabilistic events. Returning to our brief example, if we let $P_1$ be the probability of our detector detecting a single photon, then,
\[
(\Delta\phi^2)^{-1}=\frac{1}{P_1}\left(\frac{dP_1}{d\phi}\right)^2+\sum_{i}\frac{1}{P_{i+1}}\left(\frac{dP_{i+1}}{d\phi}\right)^2=
\frac{1}{P_1}\left(\frac{dP_1}{d\phi}\right)^2+\frac{1}{1-P_1}\left(\frac{dP_1}{d\phi}\right)^2,
\]
where we have used that $\sum_{i} P_{i+1}=1-P_1$. In order to obtain the probability distributions in question, we again use our Wigner functions and form projective measurements. In this case, the probability of a detector receiving exactly one photon is given by the projection of the Wigner function onto the single photon subspace. This is achieved by performing the measurement of \cite{Kim2008},
\begin{equation}
P_1=2 \int (2p_1^2+2x_1^2-1)e^{-x_1^2-p_1^2} W(x_1,p_1) dx_1 dp_1= 2 \pi \int F_1(x_1,p_1) W(x_1,p_1) dx_1 dp_1,
\label{eq:project1}
\end{equation}
where $F_1(x_1,p_1)$ is the Wigner function for the single photon Fock state, shown in Eq.~(\ref{eq:fockwig}). This implementation shows that we have the option of constructing probability distributions through projective measurements, which may be simpler or more physical than a particular quantum operator. Another important note is that of so called click detection (on/off detection), which a typical APD performs, meaning it responds to the presence of \textit{any} number of photons, but cannot discriminate the number of photons that is present. In this case, we can model this APD by the projection onto the subspace of all photon numbers states, other than vacuum, by using the projection $P_c\equiv 1-F_0$. This would give us the probability distribution of a detector receiving any number of photons, other than zero, just as the APD measures. While projection treatment is perhaps advantageous in that it does not require the second moment calculations, it instead requires probability distributions and, typically, projective measurements. It is worth noting that some quantum operators themselves are also probability distributions, such as in the case of the parity operator and therefore either method can be used to obtain the phase variance. This treatment also suffers the same problem as in the error propagation; however, it has many possible choices of probabilities and thus a reasonable question to ask is, which one gives the smallest phase variance?

In order to show that we have obtained the smallest phase variance possible, we construct the Quantum Cram\'{e}r Rao Bound (QCRB), through the Quantum Fisher Information (QFI) \cite{Caves1994,Nori,QFI1,Jing2013,Jing2015}. This treatment differs from the previous in that it will \textit{not} depend on a measurement choice, and instead gives the best phase variance bound possible with \textit{any} possible measurement choice allowed by quantum mechanics (meaning it is described by a positive operator valued measure (POVM)). In general this quantity is difficult to calculate and while it gives us the ultimate lower bound on a phase variance measurement, as mentioned, it does not depend on the measurement choice and therefore does not tell us the ``optimal" measurement directly, merely the best possible with any measurement. In order to show the optimality of a measurement then, we separately calculate the phase variance obtained from a specific measurement and compare it to the QCRB. While it is not an exclusive bound (meaning multiple measurement may achieve the QCRB), if a chosen measurement achieves the QCRB, then no other measurement can outperform this. A particular measurement which reaches the QCRB is then said to be optimal, but it may have many technical challenges in actually implementing and a full noise model is useful at this point to characterize the effect of various noise sources on various measurements, which we discuss in later sections.

While there are a variety of ways to calculate the QFI and QCRB, most apply easily to some specific class of states but not others, or may be widely applicable but very calculation intensive when exotic states of light are considered. Here we will focus on the calculation of the QFI through Gaussian information as well as directly from the Wigner function. The Gaussian information method applies to any Gaussian state, pure or mixed, while using the Wigner function applies only to pure states. In terms of Wigner functions, we may calculate the QFI of any pure state according to \cite{Braun2},
\begin{equation}
(\Delta\phi^2_{\textrm{QCRB}})^{-1}=\textrm{QFI}=2(2\pi)^M\int \left(\frac{\partial W(\textbf{X})}{\partial\phi}\right)^2 d^{2M}\textbf{X},
\label{eq:wigqfi}
\end{equation}
where $M$ is the number of spatial modes. It should be noted, as we mentioned earlier with our rule replacement method, integrals of Wigner functions can be computationally intensive at times, but the rule replacement method works equally well for this calculation. While this method seems deceivingly simple, in practice, even with the discussed methods, computing this integral with a relatively complicated Wigner function can be quiet challenging. If we further assume that our state is pure and Gaussian then we can find that,
\begin{equation}
(\Delta\phi^2_{\textrm{QCRB}})^{-1}=\textrm{QFI}=\left(\frac{\partial R}{\partial \phi}\right)^\top\sigma^{-1}\frac{\partial R}{\partial \phi}+ \textrm{Tr}\left[\left(\frac{\partial \sigma}{\partial \phi}\sigma^{-1}\right)^2\right],
\label{eq:puregaussqfi}
\end{equation}
where $R,\sigma$ are the mean and covariance, respectively. Note that this definition only differs slightly from \cite{Braun2}, as we use different definitions of quadratures, $\hat{x},\hat{p}$. If we relax the condition that the state must be pure, but retain the requirement of Gaussian form, we can still find that the QFI is manageable using Gaussian information. Specifically, the QCRB in this case takes the form \cite{Lee2014},
\begin{equation}
(\Delta\phi^2_{\textrm{QCRB}})^{-1}=\textrm{QFI}=\frac{1}{2}\textrm{Tr}\left(\partial_\phi \sigma\left[\sigma (\partial_\phi \sigma)^{-1}\sigma^\top+\frac{1}{4}\Omega(\partial_\phi \sigma)^{-1}\Omega^\top\right]^{-1}\right)+(\partial_\phi R)^\top \sigma^{-1}\partial_\phi R,
\end{equation}
where $\Omega$ is the symplectic matrix defined by
\begin{eqnarray}
\Omega=\bigoplus_{j=1}^M \left(
\begin{array}{cc}
0 & 1\\
-1 & 0
\end{array}
\right) .
\end{eqnarray}
We should also note that this construction is valid after a change of basis, as it proves to be to a computational advantage, with the change of basis according to,
\begin{equation}
\begin{split}
\sigma &\rightarrow \frac{1}{2}H\cdot\sigma_{x,p}\cdot H^{\dagger}\\
R &\rightarrow H\cdot R_{x,p}\\
H &=\frac{1}{\sqrt{2}}\bigoplus_{j=1}^M  \left(
\begin{array}{cc}
1 & i\\
1 & -i
\end{array} \right) .
\end{split}
\end{equation}
This calculation of the QCRB is particularly useful since the only requirement is the Gaussian form, but can accommodate mixed states, which include the classical thermal state. This thermal state is crucial in some noise models \cite{DeSalvo2012,Zmu2003} and necessarily takes any pure state to a mixed state when thermal noise is considered in the noise model and therefore this calculation is particularly useful when considering realistic noise models.

\section{Noise Modeling}
\subsection{Photon Loss}
\label{sec:photonloss}
For any accurate model, one must always consider the effect of various sources of noise. Here, we will discuss the modeling of various sources of loss and noise, typically found in interferometers. This includes photon loss to the environment \cite{Lassen2010}, inefficient detectors \cite{Hogg2014}, phase drift, and thermal noise \cite{DeSalvo2012,Zmu2003}. In general, each of these can be mitigated through various techniques, but not completely eliminated and therefore, in our attempt to model a realistic interferometer, we must have a way to model these unavoidable effects.

\begin{figure}[!htb]
\centering
\includegraphics[width=0.45\columnwidth]{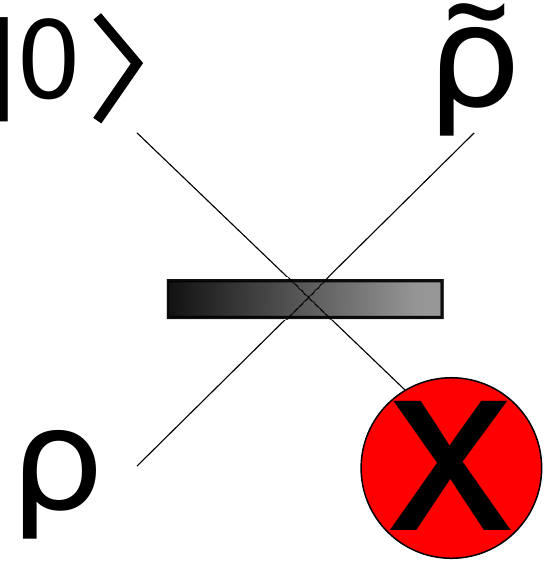}
\caption{A model of linear photon loss by passing a general state of light, $\rho$, through a variable transmissivity beam splitter, with vacuum as the second input. One of the output modes is then traced over to represent the loss of photons which then results in a lossy state, $\overset{\sim}{\rho}$.}
\label{fig:loss}
\end{figure}

First we consider photon loss in the model by way of two mechanisms, photon loss to the environment inside the interferometer and photon loss at the detectors, due to inefficient detectors \cite{bib:GerryKnight05}. Both of these can be modeled by placing a fictitious beam splitter, of variable transmissivity, inside the interferometer with vacuum and a interferometer arm as input and tracing over one of the output modes, to mimic loss of photons to the environment. This process is shown in Figure~\ref{fig:loss}.

These linear photon loss mechanisms manifest themselves as a simple change of variables with respect to the average photons. Using our example input states from before, this amounts to a change of variables of $|\alpha|\rightarrow \sqrt{D(1-L)} |\alpha|$ and $r \rightarrow \sinh^{-1}(\sqrt{D(1-L)} \sinh{r})$, where $0\leq D \leq 1$ is the detector efficiency (in decimal), is assumed to be equal for all detectors and $0\leq L \leq 1$ is the photon loss inside the MZI, assumed to be equal in both arms. This variable replacement greatly simplifies calculations over the full model of inserting many fictitious beam splitters and our assumption of equal losses and equal detector efficiencies is fairly reasonable if one reasons that both arms of the interferometer are in similar media and both detectors are identical. It is clear that the value of the transmissivity of the variable beam splitter, $L$, controls the amount of loss, in decimal, to the environment. This process of variable replacement can be generalized to other input states with a similar variable replacement condition of simply, $\bar{n}\rightarrow D(1-L)\bar{n}$, where $\bar{n}$ is the average photon number in each spatial mode. If our assumption of equal losses in both arms and identical detector inefficiencies is relaxed, then a full calculation becomes necessary. 

\subsection{Phase Drift}
\label{sec:phasedrift}

Another common effect in MZI's is the random drift of phase. The way in which we treat phase drift, to the best of our knowledge, has not been presented elsewhere. Normally, we assume the unknown phase to be a fixed value, but in practice it may vary slightly over many experimental trials; we call this effect, phase drift. Generally, when we calibrate our MZI, we would place a control phase in one arm of the interferometer, which allows us to tune the interference between the two arms. We could assume we have infinite precision with our control phase, but in practice, whatever mechanism controls the value of our control phase, it can drift slightly over many experimental trials. In principle, we attempt to set this control phase to an optimal value; in order to give rise to a measurement with the best statistics, however, the control phase value will vary around this optimal phase setting. For this reason we aim to show this phase draft in a more mathematical way and therefore we can use the analytical forms of the various measurement phase variances, as a function of unknown phase, $\phi$, and simulate phase drift by computing a running average of the phase variance, with a pseudo-randomly chosen phase, near the optimal phase, for each measurement. This is accomplished in the following way: we find the true optimal phase (typically a multiple of $\pi$) and allow a pseudo-random number generator to choose a phase near this optimal phase (within 20\% above or below the optimal value), which we use in place of the optimal phase. We perform this pseudo-random process over many trials, each trial forming the average of the choices from previous trials. In this way, after many trials, our random choosing approaches the true optimal phase. This mimics the idea of the experiment in that, a single measurement provides very little information on the unknown phase, $\phi$, and it is only with many measurements that we can say we have obtained a good estimate of the unknown parameter. The behavior of various detection schemes under the effects of phase drift are not necessarily identical, as we show in later sections, some measurement schemes, specifically those that utilize multiple spatial modes, perform better than single mode measurement schemes, under phase drift.

\subsection{Thermal Noise}
\label{sec:thermalnoise}

\begin{figure}[!htb]
\centering
\includegraphics[width=0.5\columnwidth]{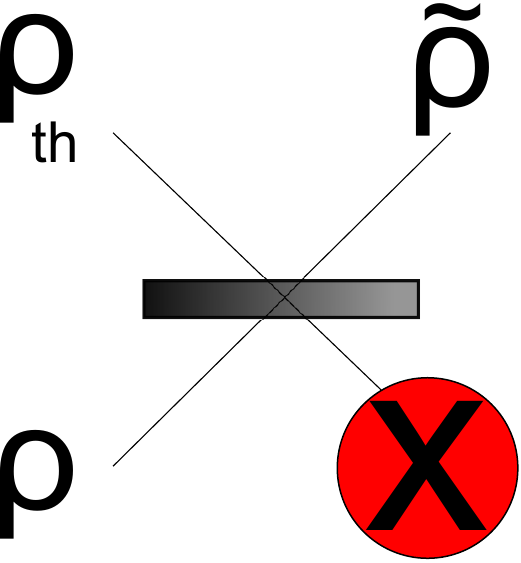}
\caption{Model of the interaction of a general state $\rho$, inside the MZI, with the environment, represented by a thermal state, $\rho_{th}$.}
\label{fig:thm1}
\end{figure}

In addition to photon loss, detector efficiency, and phase drift, we also model the inevitable interaction with thermal noise from the environment \cite{DeSalvo2012,Zmu2003}. This is accomplished much in the same way as a photon loss model, but here we consider a thermal state incident on a fictitious beam splitter, on both arms of the interferometer, inside the interferometer and trace out one of its output modes, shown in Figure~\ref{fig:thm1}. This allows a tunable amount of thermal noise (by changing the average photon number in the thermal state), into the interferometer. Again, as we will see in a later section, the effects of this injection of thermal photons can vary, depending on the measurement scheme that is considered.

\pagebreak
\singlespacing
\chapter{Photon Addition and Subtraction}
\doublespacing
\section{Photon Addition and Subtraction}
\subsection{Photon Addition and Subtraction with Wigner functions}
\label{sec:photonsub}
We have discussed much of the general way one can use Wigner functions to model an MZI and the various benchmarks one uses to qualify many of the typical measurement choices. We will now discuss the modeling of an interesting, but less typical consideration in MZIs, the use of addition or subtraction of photons from a beam of light. First proposed by Agarwal and Tara \cite{Agar1,Agar2}, an implementation of noiseless amplification, photon addition and subtraction has received much attention over the past decade \cite{Zavatta2004a,Zavatta2005a,Braun,Gerry2012a,Gerry2014a,Zavatta2011,Josse2006}. While there is some choice on how to exactly model the operation of photon addition and subtraction, we first suggest a word of caution. As this is the realm of \textit{probabilistic} noiseless amplification, we should not overlook the fact that we must consider the probabilistic nature of this process.

One way that some have chosen to model the operator of photon addition and subtraction is with the use of the photon creation and annihilation operators. Using these operators, we can write the addition of a photon, in terms of a density matrix, as,
\begin{equation}
\hat{\rho}_+=\hat{a}^\dagger\hat{\rho}\hat{a}/N_+,
\end{equation}
where $N_+=\textrm{Tr}(\hat{a}\hat{a}^\dagger\hat{\rho})=N+1$ and $N$ is the average photon number, prior to photon addition, and shows that this process must also be properly re-normalized. Similarly, we can write photon subtraction as,
\begin{equation}
\hat{\rho}_-=\hat{a}\hat{\rho}\hat{a}^\dagger/N_-,
\end{equation}
where $N_-=\textrm{Tr}(\hat{a}^\dagger\hat{a}\hat{\rho})=N$. However, we would of course like this representation, in terms of Wigner functions, therefore with some work we obtain, \cite{Braun}
\begin{equation}
W(\textbf{X})_{\pm}=\frac{1}{2}\left(x_i^2+p_i^2\mp x_i \partial_{x_i} \mp p_i \partial_{p_i}+\frac{1}{4}(\partial_{p_i}^2+\partial_{x_i}^2)\pm1\right)W(\textbf{X})/N_\pm,
\end{equation}
with $W(\textbf{X})$ representing the Wigner function to be photon added or subtracted in the $i^{th}$ spatial mode. All the previously presented methods are valid for any state of light, but we shall refer to this construction as the mathematical treatment of photon addition or subtraction, as it is mathematically correct, but does not necessarily faithfully reproduce all the aspects of photon addition and subtraction, as we will show in the following discussion.

The use of the mathematical description of photon addition and subtraction may be sufficient, depending on the specific model one is considering, but if a full model of the effects of this process are desired, then an alternative method should be used. This alternative method consists of using a physical process, along with projective measurements, to implement photon addition and subtraction.

\begin{figure}[!htb]
\centering
\begin{subfigure}
{\includegraphics[width=0.3\columnwidth]{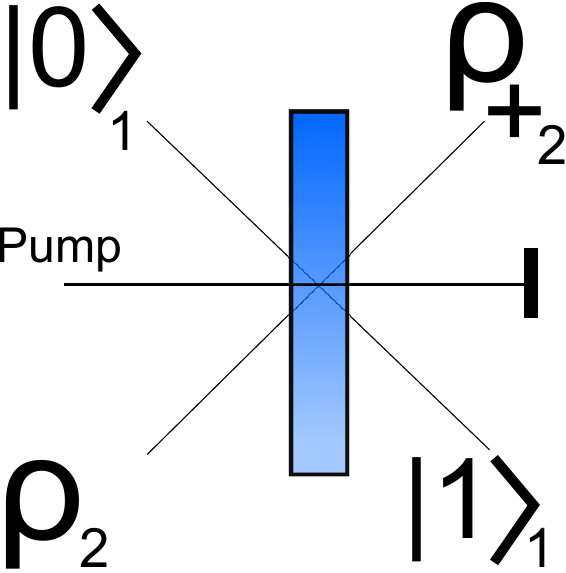}}
\caption{SPDC model of photon addition. A general state, $\rho$ and vacuum are incident on a pumped non-linear crystal. Occasionally, depending on the pump strength, this crystal emits photon pairs into the two output modes. One output mode is used to herald that the other mode has been photon added.}
\label{fig:SPDC}
\end{subfigure}
\vspace{1em}
\begin{subfigure}
{\includegraphics[width=0.3\columnwidth]{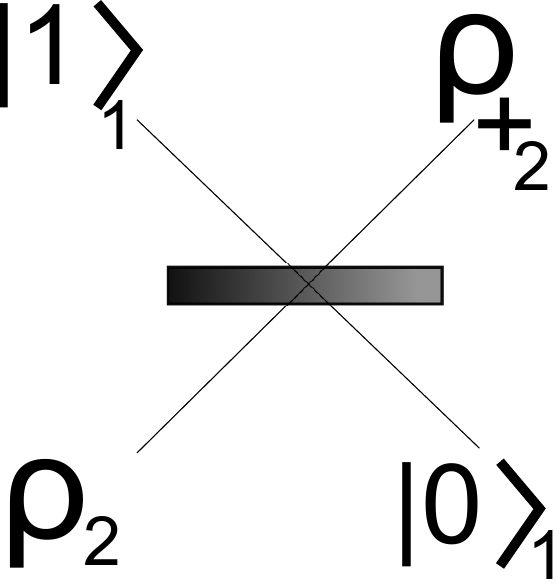}}
\caption{Beam splitter model of photon addition. A general state, $\rho$ and a single photon Fock state are incident on a variable transmissivity beam splitter. On the condition that one output receives no photons, the remaining mode is known to be photon added.}
\label{fig:BSadd}
\end{subfigure}
\end{figure}

For photon addition we can consider a few different implementations, but here briefly highlight two implementations. One uses Spontaneous Parametric Down Conversion (SPDC) \cite{Burnham1970,Ghosh1987,Kwiat1995} to facilitate photon addition while the second model uses a variable transmissivity beam splitter, along with a single photon source. While these two processes involve physically different optical elements, both of these models accurately reproduce the mathematical treatment described above, in the limit of vanishing interaction. Both implementations of these models are shown in Figure~\ref{fig:SPDC} and Figure~\ref{fig:BSadd}.

Shown in Figure~\ref{fig:SPDC}, we see that the first of these models uses an active optical element, the SPDC process, to probabilistically generate photon pairs. One individual from the pair strikes a detector, which confirms that the other spatial mode has the remaining of the pair. This outgoing beam is then known to have been photon added. We have presented the framework needed to model this process, in terms of Wigner functions, but for clarity the specific transforms one requires are given by, the preperation of the initial two mode state, given by,
\begin{equation}
W(x_1,p_1,x_2,p_2)= F_0(x_1,p_1)\times W_\rho(x_2,p_2),
\end{equation}
where $F_0(x_1,p_1)$ is the Wigner function of the vacuum state in spatial mode one. The action of the two mode squeezer is then,
\begin{eqnarray}
S_2(r,\theta)\cdot \left(
\begin{array}{c}
x_1\\
p_1\\
x_2\\
p_2
\end{array} \right)
=
\left(
\begin{array}{cccc}
\cosh r&0&\gamma&\delta\\
0&\cosh r&\delta&-\gamma\\
\gamma&\delta&\cosh r&0\\
\delta&-\gamma&0&\cosh r\\
\end{array} \right) \cdot  \left(
\begin{array}{c}
x_1\\
p_1\\
x_2\\
p_2
\end{array} \right),
\end{eqnarray}
where $\gamma=\sinh r \cos\theta$, $\delta=\sinh r \sin\theta$ and we use the convention that the upper spatial mode is labeled by the subscript one. The action of the condition that a single photon must be emitted into one of the output modes and detected is then modeled as a projection of that mode onto the single photon subspace, given by,
\begin{equation}
 W(x_2,p_2,r,\theta)_{\rho_+}=2 \pi \int F_1(x_1,p_1) W(x_1,p_1,x_2,p_2,r,\theta) dx_1 dp_1,
\label{eq:proj1}
\end{equation}
where $F_1(x_1,p_1)$ is the Wigner function for a single photon Fock state, given by Eq.~(\ref{eq:fockwig}), in the  lower mode, as we also use the convention that mode labels transfer through optical elements in the transmitted labels are kept; i.e. , across an optical element, the lower mode becomes the upper and the upper becomes the lower. It is also important to note that since we have performed a projective measurement, it now requires proper normalization, performed simply by,
\begin{equation}
N=\int W(x_2,p_2,r,\theta)_{\rho_+} dx_2 dp_2 .
\label{eq:prob1}
\end{equation}
We use this normalization constant to enforce the condition that our Wigner function is normalized at all times, \[
\frac{1}{N}\int W(x_2,p_2,r,\theta)_{\rho_+} dx_2 dp_2=1.
\]
It is also important to note that this normalization constant is also the probability that the detector in the herald mode, detects a single photon. This is particularly useful later when we account for the probabilistic nature of this scheme.

The transforms for modeling the photon addition process with the beam splitter model follows much the same construction, shown in Figure~\ref{fig:BSadd}. Specifically, the initial state,
\begin{equation}
W(x_1,p_1,x_2,p_2)= F_1(x_1,p_1)\times W_\rho(x_2,p_2),
\end{equation}
the action of the beam splitter of arbitrary transmissivity,
\begin{eqnarray}
BS(T)\cdot \left(
\begin{array}{c}
x_1\\
p_1\\
x_2\\
p_2
\end{array} \right)
=
\left(
\begin{array}{cccc}
\sqrt{T}&0&\sqrt{1-T}&0\\
0&\sqrt{T}&0&\sqrt{1-T}\\
\sqrt{1-T}&0&-\sqrt{T}&0\\
0&\sqrt{1-T}&0&-\sqrt{T}\\
\end{array} 
 \right) \cdot  \left(
\begin{array}{c}
x_1\\
p_1\\
x_2\\
p_2
\end{array} \right),
\label{eq:bs2}
\end{eqnarray}
 and the projection,
\begin{equation}
 W(x_2,p_2,T)_{\rho_+}=2 \pi \int F_0(x_1,p_1) W(x_1,p_1,x_2,p_2,T) dx_1 dp_1,
\label{eq:proj2}
\end{equation}
while the normalization constraint is much the same as previously described.

Clearly these two physical processes appear different in the way in which they function, as they have different initial states, different parameters and projections. However, one will find that in the limit of vanishing interaction, both of these models exactly match that of the mathematical treatment of photon addition. Specifically this means that, for the SPDC model, as $r\rightarrow 0$ the final state will exactly match the action of the mathematical treatment. Similarly, for the beam splitter model, this condition is satisfied for $T\rightarrow 1$. This leads us to why the mathematical model is somewhat limiting. Also in these limits, as we will see later, the probability of these events (given by $N$) approach zero. This means that while these treatments reduce to the mathematical model, physically the process is extremely unlikely to ever occur, something not very useful in experiments!

\begin{figure}
\centering
\includegraphics[width=0.45\columnwidth]{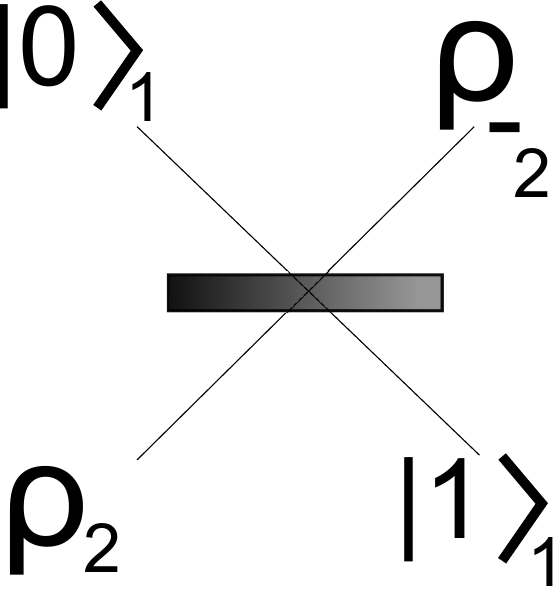}
\caption{Beam splitter model of photon subtraction. A general state, $\rho$ and a vacuum state are incident on a variable transmissivity beam splitter. On the condition that one output receives one photon, the remaining mode is known to be photon subtracted.}
\label{fig:BSsub}
\end{figure}

In the case of photon subtraction, we use a similar model to the beam splitter model of addition, though with different conditions. Shown in Figure~\ref{fig:BSsub}, this process utilizes  the same beam splitter of variable transmissivity, but different initial states and projective measurements. Following the same process as before we would perform this photon subtraction following the regimen of preparing the initial state,
\begin{equation}
W(x_1,p_1,x_2,p_2)= F_0(x_1,p_1)\times W_\rho(x_2,p_2),
\end{equation}
utilizing the beam splitter transform shown in Eq.~(\ref{eq:bs2}), a similar projection as in Eq.~(\ref{eq:proj1}) and proper normalization as in Eq.~(\ref{eq:prob1}). As with the models of photon addition, this model also agrees with the mathematical model in the limit that $T\rightarrow 1$, but again $N\rightarrow 0$, in this limit.

These processes can both be generalized to the case of $m$ photon addition and subtraction by generalizing the initial state and post selected projective measurement in a straight forward way. Specifically, for $m$ photon additions we modify the SPDC model to the case where $m$ photons are detected in the herald mode, confirming that $m$ photons have been added to the remaining mode. In the case of the beam splitter photon addition model, we consider the $m$ photon Fock state as the initial state but still condition on receiving no photons in the output mode, confirming $m$ photons have been added to the remaining output mode. For the case of $m$ photon subtraction we consider a projection onto the $m$ photon Fock state, which projects the remaining state into the $m$ photon subtracted subspace. As one may expect, while these generalize fairly easily, the probabilities of these $m$ photon additions and subtractions also decrease, with increasing $m$, as we will show later.

\subsection{Photon Addition and Subtraction Statistics}

\begin{figure}
\centering
\includegraphics[width=0.75\columnwidth]{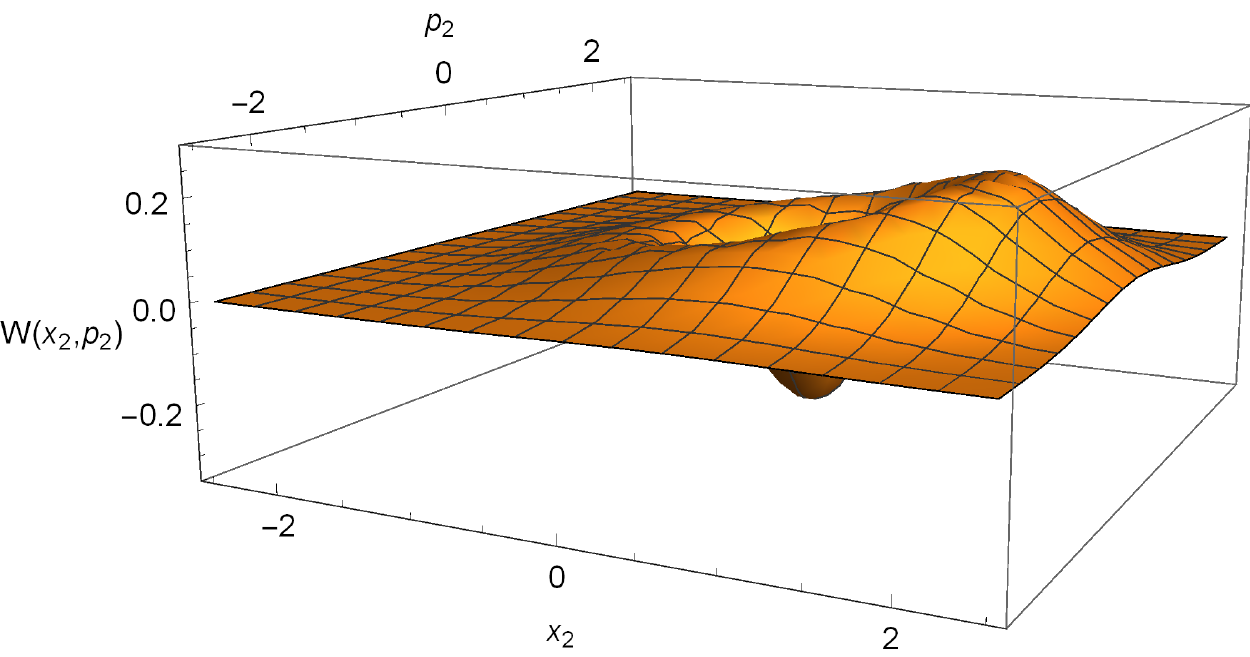}
\caption{Wigner function for a SPACS. Notice that a typical coherent state has a particularly simple Gaussian Wigner function, but our SPACS is clearly non-Gaussian and even attains negativity in its Wigner function. The relevant parameters have been fixed at values of, $|\alpha|^2=0.1225 , T=0.95, \theta_{\textrm{coh}}=0 .$}
\label{fig:wigadd}
\end{figure}

\begin{figure}
\centering
\includegraphics[width=0.75\columnwidth]{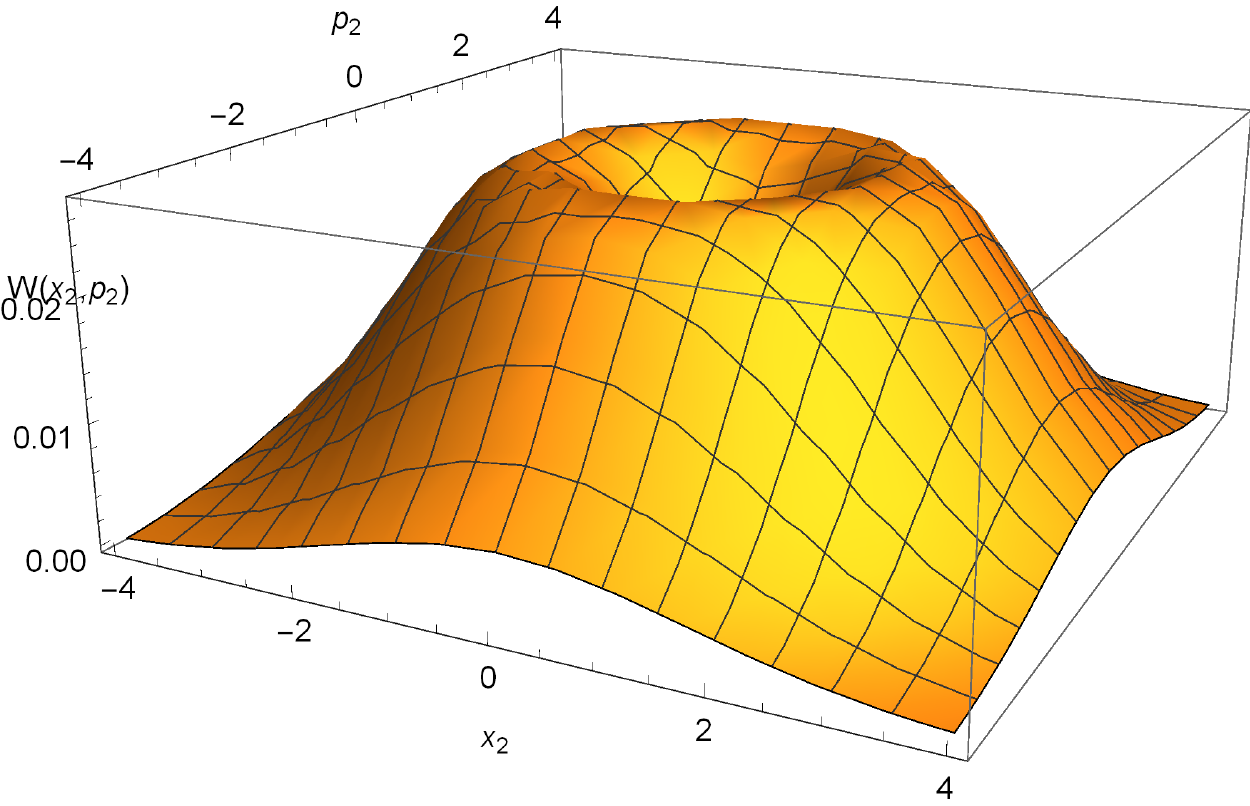}
\caption{Wigner function for a SPSTS. Notice that a typical thermal state has a particularly simple Gaussian Wigner function, but our SPSTS is clearly non-Gaussian and contains a ``dip" at the origin. The relevant parameters have been fixed at values of, $\bar{n}_{\textrm{th}}=2$ and $T=0.95 .$}
\label{fig:wigsub}
\end{figure}

Now that we have described the ways in which to model photon addition and subtraction, we can investigate some of the interesting properties of these processes. While their models are perhaps fairly intuitive to follow, we will show that their effects are not exactly as one may expect. As a first step, shown in Figure~\ref{fig:wigadd} and Figure~\ref{fig:wigsub} are the Wigner functions for a photon added coherent state and a photon subtracted thermal state, respectively. Note that a coherent state is invariant under photon subtraction since it obeys the relation $\hat{a}|\alpha \rangle=\alpha |\alpha \rangle$, meaning a coherent state under photon subtraction, returns the same coherent state. These states are generated following the prescription described above. We can notice that in the case of the Single Photon Added Coherent State (SPACS), even though we have only added a single photon, the Wigner function has changed drastically as a result, attaining negativity near the origin, confirming that this is a quantum state. 

In the case of a Single Photon Subtracted Thermal State (SPSTS), shown in Figure~\ref{fig:wigsub}, we see a drastically different distribution, which is positive everywhere, but has a small dip at the origin, caused by the photon subtraction process.


\begin{figure}
\centering
\includegraphics[width=0.75\columnwidth]{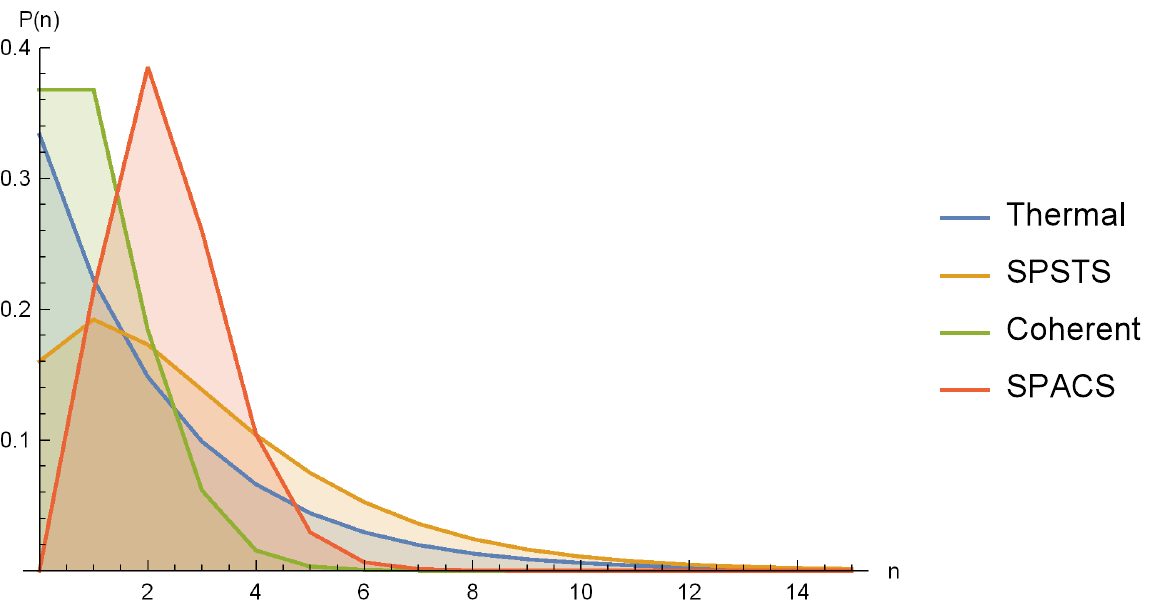}
\caption{Photon number distributions for normal coherent and thermal states and their photon added and subtracted transformations. Parameters have been fixed at, $|\alpha|^2=1, \bar{n}_{\textrm{th}}=2, T=0.9$.}
\label{fig:pnums1}
\end{figure}

While these Wigner functions serve as a useful way to observe these states, we can analyze them further by constructing their photon number distributions as discussed in Eq.~(\ref{eq:numdist}). In Figure~\ref{fig:pnums1}, we show the photon number distributions for a, normal coherent state and thermal state, along with the SPACS and SPSTS. We can notice that in both cases, the photon number distribution has shifted to the higher photon numbers and the probabilities of lower photon number has decreased. In the case of photon addition, this is fairly expected, as we are adding photons to the state. However, in the case of photon subtraction, the photon number has also \textit{increased}, a somewhat surprising claim. This can be explained by the fact that, under the assumption of photon subtraction, we gain partial knowledge of the photon number distribution. Specifically, consider the thermal state of low average photon number ($\bar{n}_{\textrm{th}}\leq5$); if the photon subtraction process has succeeded, then we \textit{know} that the state must have contained at least a single photon, which partially collapses the photon number distribution near the origin. This has the effect of shifting the distribution to the higher photon numbers. A similar effect happens in the case of photon addition. Under the assumption of successful photon addition, we are sure that the distribution cannot contain zero photons; thus our distribution shifts to the higher photon states. This is a visual description of the effects of the simple cases of photon addition and subtraction, but in general it is advantageous to have a qualitative way to describe the effect of photon addition and subtraction. Using the process described earlier to model the process of $m$ photon addition and subtraction, we can see that a pattern emerges in the average photon numbers for the addition of $m$ photons to a coherent state as well as $m$ subtraction of photons from a thermal state. These expressions can be found to be,
\begin{equation}
\langle \hat{n}_{\textrm{coh}} \rangle_{m+}=T|\alpha|^2+2m-\frac{m L_{m-1}(-T|\alpha|^2)}{L_m(-T|\alpha|^2)},
\label{eq:ncoh+}
\end{equation}
where $0\leq T \leq1$ is the decimal value of the variable beam splitters transmissivity, $L_m$ is the Laguerre polynomial of $m^{th}$ order and $m$ is the number of added photons. For the case of a $m$ photon subtracted thermal state, we can find,
\begin{equation}
\avg{\hat{n}_{\textrm{th}}}_{m-}=\frac{(m+1)\bar{n}_{\textrm{th}}T}{\bar{n}_{\textrm{th}}(1-T)+1},
\label{eq:nthm-}
\end{equation}
where $\bar{n}_{th}$ is the average photon number in the thermal state, prior to subtraction.

We can notice that if one sets $m=1,T=1$ in Eq.~(\ref{eq:ncoh+}), then this reduces to $\langle \hat{n}_{\textrm{coh}} \rangle_{1+}=|\alpha|^2+2-\frac{1}{|\alpha|^2+1}$. In this case, then we can see that the average photon number, under the addition of a single photon, increases by nearly two. Intuitively we would believe that it, of course, increases by one, the photon we added, but the story here is that we are mixing a state of \textit{definite} photon number, the single photon Fock state and the coherent state, with an \textit{average} photon number, so it need not be necessarily true that the new average photon number increase by only one. The change depends on the interaction of the two states, in terms of their photon number distributions. In the case of a photon subtracted thermal state, a perhaps even more surprising effect can be seen. Taking $m=1,T=1$ in Eq.~(\ref{eq:nthm-}), reduces it to $\avg{\hat{n}_{\textrm{th}}}_{1-}=2\bar{n}_{\textrm{th}}$, \textit{twice} its previous value. This, understandably, can seem suspicious. Under the action of subtracting a single photon, from a thermal beam, we actually \textit{double} its average photon number! We again turn to the explanation of photon number distributions to argue this counter-intuitive effect. Under the successful action of photon subtraction, we know this could have only occurred if there is \textit{at least} one photon in the thermal beam. So by assuming the subtraction event succeeds, we also gain the knowledge that the thermal beam contains at least one photon, which necessarily must shift the photon number distribution to the higher photon numbers.

While the effects of the action of photon addition and subtraction can be viewed as surprising and counter-intuitive, it is worth investigating the efficiencies of this process, since they occur in a probabilistic way. As mentioned earlier, the required renormalization process also serves to quantify the efficiency of the photon addition or subtraction event, given by Eq.~(\ref{eq:prob1}). Using our previous example of a photon added coherent state, this probability is,
\begin{equation}
P_{\textrm{coh}_{m+}}=(1-T)^m\textrm{e}^{|\alpha|^2(T-1)} L_m(-T|\alpha|^2 ),
\label{eq:paddm}
\end{equation}
where we can notice that for $T=1$ the probability reduces to $P_m=0$, a somewhat regrettable result. For the case of photon subtraction from a thermal state, we see,
\begin{equation}
P_{\textrm{th}_{m-}}=\frac{(\bar{n}_{th}(1-T))^m}{(\bar{n}_{th}(1-T)+1)^{m+1}},
\label{eq:psubn}
\end{equation}
which has the same behavior for $T=1$.

\begin{figure}
\centering
\includegraphics[width=0.65\columnwidth]{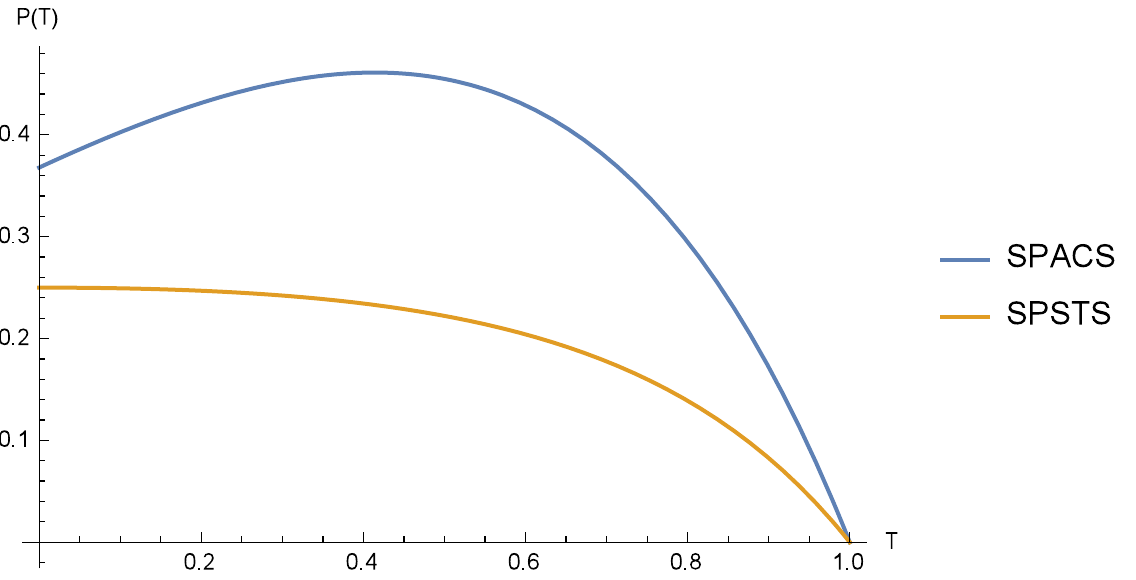}
\caption{Probability of successfully generating a SPACS and SPSTS according to the model presented in the text. Average photon number prior to addition or subtraction has been fixed to $|\alpha|^2=\bar{n}_{\textrm{th}}=1$.}
\label{fig:probs1}
\end{figure}

This realization tells us that, we achieve the greatest effect from photon addition (the largest increase in average photon number) when the likelihood of successful photon addition decreases to zero. Of course, we have the option of not only considering such an extreme case. We can also consider $T<1$, as this regime still gives us an enhanced photon number, but with a non-zero probability of success. In Figure~\ref{fig:probs1}, we see the probability, as a function of the transmissivity of our variable beam splitter, for the case of a SPACS and SPSTS.  It is this type of investigation that showcases our claim of why the use of the mathematical treatment of photon addition and subtraction may not be adequate in some applications, as it only applies when $T\approx1$, but as we have just shown, in this limit, the probability of successfully generating such a state is significantly small.

\begin{figure}
\centering
\includegraphics[width=0.65\columnwidth]{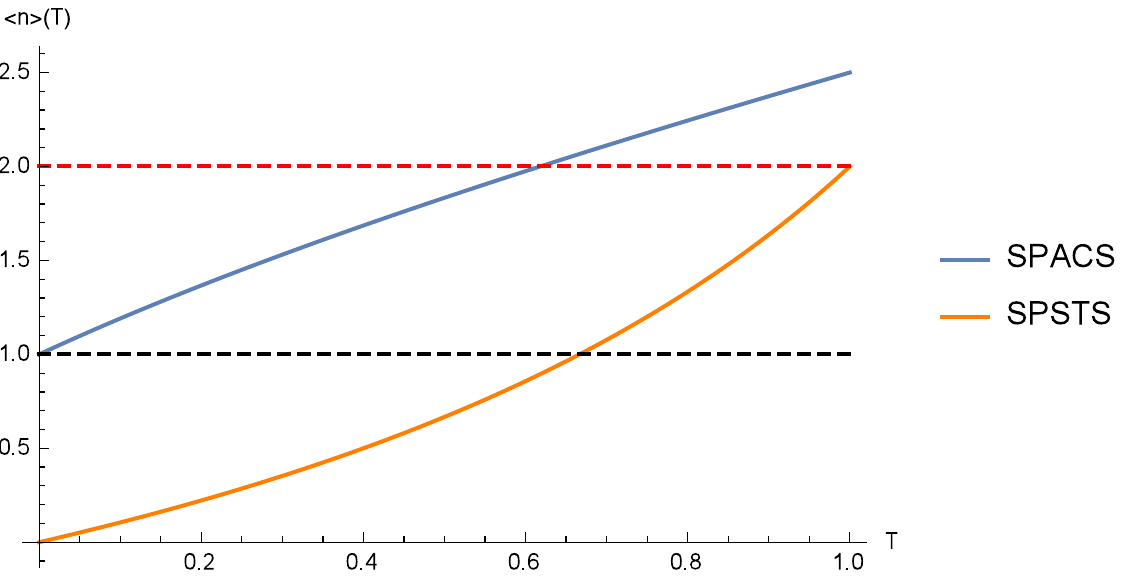}
\caption{Average photon number of a SPACS and SPSTS, after successful photon addition or subtraction, as a function of the transmissivity of the variable beam splitter. Average photon number prior to addition or subtraction has been fixed to $|\alpha|^2=\bar{n}_{\textrm{th}}=1$. Also shown are dashed horizontal lines at the expected average photon values for the subtraction case (Black) and addition case (Red).}
\label{fig:nums1}
\end{figure}

Also in Figure~\ref{fig:nums1}, we see the average photon number of these two states as a function of the transmissivity. From these two plots we can note that the number statistics generally increase with $T$, but the probabilities, for sufficiently large $T$, drop sharply. This tells us that we can still achieve an enhancement in the photon number statistics, while maintaining a respectable probability of success if we choose a value for $T$ of $0.65 \leq T \leq 0.9$. Also shown in this figure are the expected average photon numbers. We chose this description because, in the case of the SPACS, the average photon number in the coherent state is fixed at one and we inject another photon during the addition process; therefore any photon number above two, must be from a quantum mechanical effect and it is this regime that we are interested in. A similar effect can be described in the case of the SPSTS, which has an average photon number of one; therefore anything above this value is explained by a quantum effect. 

\singlespacing
\subsection{Signal to Noise Ratios with Photon Addition and Subtraction}
\begin{figure}
\centering
\includegraphics[width=0.65\columnwidth]{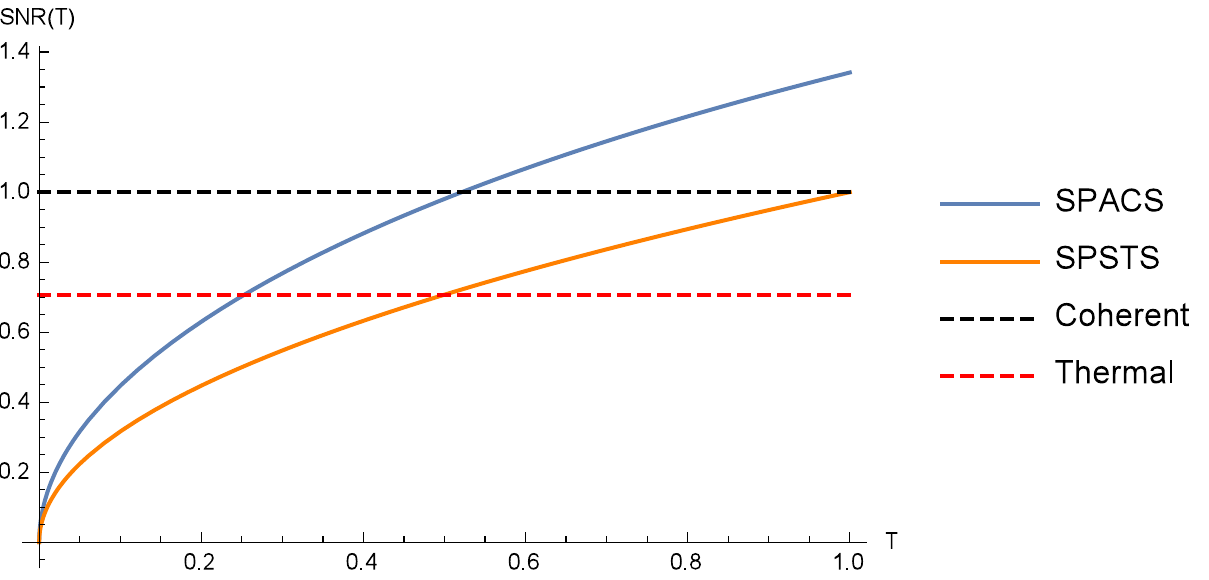}
\caption{Signal to noise ratio of the SPACS and SPSTS as compared to standard coherent and thermal states of average photon number $|\alpha|^2=\bar{n}_{\textrm{th}}=1$. A clear enhancement can be seen for values of $T\geq 0.5$.}
\label{fig:SNR1}
\end{figure}

\doublespacing
Also of interest is the effect these process has on the noise of these states. Typically this is shown by calculating the variance of the photon number and forming the Signal to Noise Ratio (SNR), defined by,
\begin{equation}
\textrm{SNR}=\frac{\avg{\hat{n}}}{\sqrt{\avg{\hat{n}^2}-\avg{\hat{n}}^2}}.
\label{eq:SNR1}
\end{equation}
For our specific example of photon added coherent states and photon subtracted thermal states with $m$ photon additions and subtractions, respectively, we can show the their SNR goes as,
\begin{equation}
\textrm{SNR}_{\textrm{coh}_{m+}}=\frac{\langle \hat{n}_{m+} \rangle-m}{\sqrt{\langle \hat{n}_{m+}^2 \rangle-\langle \hat{n}_{m+} \rangle^2}},
\label{eq:SNRcoh}
\end{equation}
where the second moment is given by,
\begin{equation}
\langle \hat{n}_{m+}^2 \rangle=\frac{ (m+2)(m+1)L_{m+2}(-T|\alpha|^2)-3(m+1)L_{m+1}(-T|\alpha|^2)+L_m(-T|\alpha|^2) }{L_m(-T|\alpha|^2)}
\end{equation}
and $\langle \hat{n}_{m+} \rangle$ is given by Eq.~(\ref{eq:ncoh+}). Note that in the definition of our SNR for photon addition we have subtracted off the number of added photons $m$. This ensures that our SNR, as a figure of merit is representative of the quantum effect of photon addition and does \textit{not} include the artificial injection of the added photons themselves, only their effect on our coherent state. In the case of a photon subtracted thermal state, we find,
\begin{equation}
\begin{split}
\textrm{SNR}_{\textrm{th}_{m-}}&=\frac{\sqrt{2Tn_{\textrm{th}}(m+1)}}{\sqrt{2n_{\textrm{th}}+1}}\\
&=\sqrt{T(m+1)} \textrm{SNR}_{\textrm{th}},
\end{split}
\label{eq:SNRthm}
\end{equation}
where $\textrm{SNR}_{\textrm{th}}$ is the SNR of a normal thermal state. Shown in Figure~\ref{fig:SNR1}, we see the SNR for a SPACS and SPSTS, as compared to a normal coherent state and thermal state, of the same fixed average photon number.

Again we can comment that, in general the SNR is best at larger values of $T$, but we recall that as $T \rightarrow 1$, our probability of successfully generating these states decreases quickly.

We have shown the behavior of two specific states, under the action of single photon addition and subtraction, but also provided analytical forms for the statistics for these states under $m$ photon addition and subtraction; however these two presented cases are just a sample of some of the effects one can observe under the action of these photon addition and subtraction operations. The behavior of other states continues to have similar effects, but in general, the effect of photon addition and subtraction can vary, depending on which state they are performed. We also stress the need for an accurate model of photon addition and subtraction as its probabilistic nature can lead to overly optimistic conclusions, which do not take into account the efficiency of these processes.

\begin{figure}[!htb]
\centering
\includegraphics[width=\columnwidth]{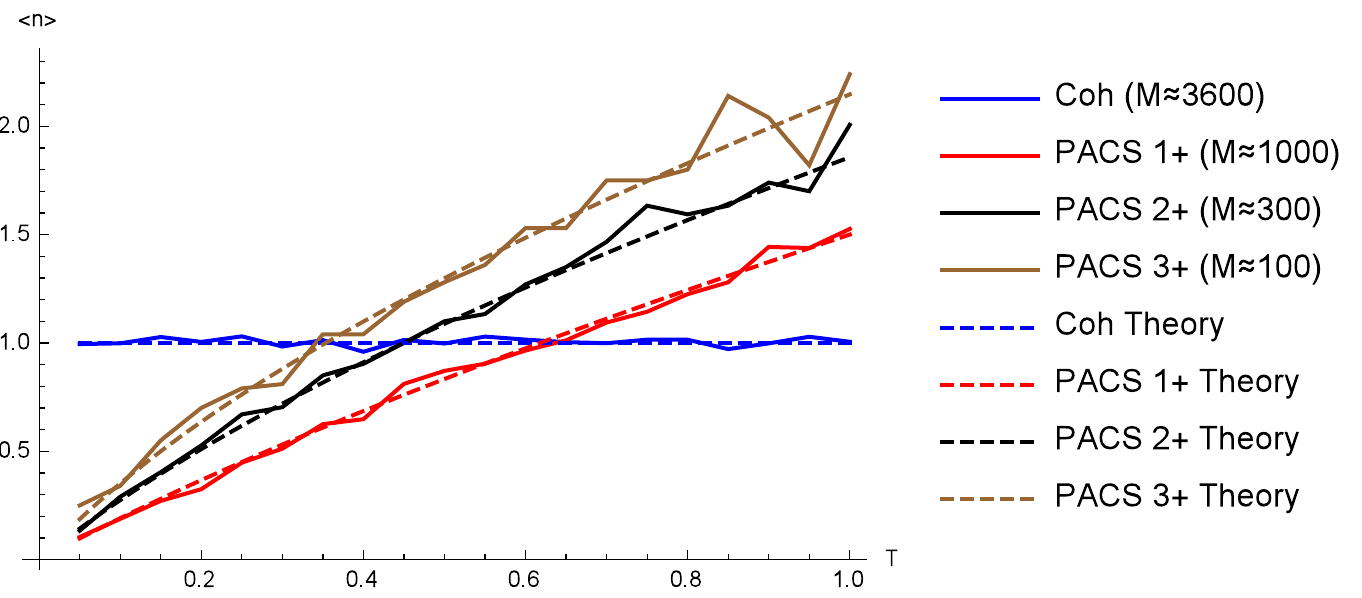}
\caption{Plot of average photon number of various PACS ($m=1,2,3$), as a function of variable transmissivity, $T$, with simulated data for a number of measurements of 3600 per value of transmissivity and the associated successful measurements (M) for each scheme. For each value of transmissivity, we have averaged over all successful measurements. Theory lines ($M\rightarrow\infty$) also shown for each case and we have fixed $|\alpha|^2=1$; therefore any curves above one (blue) can be viewed as an enhancement of the photon number.} \label{fig:data2}
\end{figure}

\begin{figure}[!htb]
\centering
\includegraphics[width=\columnwidth]{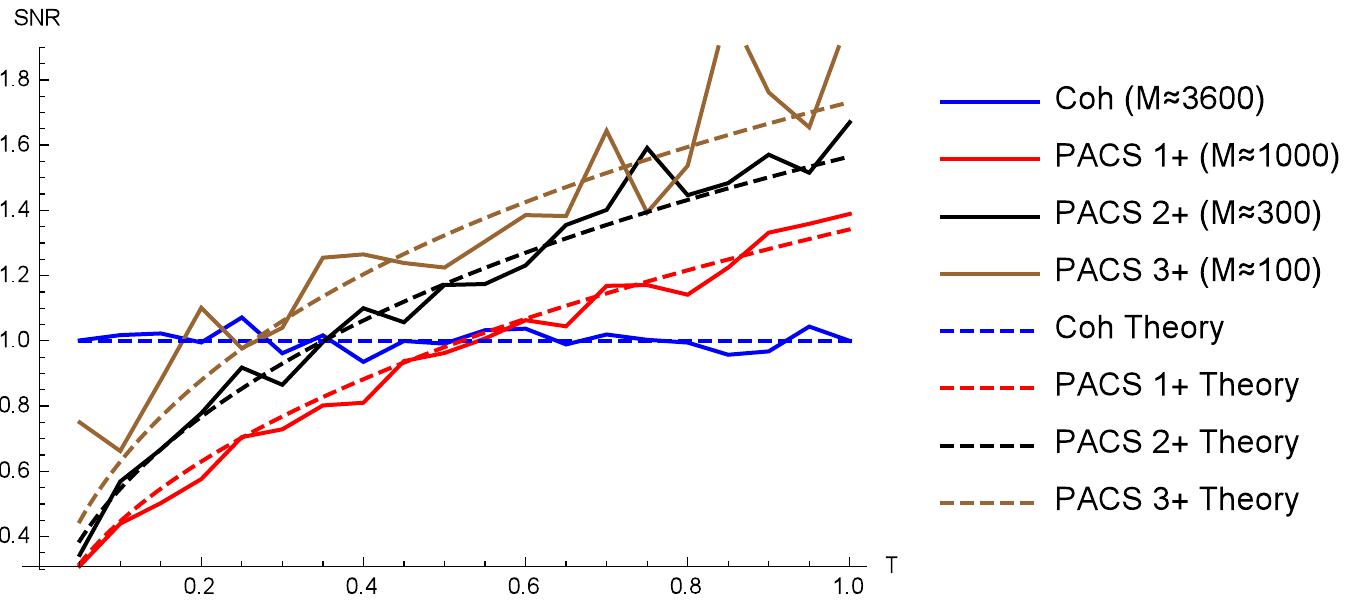}
\caption{SNR of various PACS ($m=1,2,3$), as a function of variable transmissivity, $T$, with simulated data for a number of measurements of 3600 per value of transmissivity and the associated successful measurements (M) for each scheme. Theory lines ($M\rightarrow\infty$) also shown for each case and we have fixed $|\alpha|^2=1$; therefore any curves above one (blue) can be viewed as an enhancement of the SNR.} \label{fig:data1}
\end{figure}

We will now show how the probabilistic nature of photon addition affects the statistics of the SPACS. While the previous figures show promise for the SNR as a function of the transmissivity, along with the associated probabilities for each measurement condition, it is perhaps more useful to look at simulated data, to showcase how these schemes, along with their inherent probabilities are expected to perform in an experiment. We calculate this simulated data following the prescription of, using the discrete, analytical form of the photon number distribution according to Eq.~(\ref{eq:numdist}) for each case of PACS, then average over a psuedo-random choice of photon numbers from this distribution. The amount of choices made in each case is directly dependent on their efficiencies given by Eq.~(\ref{eq:paddm}). These choices are then made for each value of transmissivity, in steps of $\Delta T=0.05$. Recall that the efficiency of each PACS drops sharply in the regime where $T\rightarrow 1$. In Figure~\ref{fig:data2} we show the result of this method for the average of various photon added coherent states ($m=1,2,3$), along with dashed curves showing the theory representations.  The total number of measurements is fixed at $3600$ measurements for all the displayed states, with $M$ showing the number of kept measurements for each case, due to the post selection requirement  (this is simply related by each of the $P_m$'s given by Eq.~(\ref{eq:paddm})). After averaging over all kept measurements for each value of transmissivity, we see that the higher photon additions do attain a higher photon number as predicted by theory, but the scatter also worsens for the higher photon additions as the kept measurements also decreases significantly. We stress that, in each case, the total measurements taken is $M=3600$, and the use of post selection discards some of these measurements, but we are comparing schemes which run in a simulated experiment for \textit{equal time} and thus we can still attain enhanced photon number, while incorporating the efficiencies of the photon addition process, with the addition of slightly nosier data. The scatter in the data could be lessened however, if one had the option performing longer experiments in the case of photon addition. We also show, in Figure~\ref{fig:data1} how this effect is modeled in terms of the SNR. Note the significantly increased scatter in the PACS $3+$ SNR as compared to the PACS $1+$; this is a direct result of our model showcasing the effect of the decreased efficiency in adding three photons when compared to adding a single photon.


We have shown a way to account for the probabilistic nature in the case of photon statistics, but we will also now discuss the effects of this probabilistic process in terms of a phase estimation.

\subsection{Photon Addition and Subtraction in Phase Estimation}
In previous sections we discussed the field of phase estimation along with its goals and benchmarks. Now we will revisit this realm with the twist of probabilistic effects, such as photon addition and subtraction. We have seen that some statistics of photon addition and subtraction can deteriorate when we consider the full probabilistic model, but we again show that proper modeling of this process needs to be considered before claims of improvement can be made. The key consideration in this field is that typically studied states are, in theory, capable of being generated deterministically, but in the case of photon added and subtracted states, we are forced to probabilistically generating these states and therefore before we compare these two kinds of processes, we must account for this change in nature.

In the case of phase estimation, consider the example of a SPACS and squeezed vacuum state as initial states in a standard MZI setup. We have shown the ways in which one can model such a system, but as the creation of the SPACS is probabilistic, its use requires some caution. The use of the SPACS, along with Fisher information as a metric, requires us to use probabilistic Fisher information, as the creation of the SPACS is nondeterministic and Fisher information initially assumes identical trials over many experimental runs. This modification is a simple one and was recently described in \cite{David2015}, where they showed coupling a probabilistic process, with Fisher information. For our example of the SPACS and squeezed vacuum, we model it following the many examples discussed previously and propagate these states through the MZI, to the final detectors. Here, we will utilize Fisher information to describe the phase variance of our chosen measurement. For simplicity, we chose a basic measurement of click detection, that is, the event that our detector (either output) detect \textit{any number} of photons, which is the event that a standard avalanche photo-diode is sensitive to. We denote the probability that our detector receives a click as $P_c$; then we must also have that the probability that our detector receive no photons is $1-P_c$. From these two possible events, we can easily see this covers all possible outcomes. We then follow Eq.~(\ref{eq:cfi1}) to construct the phase variance from the Fisher information, with the modification that we must carefully consider the effects of using the nondeterministic creation of the SPACS. The reasoning is straight forward in that, we will collect the maximum information about our unknown parameter $\phi$ if we keep \textit{all} experimental trials, even those that do not produce our desired states. Ideally, in the cases where we fail to generate our desired states, little information about $\phi$ is obtained, but realizing that Fisher information is strictly a positive quantity, we can only \textit{lose} information if we do not keep these failed cases. To accomplish this, we form the following probabilistic Fisher information,
\begin{equation}
\begin{split}
\textrm{CFI}_{P_c}=&P_+\left(\frac{1}{P_{c_s}}\left(\frac{\partial P_{c_s}}{\partial \phi}\right)^2+\frac{1}{1-P_{c_s}}\left(\frac{\partial P_{c_s}}{\partial \phi}\right)^2\right)\\
&+ (1-P_+)\left(\frac{1}{P_{c_f}}\left(\frac{\partial P_{c_f}}{\partial \phi}\right)^2+\frac{1}{1-P_{c_f}}\left(\frac{\partial P_{c_f}}{\partial \phi}\right)^2\right)\\
\end{split}
\label{eq:pcfi1}
\end{equation}
where $P_+$ is the probability of successfully generating a SPACS, $P_{c_s}$ is the probability of getting a click at the detector, when a SPACS is successfully created, $P_{c_f}$ is the probability of getting a click at the detector, when a SPACS is failed to be created. While this form can first appear cumbersome, with some explanation, each piece can be interpreted in a particularly straightforward way. The first line can be seen to be the information gathered during the event that a SPACS is created and the detector clicks (first term in parenthesis) and does not click (second term). The second line is the information gathered during the event that a SPACS is not created and the detector clicks (first term in parenthesis) and does not click (second term). Note that the probabilities of clicking are different for the case of a successful creation or failed creation of a SPACS. In principle, the detector we use to herald for successful creation of the SPACS could also contain information, but as this process is taking place at the input and does not depend on $\phi$, we can neglect this portion of the CFI$_{P_c}$. Also note that each term presented in Eq.~(\ref{eq:pcfi1}) is positive; therefore if we only consider the postselected state where SPACS is successfully generated, we are discarding information, which necessarily leads to a worse phase variance. Also of importance to note is that $P_+ \leq 1$ and therefore, as a nondeterministic process, serves to weight our information by its probability of success. It is this treatment that is typically overlooked in treatments of photon addition or subtraction. Without the inclusion of these weights, one is assuming the photon addition or subtraction can be performed deterministically, for every run of an experiment.

\singlespacing
\section{Photon Addition and Subtraction Must be Nondeterministic}
\doublespacing
A natural question to wonder is then, perhaps the previously described physical processes are poor choices of the mathematical treatment and there may exist one that models the mathematical model deterministically. With a relatively simple gedanken experiment, we can argue that this is impossible. This will show that if one believes that super-luminal communication is not possible, then so to is deterministic photon addition or subtraction. This gedanken experiment, described in terms of an Alice and Bob scenario, is as follows:

\begin{enumerate}
\item{Alice and Bob setup a simple optical system consisting of only a laser, squeezer, beam splitter and black box that performs deterministic photon addition.}
\item{They agree that if a photon added coherent state is measured, this serves as a logical 1. A normal squeezed coherent state serves as a logical 0.}
\item{Alice prepares a squeezed coherent state, splits it on a 50-50 beam splitter, keeps one output locally and sends the other output to Bob.}
\item{Bob can choose to perform $m$ photon additions to his portion of the beam. A value of $m \geq 5$ ensures that the photon statistics between a photon added squeezed coherent state and a normal squeezed coherent state, are significantly different.}
\item{Due to entanglement between the two portions of the squeezed coherent state, if Bob performs photon addition on his end, Alice also sees effects to the photon statistics on her end.}
\item{With the conditions set previously, this allows Bob to send 1 or 0 to Alice, deterministically, at super luminal speeds, as the collapse of the wave function is instantaneous.}
\end{enumerate}

\begin{figure}[!htb]
\centering
\includegraphics[width=0.8\columnwidth]{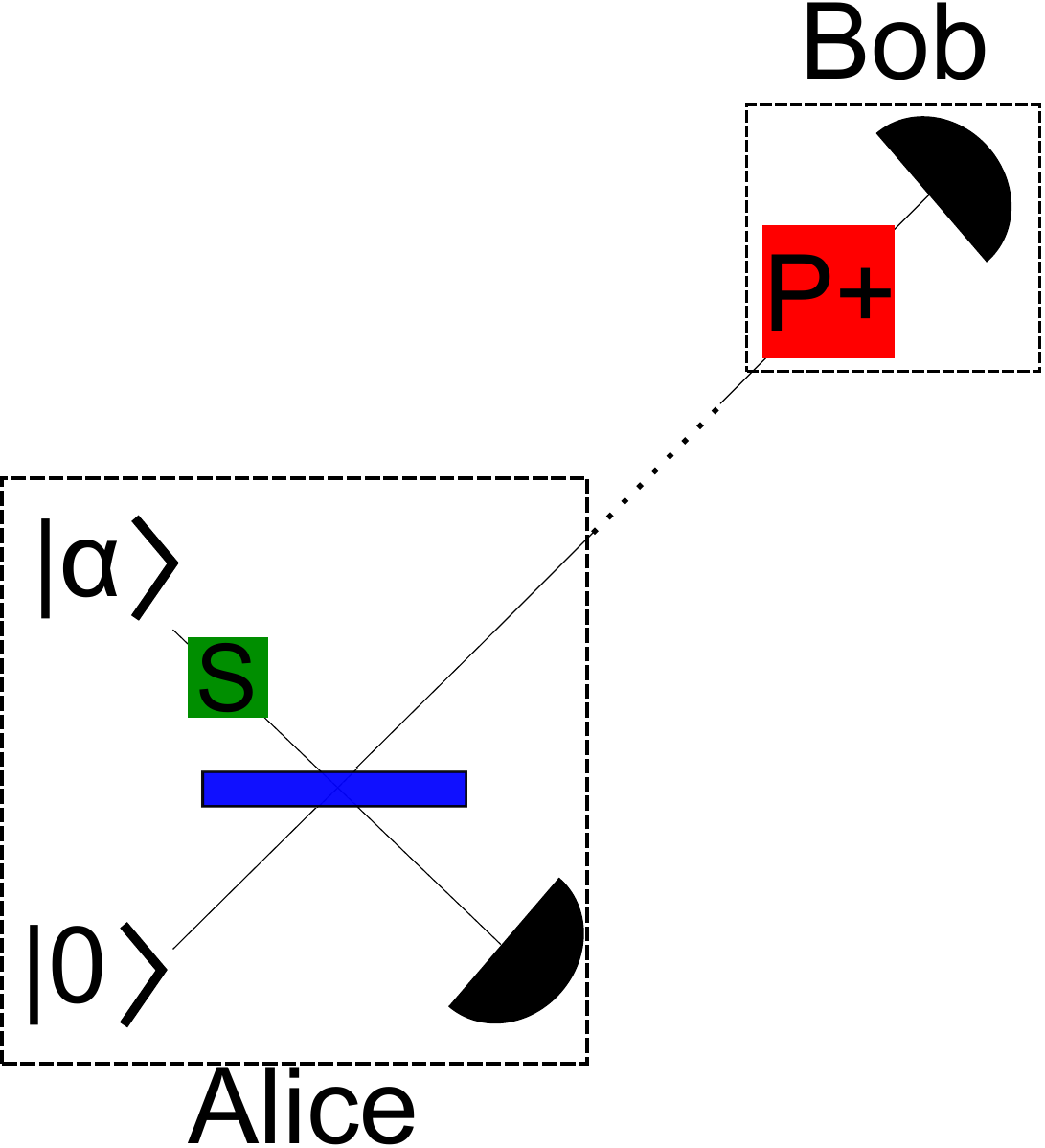}
\caption{Gedanken experiment that shows, if deterministic photon addition were possible, then such a protocol would allow for super-luminal communication. Green box labeled S denotes a single mode squeezer while the red box labeled P$_+$ denotes a deterministic photon adder. The dotted line depicts that, in principle, Alice and Bob may be separated by an arbitrary distance.} \label{fig:ged1}
\end{figure}

A depiction of this setup is shown in Figure~\ref{fig:ged1}. Clearly, something must prevent this process from being physically realizable. While we are ignoring effects such as decoherence and loss of entanglement over long distances, in principle, this should not be possible over \textit{any} distance. The issue here then lies with the fact that photon addition and subtraction \textit{must} be nondeterministic, which then would modify the previously described scenario to one in which Bob must tell Alice (via classical communication) when the photon addition process has succeeded, limiting it to classical limits. This protocol would then be very similar to a quantum teleportation protocol, which also requires classical communication. While there are likely many other explanations that one could construct to show such a contradiction, the conclusion is that a deterministic photon addition or subtraction process should not be possible. This connects to our previous argument that when modeling photon addition and subtraction, one must consider some physical model in order to account for the required nondeterministic nature of this process.

\pagebreak
\singlespacing
\chapter{Full Examples}
\doublespacing
\section{Simplified Advanced LIGO}
Throughout this dissertation, we have shown how to model various states of light, optical elements, and quantify phase measurements. For completeness, we will now fully show an example of an interesting process, the search for gravity waves. While our treatment lacks many of the technical challenges of LIGO in their effort to achieve a direct measurement of a gravity wave, this example will still serve as a simplified version of their efforts, with a conclusion that a superior measurement scheme still exists.

While the full LIGO interferometer is a large scale Michelson interferometer, we choose to model it by an equivalent model, the Mach-Zehnder interferometer (MZI) as shown in Figure \ref{fig:MZI}. Here, an input of a coherent state and squeezed vacuum is used to model that of the Advanced LIGO setup. With this input state, it is known that the phase sensitivity can be below the SNL, typically defined as $\Delta \phi^2_{\textrm{SNL}}=1/N$, where $N$ is the total number of photons entering the MZI \cite{Caves1981}.

In terms of Wigner functions, the input state can then be written as,
\begin{equation}
\begin{split}
&W(\textbf{X})=\\
&\frac{1}{\pi^2}\textrm{Exp}(-2|\alpha|^2-p_1^2+2\sqrt{2}|\alpha|x_1-x_1^2-(\textrm{e}^{2r}p_2^2+\textrm{e}^{-2r}x_2^2))
\end{split}
\end{equation}
where $\textbf{X}$ is a list of the mode labels ${x_1,p_1,x_2,p_2}$, labeling the position and momentum components of each spatial mode. The average photon number in the coherent state is $N_{\textrm{coh}}=|\alpha|^2$ and in the squeezed vacuum state $N_{\textrm{sqz}}=\sinh^2{r}$, which sets the SNL to be  $\Delta \phi^2_{\textrm{SNL}}=1/N_{\textrm{tot}}=1/(|\alpha|^2+\sinh^2{r})$. Both states have equal phases, as this gives rise to the optimal phase sensitivity (discussed later) and are taken to be $\theta_{\textrm{coh}}=\theta_{\textrm{sqz}}=0$.

The propagation of this Wigner function is accomplished by a simple transformation of the phase space variables through the MZI, dictated by its optical elements. These transformations are described by
\begin{eqnarray}
BS(1/2)=\frac{1}{\sqrt{2}}
\left(
\begin{array}{cccc}
1&0&1&0\\
0&1&0&1\\
1&0&-1&0\\
0&1&0&-1\\
\end{array} \right)
\end{eqnarray}
\begin{eqnarray}
PS(\phi)=
\left(
\begin{array}{cccc}
\cos(\frac{\phi}{2})&-\sin(\frac{\phi}{2})&0&0\\
\sin(\frac{\phi}{2})&\cos(\frac{\phi}{2})&0&0\\
0&0&\cos(\frac{\phi}{2})&\sin(\frac{\phi}{2})\\
0&0&-\sin(\frac{\phi}{2})&\cos(\frac{\phi}{2})\\
\end{array} \right)
\end{eqnarray}
where both beam splitters are fixed to be 50/50 and we have chosen to use a symmetric phase model in order to simplify calculations as well as agree with many other references \cite{Nori,Hofmann}. Using these transforms, the total transform for phase space variables is given by,
\begin{eqnarray}
\left(
\begin{array}{cc}
x_{1f}\\
p_{1f}\\
x_{2f}\\
p_{2f}
\end{array}\right)
=BS(1/2) \cdot PS(\phi) \cdot BS(1/2) \cdot
\left(
\begin{array}{cc}
x_{1}\\
p_{1}\\
x_{2}\\
p_{2}
\end{array} \right)
\end{eqnarray}
From here, the final variables (denoted by $x_{1f}$ etc.) are inserted to the initial Wigner function to obtain the Wigner function at the output.

We can also consider photon loss in the model by way of two mechanisms, photon loss to the environment inside the interferometer and photon loss at the detectors, due to inefficient detectors. Both of these can be modeled by placing a fictitious beam splitter in the interferometer with vacuum and a interferometer arm as input and tracing over one of the output modes, to mimic loss of photons to the environment \cite{photon_loss}. This linear photon loss mechanism can be modeled with the use of a relatively simple transform, since these states are of Gaussian form. Specifically this amounts to a transform of the covariance matrix according to $\sigma_L=(1-L)\mathbb{I} \cdot \sigma +L\mathbb{I}$, where $\sigma$ is the covariance matrix of the two mode Gaussian state (in $(x,p)$ phase space), $0\leq L\leq 1$ is the combined photon loss and $\mathbb{I}$ is the 4x4 identity matrix. Similarly the mean vector is transformed according to $\langle\hat{R}\rangle_L=\sqrt{(1-L)}\mathbb{I}\cdot\langle\hat{R}\rangle$.

\subsection{LIGO Measurement}

\subsubsection{Quantum Cram\'{e}r Rao Bound}
In the classical version of this setup, a coherent state and vacuum state were used as input. With these two input states, the best sensitivity one can achieve is bounded by the SNL, which is achievable with many different detection schemes, but importantly, it is achievable with a standard, single mode, intensity measurement, $I=\langle \hat{a}^{\dagger}\hat{a} \rangle=\langle \hat{x}^2+\hat{p}^2 \rangle/2$, which is implemented by simply collecting the outgoing light, directly onto a detector. It is this detection scheme that LIGO is configured for and has many technologies employed to extract the most efficiency out of this measurement scheme. The benefit of using squeezed vacuum in place of vacuum is then that the phase measurement can now reach below the previous SNL. In order to compare various choices of measurement schemes, we not only need to calculate the various measurement choices, but also need to show the best sensitivity attainable with these input states. The best phase measurement one can do is given by the Quantum Cram\'{e}r Rao Bound (QCRB) \cite{cramer1946mathematical} and is related to the Quantum Fisher Information (QFI) \cite{QFI_Caves} simply by $\Delta \phi^2_{QCRB}=QFI^{-1}$. For the input states of a coherent and squeezed vacuum, one can use the Schwinger representation and many references to calculate the QFI, since these are pure states \cite{Nori,dd1a}. Another option, and the method we use here, instead utilizes the Gaussian form of the states and can be calculated directly in terms of covariance and mean \cite{QFI1,parisgaussian}. This method applies to pure and mixed states, as long as it maintains Gaussian form. Using this formalism, the QCRB for a coherent state and squeezed vacuum into an MZI can be found to be \cite{Kaushik1}
\begin{equation}
\Delta \phi^2_{\textrm{QCRB}}=\frac{1}{|\alpha|^2 \textrm{e}^{2r}+\sinh^2(r)}.
\label{eq:qcrb}
\end{equation}
While this gives us a bound on the best sensitivity obtainable with these given input states, it does not directly consider loss or even tell us which detection scheme attains this bound. To handle these issues we proceed to model loss as described earlier and calculate the lossy QCRB. As described earlier, this is done with fictitious beam splitters, but again this has the same linear effect as before. The lossy QCRB  of this mixed state then becomes
\begin{equation}
\begin{split}
&\Delta \phi^2_{QCRB_{Lossy}}=\\
&-((A(AB-C)(AB+C)\textrm{e}^{r}(8A^3B^5C+4|\alpha|^2C^2\\
&+A(4AB^4(A-A\cosh{2r}-1)\\
&+B^2(2-A-4A|\alpha|^2+2\cosh{2r}+A\cosh{4r})\\
&-A\sinh^3{2r})))/((B-2AB+C)\\
&(1+A^2-\cosh{2r}(A^2-1))^2)),
\label{eq:qcrb}
\end{split}
\end{equation}
where, $A=(1-L),B=\sinh{r},C=\cosh{r}$.

Note that this QCRB with loss only considers linear photon loss caused by photon loss inside the interferometer and photon loss due to inefficient detectors. In reality, Advanced LIGO needs to account for very specific sources of noise \cite{LIGO_noise1,LIGO_noise2}, but our methods purpose is to show a preliminary case when simple loss models are considered.

\subsubsection{Specific Measurements}
Now that we have a bound on the best possible sensitivity, we now seek to show how various choices of measurement compare to this bound. Along with Advanced LIGO's current measurement scheme, single port intensity measurement, we consider some other typical measurement choices, homodyne, intensity difference, and parity. While each of these measurements would require a significant reconfiguration of  Advanced LIGO's setup, it is worthwhile to show how each choice impacts the resulting phase sensitivity measurement. This is accomplished for a chosen measurement operator $\langle \hat{O}\rangle$ by way of $\Delta \phi^2=\Delta\hat{O}^2/|\partial \langle \hat{O}\rangle/\partial \phi|^2$.

\begin{figure}[!htb]
\includegraphics[width=\columnwidth]{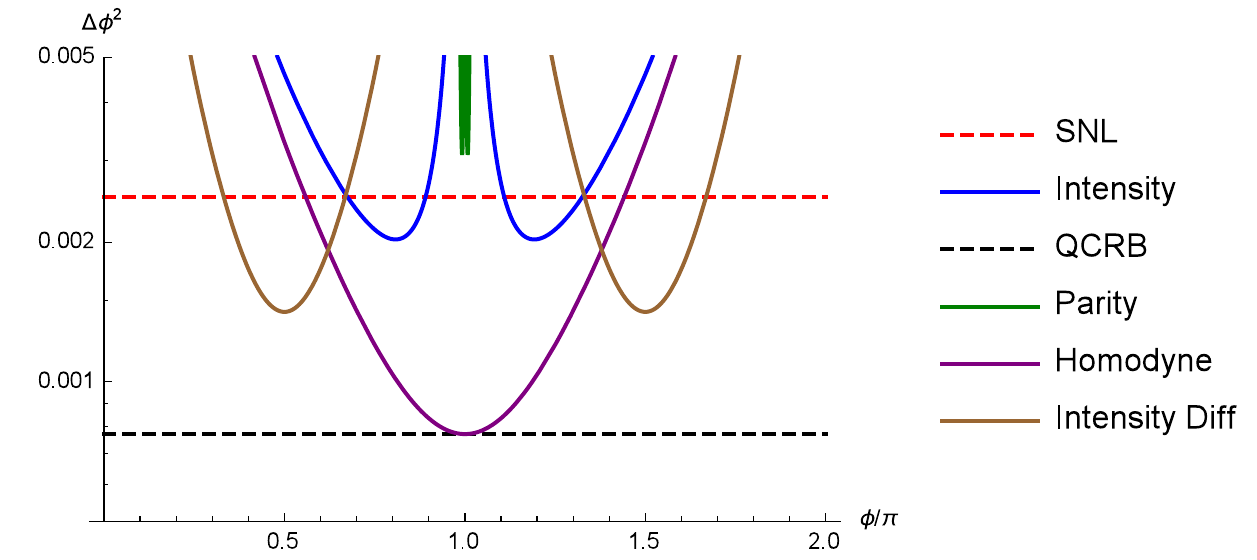}
\caption{Log plot of phase variance for various detection schemes for a coherent state and squeezed vacuum into an MZI, as a function of the unknown phase difference, $\phi$. Loss parameters have been set to, $L=20\%$. Input state parameters for each respective state are set to $|\alpha|^2=500$ and $r=1$. SNL and QCRB are also plotted with the same loss parameters.}
\label{fig:phases}
\end{figure}

For homodyne detection, $\hat{O}=\hat{x}$ (we find the optimal homodyne measurement is taken along the $x$ quadrature). For a balanced homodyne detection scheme, one would impinge one of the outgoing light outputs onto a 50/50 beam splitter, along with a coherent state of the same frequency as the input coherent state (usually this is derived from the same source) and performing intensity difference between the two outputs of this beam splitter. A standard intensity difference is simply defined as $\hat{O}= \hat{a}^{\dagger}\hat{a}- \hat{b}^{\dagger}\hat{b}$. Parity detection is defined to be $\hat{O}=(-1)^{\langle\hat{a}^{\dagger}\hat{a} \rangle}=\pi  W(0,0)\equiv \langle \hat \Pi \rangle$. While all chosen measurements can surpass the SNL, in the lossless case, to various degrees, in order of improving phase sensitivity, the current Advanced LIGO standard (intensity) performs the \textit{worst}, followed by intensity difference, homodyne, and finally parity. Shown in Figure~ \ref{fig:phases}, we show a logplot of the phase variances obtainable from various measurement choices. Intensity difference, homodyne, and parity each almost nearly perform the same, and their different performance is nearly impossible to see in Figure~\ref{fig:phases}. In this figure however, it is clear at which value of phase the various measurements attain their lowest value. It is this value of phase that one attempts to always take measurements at with the use of a control phase inside the interferometer. The width of each of curve then can be interpreted as the chosen measurement schemes resistance to phase drift, fluctuations in our ability to fix the control phase. From this viewpoint then, its clear that while the performance of various measurement may attain nearly the same phase variance, parity would appear to be quite limited in its ability to maintain enhancement in the presence of significant phase drift. The modeling and discussion of phase drift is shown in the appendix. Evaluating each phase variance at the optimal phase value then gives us analytical results for each measurement as a function of $r$ and $|\alpha|$. We confirm that, under ideal (lossless) conditions, parity attains the best sensitivity and exactly matches the lossless QCRB \cite{Kaushik1}, 
\begin{equation}
\Delta \phi^2_{\hat{\Pi}}=\frac{1}{|\alpha|^2 \textrm{e}^{2r}+\sinh^2(r)}.
\end{equation}
homodyne attains,
\begin{equation}
\Delta \phi^2_{\hat{x}}=\frac{1}{|\alpha|^2 \textrm{e}^{2r}},
\end{equation}
 and intensity difference attains,
\begin{equation}
\Delta \phi^2_{\hat{a}^{\dagger}\hat{a}- \hat{b}^{\dagger}\hat{b}}=\frac{\textrm{e}^{-2r}(4|\alpha|^2+(\textrm{e}^{2r}-1)^2)}{(\cosh(2r)-2|\alpha|^2-1)^2}
\end{equation}
while a single mode intensity measurement attains a minimum of,
\begin{equation}
\Delta \phi^2_{\hat{a}^{\dagger}\hat{a}}=\frac{4|\alpha|^2\textrm{e}^{-2r}+2\cosh(2r)+4\sqrt{2} |\alpha|\sinh(2r)-2}{(\cosh(2r)-2|\alpha|^2-1)^2}.
\end{equation}

At this point we can note however, in the case of Advanced LIGO, the powers in which they operate fixes $|\alpha|^2 \approx 10^{24}$ (while LIGO operates in continuous wave mode, we assume an integration time of one second, a circulating power of $P=800$ kW and wavelength $\lambda=1064$ nm, throughout), in which case, \textit{all} the analyzed detection schemes asymptote to the QCRB, so that there is no significant advantage in utilizing a detection scheme other than their current intensity measurement, showing that their current detection scheme is nearly optimal in this high power regime. Specifically, for high powers, each detection schemes leading term in the phase variance is given by $\Delta \phi ^2_{\textrm{all}}\approx (|\alpha|^2 \textrm{e}^{2r})^{-1}$, which is nearly optimal since the $\sinh^2(r)$ term in the QCRB is negligible compared to large $\alpha$. The phase variances shown above, are at their respective minima, in terms of optimal phase. In all but the intensity measurement scheme, this optimal phase is a constant value and therefore should not prove overly difficult to stabilize. In the case of intensity measurement however, this optimal phase depends on both the squeezing strength ($r$) and the amplitude of the coherent state ($|\alpha|$). Therefore, fluctuations in the source will actually affect the optimal phase setting and in general degrade the phase measurement in this measurement scheme. Specifically, the optimal phase for a single mode intensity measurement is given by, $\phi_{\hat{a}^{\dagger}\hat{a}} \rightarrow 2\tan^{-1}(2^{1/4}\sqrt{|\alpha|/\sinh(2r)})$. The other measurements attain their minimum phase variance at optimal phases of, $\phi_{\hat{\Pi}} \rightarrow \pi,\phi_{\hat{x}} \rightarrow \pi ,\phi_{\hat{a}^{\dagger}\hat{a}- \hat{b}^{\dagger}\hat{b}} \rightarrow \pi/2$. Note that, in practice, typical experiments use an offset to remain near these optimum values, but purposely remain slightly away from the minimum, due to noise considerations.  The ranking of the various measurements phase variance can be listed as $\Delta \phi^2_{\hat{\Pi}}< \Delta \phi^2_{\hat{x}}<\Delta \phi^2_{\hat{a}^{\dagger}\hat{a}- \hat{b}^{\dagger}\hat{b}}<\Delta \phi^2_{\hat{a}^{\dagger}\hat{a}}$. Some measurements appear similar on plots due to the fact that Advanced LIGO is typically run at very large photon number ($|\alpha|^2 \approx 10^{24}$) but realistic squeezing strength limits $r \simeq 1, N_{\textrm{sqz}}=\sinh^2r\simeq 1.38$. Each of these phase variances can be generalized to the photon loss case by using the  process described earlier and severly worsens the parity measurement, but the remaining measurements maintain their rankings from the lossless case. From these forms then, we can say that in the low photon number regime ($|\alpha|^2<500$), the difference in these detection schemes can be significant, but in the high photon number regime ($|\alpha|^2>10^{5}$), there is little difference between the various detection schemes. This means that the current setup of Advanced LIGO is near optimal and no significant improvement can be made by changing detection schemes, but in small scale LIGO-type setups that operate at lower powers, utilizing a detection scheme such as homodyne, may be advantageous. This is still considering that each detection scheme may be optimized perfectly, but as we argue, is likely more difficult for an intensity measurement than the other presented measurements.

\begin{figure}[!htb]
\includegraphics[width=\columnwidth]{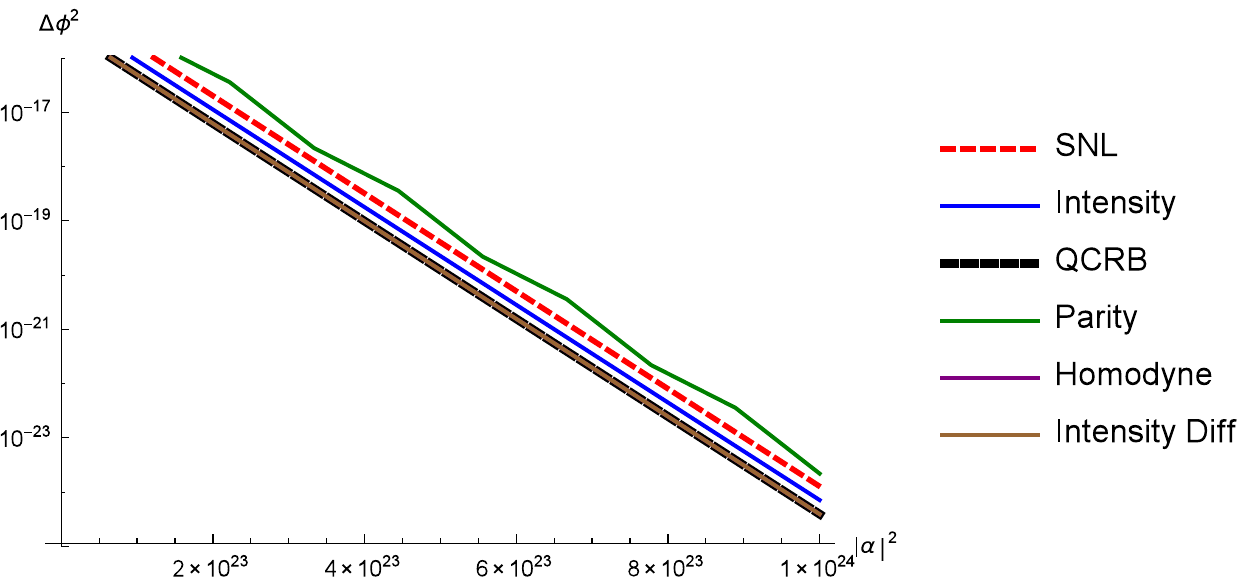}
\caption{Log plot of phase variance for various detection schemes for a coherent state and squeezed vacuum into an MZI, as a function of the average coherent photon number, $|\alpha|^2$. Loss parameters have been set to, $L=20\%$. We have assumed one can set the control phase to its optimal value, to obtain the best phase variance in each measurement choice. Squeezing strength in the squeezed state is set to $r=1$. Note that Parity is now not able to achieve even sub-SNL, due to loss, while homodyne and intensity difference quickly approach the QCRB (appear on top of one another). SNL and QCRB are also plotted with the same lossy parameters.}
\label{fig:nums}
\end{figure}
However, as shown in Figure~\ref{fig:phases}, when we consider a total photon loss of $L=20\%$, the effects on each measurements performance is significant. We can then also plot the phase variance as a function of average photon number, shown in Figure~\ref{fig:nums}, which can be related to the lights optical frequency and power by $|\alpha|^2=P/(\hbar \omega_0)$ \cite{GEO_bound}. In this form, its clear that a parity measurement suffers greatly, under lossy conditions. Parity may also be difficult to implement in a setup like Advanced LIGO as it either involves number counting (which is not feasible at the powers at which Advanced LIGO operates) or several homodyne measurements \cite{Plick2010}. Alternatively, a single homodyne measurement is nearly optimal in this lossy case and still only requires measurement on a single mode, is simpler to implement than parity, and is not nearly as sensitive to phase drift and loss. While intensity difference is also close in phase variance to a homodyne measurement (when $|\alpha|^2>100$) it requires utilization of both output modes for phase measurement, which may not be feasible in some setups.

We suggest that a homodyne measurement is likely the most realistic, optimal measurement choice for a setup like Advanced LIGO, as it is a typical measurement choice in interferometer experiments, as well as being a single mode measurement, likely resistant to photon loss, detector efficiency, and phase drift, but shows its main benefits in the low power regime. If we instead operate in the high power regime, then a homodyne measurement only achieves a factor of two improvement over an equally optimized standard intensity measurement. 

\subsubsection{Noise Model}

\begin{figure}[!htb]
\includegraphics[width=0.9\columnwidth]{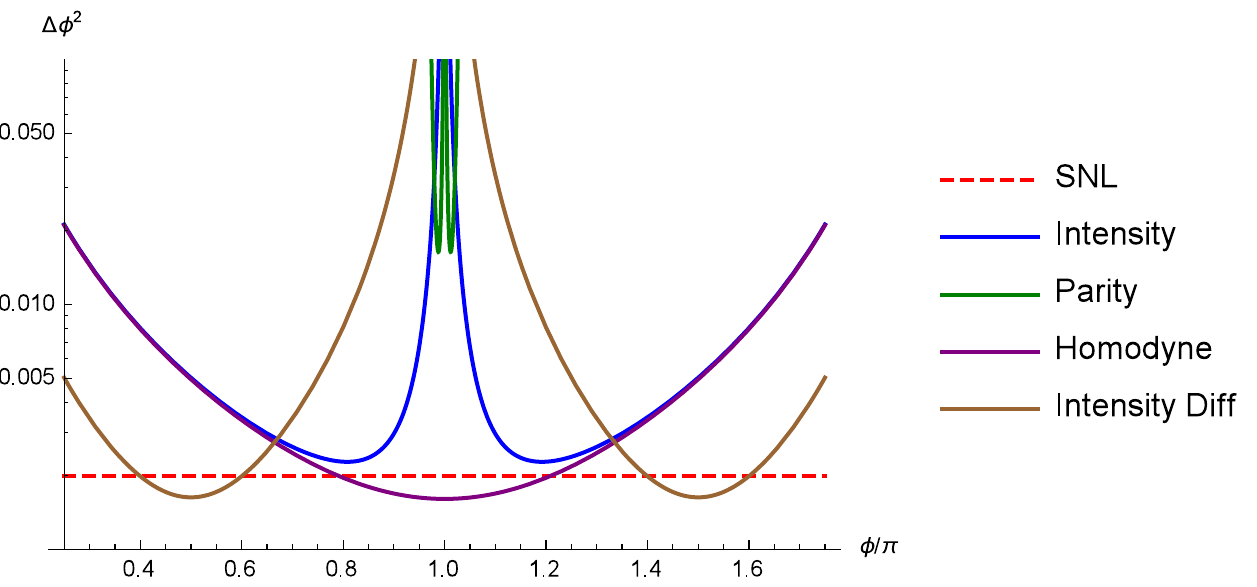}
\caption{Log plot of phase variance of the various detection schemes, with introduction of thermal noise into the signal beam, of total average photon number of $n_{\textrm{th}}=1/3$. Strength of the two input sources are set to $|\alpha|^2=500, r=1$.}
\label{fig:thermal}
\end{figure}

\begin{figure}[!htb]
\centering
\begin{subfigure}
{\includegraphics[width=0.4\columnwidth]{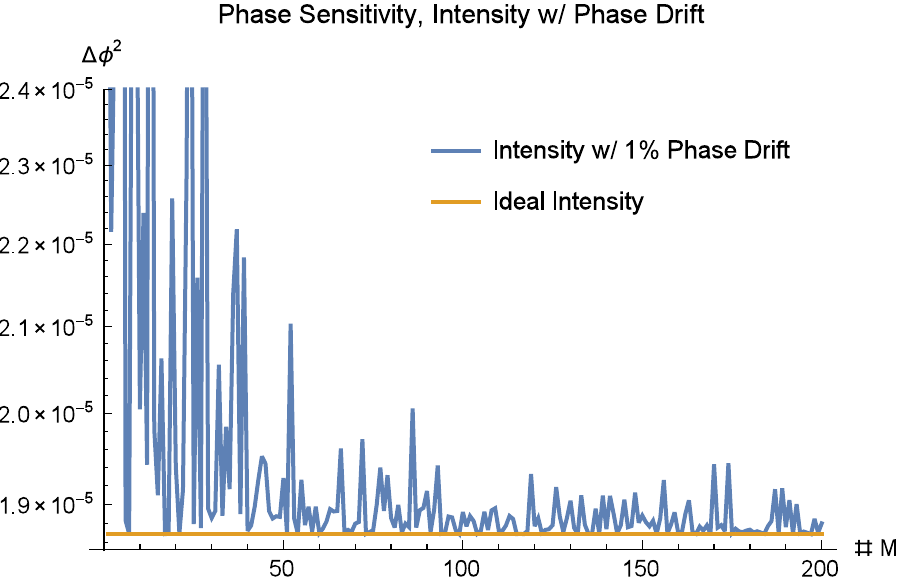}}
\end{subfigure}
~
\begin{subfigure}
{\includegraphics[width=0.4\columnwidth]{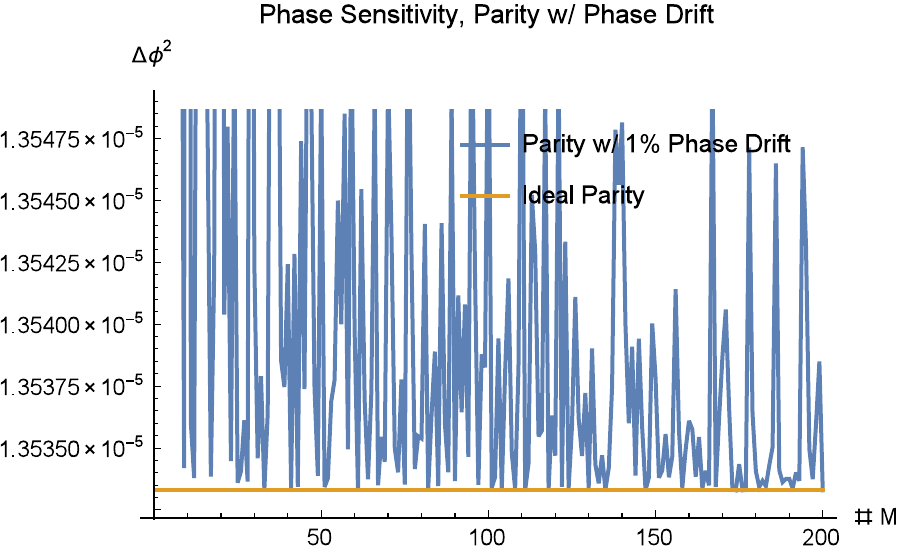}}
\end{subfigure}
~
\begin{subfigure}
{\includegraphics[width=0.4\columnwidth]{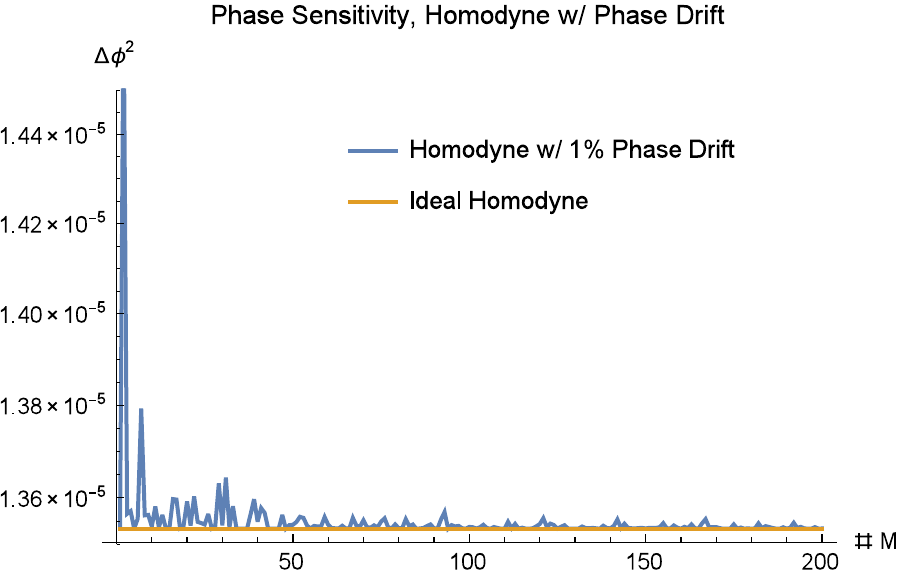}}
\end{subfigure}
~
\begin{subfigure}
{\includegraphics[width=0.4\columnwidth]{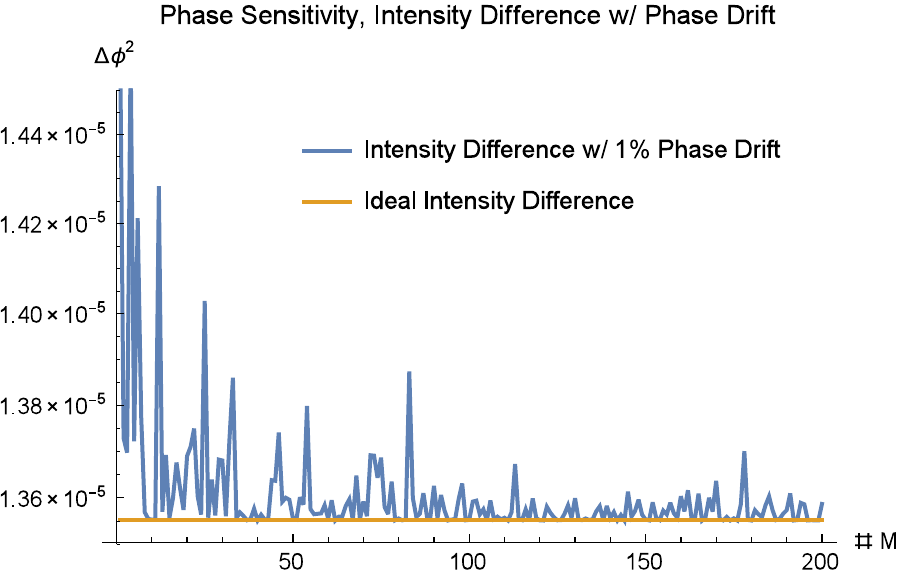}}
\end{subfigure}
\caption{Log plots of phase variance as a function of number of measurements ($M$). For all plots shown, $|\alpha|^2=100$, $r=1$ and the maximum possible phase drift, away from the optimal, for each measurement, is shown. Note that some plots have different ranges and phase drift percentages.}
\label{fig:phasedrift}
\end{figure}

In addition to photon loss, detector efficiency, and phase drift, we also model the inevitable interaction with thermal noise from the environment. This is accomplished much in the same way as a photon loss model, but here we consider a thermal state incident on a fictitious beam splitter, on both arms of the interferometer, inside the interferometer and trace out one of its output modes. This allows a tunable amount of thermal noise (by changing the average photon number in the thermal state), into the interferometer. The effects of this unwanted thermal noise, to the various measurements phase variance is shown in Figure~\ref{fig:thermal}. From this, we can see that even in the regime of introducing a relatively low photon number of thermal noise, it significantly degrades the phase variance of each scheme, but drastically affects the parity scheme, making it significantly above the SNL. Also in this regime, a standard single mode intensity measurement now does not acheive sub-SNL phase variance, but homodyne and intensity difference barely manage to achieve. We also note that the advantage of homodyne over intensity measurement is significantly decreased under thermal noise, but homodyne still maintains its superiority. An introduction of larger thermal photon number continues to degrade all measurements so that they no longer beat the SNL, but this example showcases their behavior under this noise model. It should be noted that in the optical regime, the occupation of a thermal state, at room temperature is approximately $n_{\textrm{th}}\approx10^{-20}$ and therefore, Advanced LIGO does not deal with significant contribution from this model of thermal noise, but experiments in the microwave frequencies can have $n_{\textrm{th}}\approx1$, where this model is more applicable. This model of thermal noise is not to be confused with other models of thermal noise, such as in the case of LIGO, where some references of thermal noise refer to thermal excitation of mechanical degrees of freedom, which is not considered here.
We have argued that the width of each measurements phase variance curve, shown in Figure~\ref{fig:phases}, gives an idea of each measurements resistance to phase drift. However, we will however be considering the lossless case in this section The mechanism of phase drift comes about due to the limited ability to set control phases in the interferometer with infinite precision. In general, the control phase value will vary around the optimal phase setting. For this reason we aim to show this phase drift in a more mathematical way and therefore we use the analytical forms of the various measurement phase variances, as a function of unknown phase, $\phi$, and simulate phase drift by computing a running average of the phase variance, with a pseudo-randomly chosen phase, near the optimal phase, for each measurement. This psuedo-random choice is made from a Gaussian distribution, whose mean is fixed at the optimal phase choice and has a chosen variance, shown in each plot in Figure~\ref{fig:phasedrift}. As predicted in the text, this gives a clearer picture of each measurements behavior under phase drift. 

Shown in Figure \ref{fig:phasedrift}, we see the phase variance for each measurement scheme, as a function of number of measurements. As the number of measurements is increased, the phase variance asymptotes to the ideal measurement case, given by the phase variance at the optimal phase. Note that the ranges for some plots are different as they attain different minimum phase variances. Also of significant difference is the phase drift for parity, which uses a psuedo-randomly chosen phase from a Gaussian distribution, centered on the optimal phase but with a standard deviation of only, $\sigma_{\hat{\Pi}}=0.001$, yet, still has a large range, and still performs fairly poorly as compared to the other measurement schemes. In the case of intensity, homodyne, and intensity difference measurements, which use $\sigma_{\hat{x}}=\sigma_{\hat{a}^{\dagger}\hat{a}}=\sigma_{\hat{a}^{\dagger}\hat{a}- \hat{b}^{\dagger}\hat{b}}=0.15$, its clear that homodyne and intensity difference attain a small phase variance, while also being more tolerant of phase drift, as compared to a standard intensity measure. In principle, all of these different measurement schemes will each attain their respective phase variance minimum, as the number of measurements increases to infinity, but it is instructive to see how quickly a finite number of measurements approaches the ideal phase variance minimum. We note that this model of phase drift, while general, may not necessarily apply if one has significant control over these fluctuating parameters, so that their drift effects are rendered insignificant, which is likely the case in Advanced LIGO.

\section{Photon Subtracted Thermal State}
\subsection{Photon Subtraction in Phase Measurement: The Bad}
\subsubsection{State Preparation and Photon Subtraction}
\begin{figure}[!htb]
\centering
\includegraphics[width=\columnwidth]{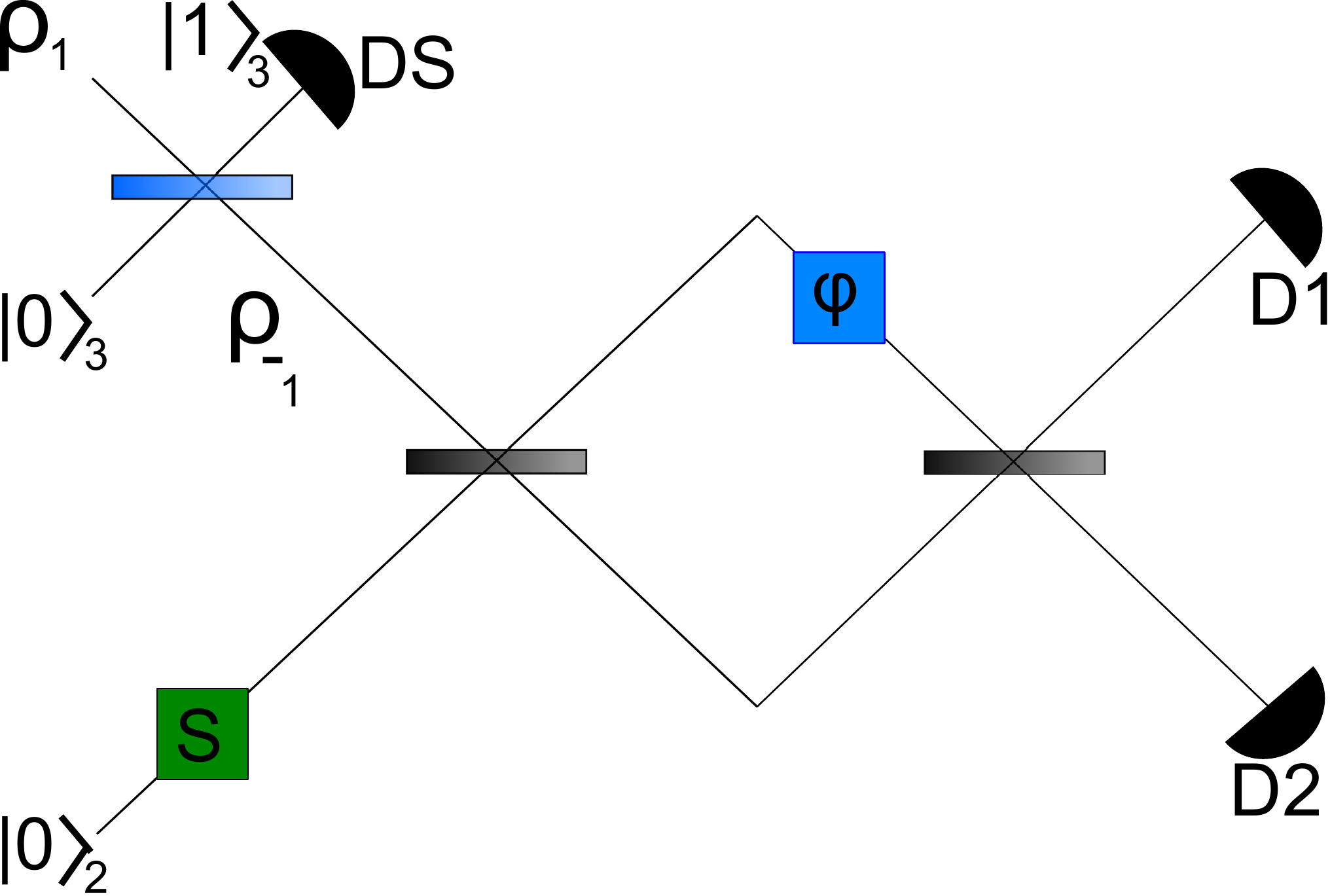}
\caption{MZI with the generation of a squeezed vacuum (lower) and a photon subtracted thermal state (upper) as input states. Blue beam splitter denotes a variable transmissivity, while black beam splitter are fixed to 50-50. Detector DS is used to herald the subtraction event in the thermal state. These two states mix, experience a phase difference $\phi$, and exit the interferometer at detectors D1 and D2.}
\label{fig:thmsubMZI}
\end{figure}

To complete our discussion of full examples, we will also show a complete working of a photon subtracted thermal state (PSTS) and squeezed vacuum, into an MZI, shown in Figure~\ref{fig:thmsubMZI}. While the thermal state itself is not an ideal choice for use in phase estimation, as it contains significant noise, we wish to investigate the effects of photon subtraction and therefore a coherent state is not valid.

We begin much in the same way as our previous example and construct the initial Wigner function from the product of a thermal state and squeezed vacuum, which takes the form of,
\begin{equation}
W(x_1,p_1,x_2,p_2)=\frac{1}{(1+4n_{\textrm{th}})\pi^2}e^{p_2^2(1-2G-2\sqrt{G(G-1)})-\frac{1}{1+4n_{\textrm{th}}}(x_1^2+p_1^2)+x_2^2(1-2G+2\sqrt{G(G-1)})},
\end{equation}
where $G \geq1$, is the gain of the squeezer, $\bar{n}=2n_{\textrm{th}}$ is the average photon number in the thermal state and we have fixed the squeezing angle, $\theta_{\textrm{sqz}}=0$. With this initial state, we then proceed to perform photon subtraction on the thermal state following the prescription shown in Section \ref{sec:photonsub}. This process involves the transform,
\begin{eqnarray}
\left(
\begin{array}{c}
x_{1_\textrm{s}}\\
p_{1_\textrm{s}}\\
x_{3_\textrm{s}}\\
p_{3_\textrm{s}}\\
\end{array} \right)= BS(T)\cdot \left(
\begin{array}{c}
x_{1}\\
p_{1}\\
x_{3}\\
p_{3}\\
\end{array} \right),
\label{eq:bssub}
\end{eqnarray}
where the ``$\textrm{s}$" subscript denotes that these phase space variable are after the subtraction beam splitter. We also need a projection operation for confirmation of the photon subtraction process; this is realized as a projection onto the single photon state, given by Eq.~(\ref{eq:project1}). After this process, we also need to re-normalize our state, following Eq.~(\ref{eq:prob1}). This process then needs to be repeated for the case where photon subtraction fails. To do this, we simply consider the complement of the projection onto the single photon state. Specifically this projection takes the form,
\begin{equation}
W_f(x_1,p_1,x_2,p_2)=\int (1-2\pi F_1(x_3,p_3))W(x_1,p_1,x_2,p_2,x_3,p_3) dx_3 dp_3,
\label{eq:projectf1}
\end{equation}
indicating that we project on the subspace corresponding to our heralded detector receiving any number of photons, other than a single photon. In this case, we say that we failed to subtract exactly one photon. These two cases, in general, lead to different Wigner functions and therefore both cases must be considered separately. From here we can now propagate our two states of light through the MZI according to Eq.~(\ref{eq:MZI1}). 
\subsubsection{Analysis}
\begin{figure}[!htb]
\centering
\includegraphics[width=\columnwidth]{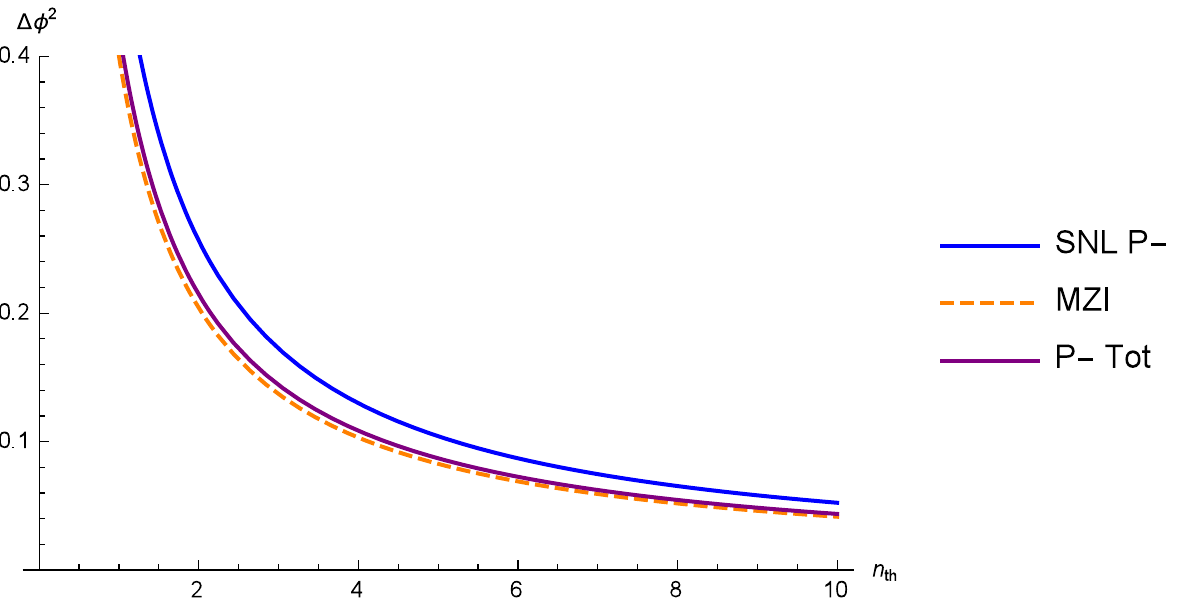}
\caption{Phase variance of a standard thermal state and squeezed vacuum into an MZI (MZI), compared to a weighted total phase measurement of a photon subtracted thermal state and squeezed vacuum into an MZI (P-Tot). While both can surpass the SNL, it's clear that photon subtraction provides no benefit over a standard setup. Relevant parameters have been fixed to $T=0.95, G=1.1, \phi=0$.}
\label{fig:pthmsubsens}
\end{figure}

We now have our photon subtracted thermal state and squeezed vacuum, at the output of the MZI. At this point we have many choices on how to proceed with detection or to attempt a calculation of the QCRB. While the ideal case would be to calculate the QCRB and find a matching detection scheme, calculating the QCRB proves extremely difficult as we are dealing with a mixed, non-Gaussian state (due to the photon subtracted thermal state). Therefore, for simplicity with this example, we choose a specific measurement, a parity measurement, and calculate its phase variance, carefully accounting for the nondeterministic nature of the photon subtraction process. Since the parity measurement is a single mode measurement, we are free to trace over the secondary mode and calculate the parity measurement on the remaining mode according to,
\begin{equation}
\avg{\hat{\Pi}}=\pi W(0_2,0_2),\quad\avg{\hat{\Pi}^2}=1,
\label{eq:par}
\end{equation}
where we have used the subscript to indicate that we have performed the parity measurement on mode two. We can now calculate the \textit{total} information of this measurement according to,
\begin{equation}
I_{\textrm{tot}}=\left(\frac{1}{P_\textrm{s}}\frac{1-\avg{\hat{\Pi_\textrm{s}}}^2}{(\partial \avg{\hat{\Pi_\textrm{s}}}/\partial \phi)^2}\right)^{-1}+\left(\frac{1}{1-P_\textrm{s}}\frac{1-\avg{\hat{\Pi_\textrm{f}}}^2}{(\partial \avg{\hat{\Pi_\textrm{f}}}/\partial \phi)^2}\right)^{-1},
\end{equation}
where the subscript ``s (f)" indicates that we are performing the parity measurement on the state resulting from a success (failure) of photon subtraction and we note we have included the probability of subtraction success (failure), in each case. We now have a weighted information for the estimate of $\phi$, in the case of successful photon subtraction and failed photon subtraction. This method ensures we are not discarding \textit{any} information, as this accounts for all outcomes with this choice of projection and measurement scheme. To obtain the phase variance of this scheme, we simply use the relation, $\Delta\phi^2=I_{\textrm{tot}}^{-1}$. In Figure~\ref{fig:pthmsubsens}, we show the phase variance, as a function of $n_{\textrm{th}}$ for the case of a standard thermal state and squeezed vacuum (MZI) and from a photon subtracted thermal state and squeezed vacuum (P- Tot), when we consider the total phase measurement. We can see that they perform nearly identically, with the photon subtracted scheme actually performing slightly \textit{worse} than the standard scheme. The two phase variances match exactly in the limit that $T\rightarrow 1$ and P-Tot never outperforms MZI, for any choice of parameters. We take from this that, given the ability to perform such an experiment, we should actually not perform subtraction at all and instead use a standard thermal state! We caution that if we instead do not consider this weighted total phase measurement, but instead \textit{only} consider the successful measurement, \textit{without} weighting it by its probability of success, then one may conclude a result that shows a photon subtracted state performs better than a standard thermal state, but this treatment assumes one can perform \textit{deterministic} photon subtraction, which we argued is impossible. This is not definitive that photon addition and subtraction is not useful at all for phase measurements, but at least illustrates that for this choice of input states and parity measurement, that photon subtraction does not provide a superior phase measurement.

\subsection{Photon Subtraction in SNR: The Good}
\subsubsection{Distant Source Model and State Preparation}
\begin{figure}[!htb]
\centering
\includegraphics[width=0.65\columnwidth]{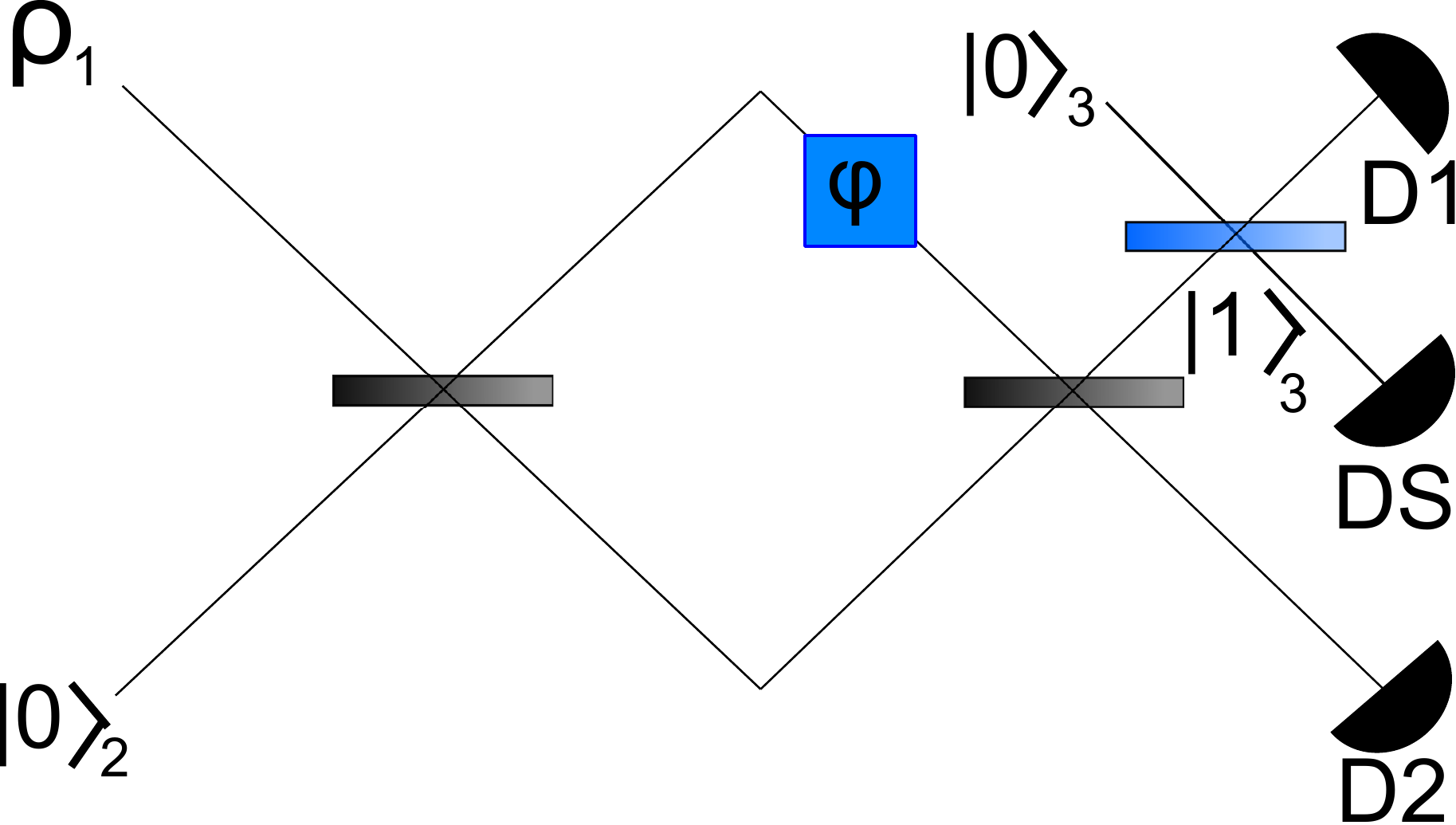}
\caption{A standard Mach-Zehnder Inteferometer where $\phi$ encompasses a phase difference between the two arms, as well as a control phase (not shown). Photon subtraction is modeled in the top output arm by a variable transmissivity (T) beam splitter (blue) where other beams splitters are standard 50-50 (black). Post selection is conditioned on the successful detection of a single photon, registered at detector DS.} 
\label{fig:MZIthm}
\end{figure}

While the previous result may imply that one should dismiss the idea of photon subtraction from thermal states, we will now present a regime where they retain a useful character. Instead of the setup discussed in the previous section, we instead consider performing photon subtraction at the \textit{output} of an MZI which has input states of a thermal state and vacuum state. Shown in Figure~\ref{fig:MZIthm}, we see the proposed configuration. The reasoning behind the placement of the photon subtraction stage, at the output, is a simple model of limited control. We assume that we are interested in measuring the statistics of a distant thermal source, to which one does not have physical access. In this case, one is limited to modification of the interferometer, at the output. With this restriction in place, the goal is now to enhance the photon statistics of such a setup. A deterministic amplifier, such as a quantum mechanical squeezer proves useless for this setup as it always amplifies the noise of the state, along with its photon number, leaving the SNR unchanged, at best. A nondeterministic amplifier, such as photon subtraction functions as a so called noiseless amplifier and showcases enhanced SNR.

We begin with the initial state of a thermal state and vacuum state, written as,
\begin{equation}
W_{in}(x_1,p_1,x_2,p_2)=\frac{1}{\pi^2(1+4n_{\textrm{th}})}e^{\frac{-x_1^2-p_1^2}{1+4n_{\textrm{th}}}}e^{-x_2^2-p_2^2},
\end{equation}
where $\bar{n}=2n_{\textrm{th}}$ is the average number of photons in the thermal mode (this choice is a simple matter of convenience). The propagation of this two mode Wigner function through the various linear optical elements of the MZI is again given by Eq.~(\ref{eq:MZI1}). It is important to note that once we have a form of the output state, prior to our probabilistic amplification, all operations are Gaussian preserving, but the use of photon addition or subtraction breaks this preservation (due to its projective measurements). 
\subsubsection{Photon Statistics}
\begin{figure}[!htb]
\centering
\includegraphics[width=0.7\columnwidth]{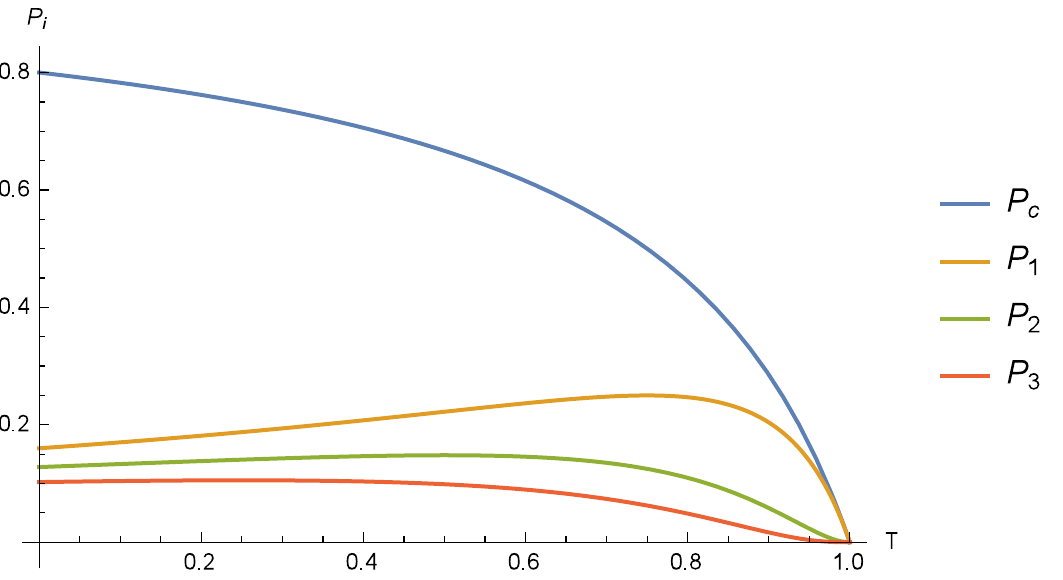}
\caption{Probability of successfully heralding a photon subtraction event for using click detection ($P_c$) and various numbers of photon counting ($P_i$), for $\bar{n}=2n_{\textrm{th}}=4$. One can note that click detection generally has the best chance of succeeding but as seen later, comes at the cost of less than ideal number statistics.} \label{fig:probs2a}
\end{figure}

Once we have obtained the state at the output, we can then perform photon subtraction as given by Eq.~(\ref{eq:bssub}) and (\ref{eq:projectf1}), applied on the upper mode. The probability of successfully subtracting a single photon is given by,
\begin{equation}
P_1=\frac{2n_{\textrm{th}}\cos^2{(\phi/2)}(1-T)}{(2n_{\textrm{th}}\cos^2{(\phi/2)}(1-T)+1)^2}.
\label{eq:psub1a}
\end{equation}
For an arbitrary condition of using full number counting detection for $m$ subtracted photons, one can find that the probability of this event goes as

\begin{equation}
P_m=\frac{(2n_{\textrm{th}}\cos^2{(\phi/2)}(1-T))^m}{(2n_{\textrm{th}}\cos^2{(\phi/2)}(1-T)+1)^{m+1}}.
\label{eq:psubn}
\end{equation}
We also consider a simpler conditional measurement, that of click detection, where an APD clicks whenever it receives any number of photons, but is not sensitive to the specific number of photons it receives. In this case,
\begin{equation}
P_c=1+\frac{1}{2n_{\textrm{th}}\cos^2{(\phi/2)}(T-1)-1}.
\label{eq:psubc}
\end{equation}
These various probabilities of successfully generating a photon subtracted thermal state are shown in Figure~\ref{fig:probs2a}.

For a postselection condition of single photon counting, we now have an average photon number of \[\bar{n}_{\textrm{th}_{1-}}=\frac{4 T n_{\textrm{th}}}{(2n_{\textrm{th}} (1-T)+1},\] and for click conditions, \[\bar{n}_{\textrm{th}_{c-}}=\frac{4 T n_{\textrm{th}} (n_{\textrm{th}} (T-1)-1)}{(2n_{\textrm{th}} (T-1)-1},\] One can see that for $T=1$, both of the previous expressions reduce to, $\bar{n}_{\textrm{th}_-}=2\bar{n}$, twice the previous value. This result, while surprising should also be considered with caution as from Eq.~(\ref{eq:psub1a}) and (\ref{eq:psubc}) we see that the probability of this state being generated tends to zero. 

\begin{figure}[!htb]
\centering
\includegraphics[width=\columnwidth]{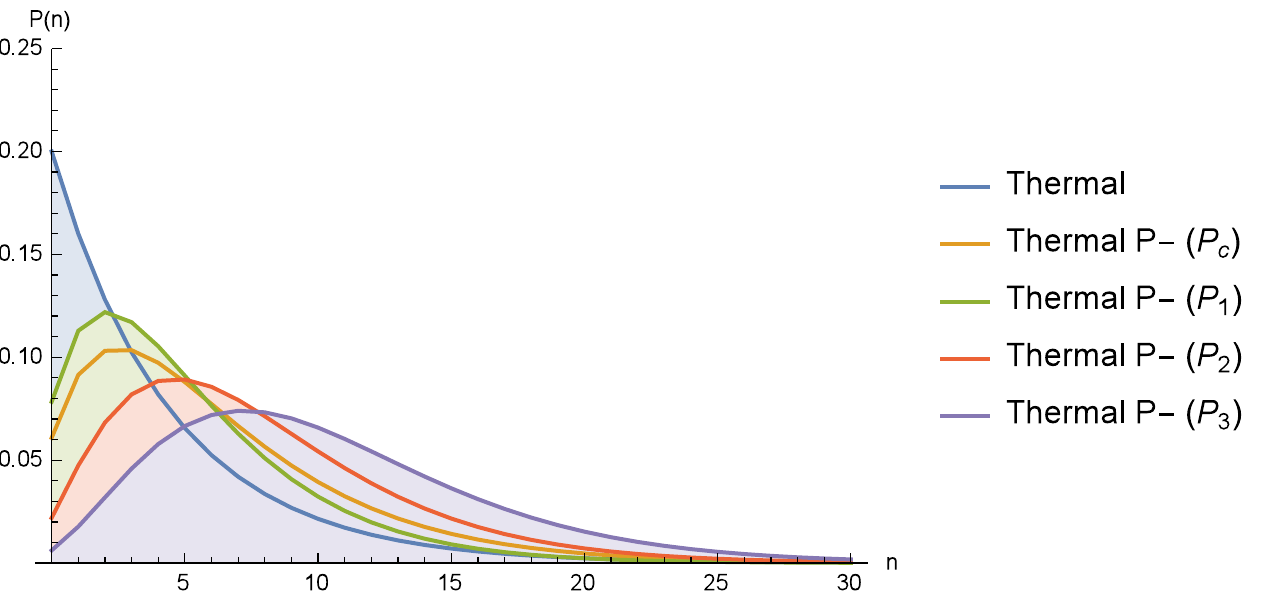}
\caption{Photon number distributions for a normal thermal state along with various photon subtracted thermal states. The average photon number for the normal thermal state is $\bar{n}=4$, while for the photon subtracted thermal states $\bar{n}\approx 4(m+1)$, where $m$ is the number of subtracted photons, significantly larger than its initial value. Transmissivity of the beam splitter is set to $T=0.9$.} \label{fig:pminus}
\end{figure}

\begin{figure}[!htb]
\centering
\includegraphics[width=\columnwidth]{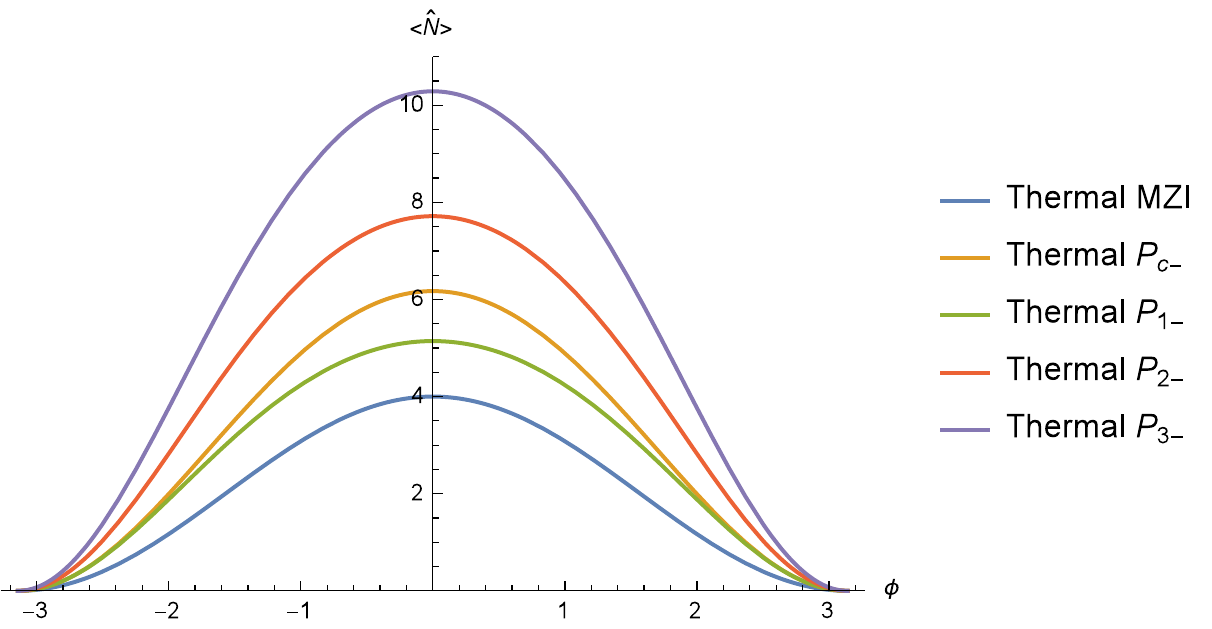}
\caption{Average photon number for a normal thermal state and photon subtracted thermal states ($\bar{n}=2n_{\textrm{th}}=4$). Various versions of the photon subtracted thermal state are shown, dependent on the choice of projective post selection (number of subtracted photons). Using a simple click detection scheme gives a modest improvement over single photon counting, but is quickly overtaken by higher photon counting schemes. The subtraction stage beam splitter transmissivity is set at $T=0.9$} \label{fig:numcount}
\end{figure}

A general form for the average photon number in a photon subtracted thermal state resulting from $m$ subtracted photons is given by,
\begin{equation}
\begin{split}
\bar{n}_{\textrm{th}_{m-}}&=\frac{2(m+1)n_{\textrm{th}}T\cos^2{(\phi/2)}}{2n_{\textrm{th}}(1-T)\cos^2{(\phi/2)}+1}\\
&=\frac{T(m+1)}{2n_{\textrm{th}}(1-T)\cos^2{(\phi/2)}+1}\bar{n}_{\textrm{MZI}},
\end{split}
\end{equation}
where $\bar{n}_{\textrm{MZI}}$ is the average photon number for a normal thermal state sent through an MZI. Note that the average photon number increases linearly with the number of subtracted photons but as shown in Eq.~(\ref{eq:psubn}) or Figure~\ref{fig:probs2a}, the probability of generating these states decreases with $m$. Also shown in Figure~\ref{fig:pminus} and \ref{fig:numcount}, we can see for successive photon subtraction, we do increase the average photon number in the resulting state.

An important point is that when $T=1$, generally the resulting state under this condition has the best characteristics. However, it's clear that generating a photon subtracted state under these conditions is extremely unlikely, as, if the beam splitter transmissivity is set to unity, no photon addition or subtraction event is likely to occur. This is characterized and shown in Figure~\ref{fig:probs2a}. One can also see this from the form of Eq.~(\ref{eq:psubc}) that $P_c \rightarrow 0$ for $T \rightarrow 1$. It is also in this limit that this beam splitter model for photon subtraction converge back to the mathematical treatment of the annihilation operator. For values of $T \neq1$ we then create a photon subtracted state with some less than ideal statistics. Of course the resulting state itself also is modified as $T$ changes and, in general, the characteristics of the state worsen as $T$ decreases to zero. However, the question remains, is there a region where the probability of creating some subtracted state remains significantly large and the resulting state also contains some useful character? In order to answer this question, we investigate various metrics. Many different metrics may be used when characterizing a quantum metrology topology. Here we will discuss phase information (phase sensitivity) through Fisher information as well as signal-to-noise ratio. Generally the phase estimation route is viewed as more robust and is typically the chosen metric in quantum metrology, but we will see that SNR, while being perhaps a more limited metric, has some aspects not possible when phase estimation is considered.

\subsubsection{Phase Measurement}
As a general improvement, even a probabilistic amplification is bounded by the Quantum Cram\'{e}r-Rao Bound (QCRB), which, for a fixed MZI topology, is calculated as a function of the input states only, and minimizes (maximizes for the Quantum Fisher Information (QFI)) over all possible detection schemes \cite{Caves1994}. Here, since our only modification of the MZI is after the unknown phase, $\phi$, it can be viewed as a particular measurement scheme, and so the QCRB is most directly calculated as a function of the total state, immediately after the unknown phase $\phi$, depicted in Figure~\ref{fig:MZI}. When this is calculated for any classical states as input states, the full state is Gaussian and thus can be calculated following \cite{QFI1}. As expected, since the states at this point are purely classical, the QCRB is simply the shot-noiselimit (SNL), which is given by $I_Q^{-1}=\frac{1}{\nu \bar{n}}=QCRB=\Delta \phi_{min}^2$, where $\bar{n}$ is the average photon number in the initial state and $\nu$ is the number of experimental trials. This result seems discouraging, as it shows that one cannot improve overall phase variance with this probabilistic amplification. However, since this probabilistic amplification is a form of weak value amplification and it is known that weak value amplification schemes show their benefit when technical noise is considered, one may still be able to use this implementation under the conditions of certain conditions of technical noise \cite{Howell2014}.

One can generalize the Fisher Information formalism to the probabilistic case, as discussed in \cite{Caves2014,David2015}, which provides a proper description of the probabilistic portion of this scheme and now no longer assumes all trials are identical, but properly weighs all trials by their associated probability of success. This adaptation is key when using any probabilistic process that claims to show improvement on the previously described, deterministic limits. This describes the CFI in two pieces, namely the information obtained at the meter (measuring the state itself) and the conditional probabilistic event. The meter information portion is given by
 \begin{equation}
F_m(\phi)=P_c(\phi)\sum_{j=1}^n\frac{1}{P_j(\phi)}\left[\frac{d P_j(\phi)}{d \phi}\right]^2,
 \end{equation}
with $P_j(\phi)$ the chosen POVM of the output state. The probabilistic measurement information is given by,
 \begin{equation}
F_{P_c}(\phi)=\frac{1}{P_c(\phi)[1-P_c(\phi)]}\left[\frac{d P_c(\phi)}{d \phi}\right]^2,
 \end{equation}
 where $P_c(\phi)$ is the probability of success for the probabilistic measurement and $\phi$ is the parameter to be estimated. Also in Ref.~\cite{Caves2014}, a nice argument points out that, post selection alone must necessarily discard some of the information about the parameter to be estimated, ideally this discarded information is small, but nevertheless must decrease the total information.  Then there are clear bounds that show that the postselected space must contain less information than the whole set of results. It remains an open consideration however, if postselection can provide some benefit in technical noise cases. The key argument here is that each experimental trial obtains information about the parameter $\phi$. So $\nu$ trials obtain $\nu$ times as much information about the parameter as one trial. Once postselection schemes are taken into account however, it is now $P_c \times \nu$ number of trials that are kept and since $P_c \leq 1$, it is clear that postselection when weighted by its probability of success sees a reduction in number of kept trials. Of course, there is an increase in information in the cases of successful postselection, but this increase is exactly countered by the reduction mentioned earlier, leaving the final total information, still bounded at the classical limit. The trouble with postselection can be that if one only considers the successful trials, without weighting them by their associated probabilities, it can lead to a total information beyond that of the classical limit. This pitfall is easy to believe but must not be misunderstood. Postselection when a metric of phase measurement is chosen, must always account for probabilities in the metric itself, due to the inherent conditions under which the typical bounds are considered. A phase uncertainty, described through Fisher Information is an asymptotic bound in the limit of many measurements. That is, one can approach the true phase uncertainty with a chosen POVM (or without a pre-chosen POVM if the QFI is considered), in the limit of many experimental trials and all bounds referenced from this argument assume this many trials case. When postselection enters the consideration, the number of trials is now reduced and properly accounting for this must be done before comparisons with typical metrology limits can be done. Note that in the case considered here, performing photon subtraction at the output of an MZI, limits us to the SNL, but this does not claim that photon addition/subtraction as a whole has no use at all in metrology, merely that it must always come with proper probabilities if phase information is the chosen metric.

As an example, we examine the case of a thermal state and vacuum into an MZI and perform photon subtraction on either output arm. The total CFI in this case, considering all events and postselection probabilities, with a chosen detection scheme of click detection the following CFI is simply the sum of the the CFI obtained at each detector. This expression is given by
\begin{equation}
\begin{split}
&F=P_c\left(\frac{1}{P_{cs_1}(1-P_{cs_1})}P_{cs_1}'^2 +\frac{1}{P_{cs_2}(1-P_{cs_2})}P_{cs_2}'^2 \right)\\
+&(1-P_c)\left(\frac{1}{P_{cf_1}(1-P_{cf_1})}P_{cf_1}'^2 +\frac{1}{P_{cf_2}(1-P_{cf_2})}P_{cf_2}'^2 \right)\\
+&\frac{1}{P_c(1-P_c)} P_c'^2 ,
\end{split}
\end{equation}
where $P_c$ is given by Eq.~(\ref{eq:psubc}) and $P_{cs_i}$ is the probability distribution of the $i^{th}$ detector clicking, $P_{cf_i}$  is the probability distribution of the $i^{th}$ detector ``not clicking", and derivatives are taken with respect to $\phi$. This construction is easily seen in pieces given by the first piece representing the information acquired at both detectors, weighted by the probability of photon subtraction succeeding. The second piece (second line) is the information acquired at both detectors, weighted by the probability of photon subtraction failing. The final piece is the information acquired by the photon subtraction herald mode itself. It's clear that the first and second piece necessarily involve different probability distributions, as postselection on vacuum or postselection on not vacuum results in different states. Once this total Fisher information is obtained, we maximize it as a function of $\phi$ and take its inverse to obtain a minimum phase variance $\Delta \phi^2$. In this case this minimum phase variance is given exactly by $\Delta \phi^2=1/(2n_{\textrm{th}})=$SNL, indeed showing agreement with the QCRB for these classical input states. While this is a classical Fisher information calculation, meaning it does not explicitly show that other measurements must be limited to this same bound, we use it as an indication that this scheme is likely limited to the classical SNL, as we would expect.

\subsubsection{SNR Measurement}
Another metric of interest is the SNR, defined by $\textrm{SNR}=\frac{\langle \hat{a}^\dagger\hat{a} \rangle}{\sigma_{\hat{a}^\dagger\hat{a}}}$, that is, the average photon number divided by its standard deviation, also typically described as intensity and $\sigma_{\hat{a}^\dagger\hat{a}}=\sqrt{\langle (\hat{a}^\dagger\hat{a})^2 \rangle-\langle \hat{a}^\dagger\hat{a} \rangle^2}$ . Since this metric is not constructed in the same way as Fisher Information, repeated trials and asymptotic limits are not necessarily incorporated into this metric. Therefore, while we should still note the associated probabilistic nature of the processes here, they are not directly incorporated into the metric. As with previous discussions, we can describe the effect on the SNR for an arbitrary number of subtracted photons from a thermal state, given by,
\begin{equation}
\begin{split}
SNR_{m-}&=\frac{|\cos{(\phi/2)}|\sqrt{2Tn_{\textrm{th}}(m+1)}}{\sqrt{2n_{\textrm{th}}\cos^2{(\phi/2)}+1}}.\\
&=\sqrt{T(m+1)} SNR_{MZI}
\end{split}
\label{eq:SNR}
\end{equation}

\begin{figure}[!htb]
\centering
\includegraphics[width=\columnwidth]{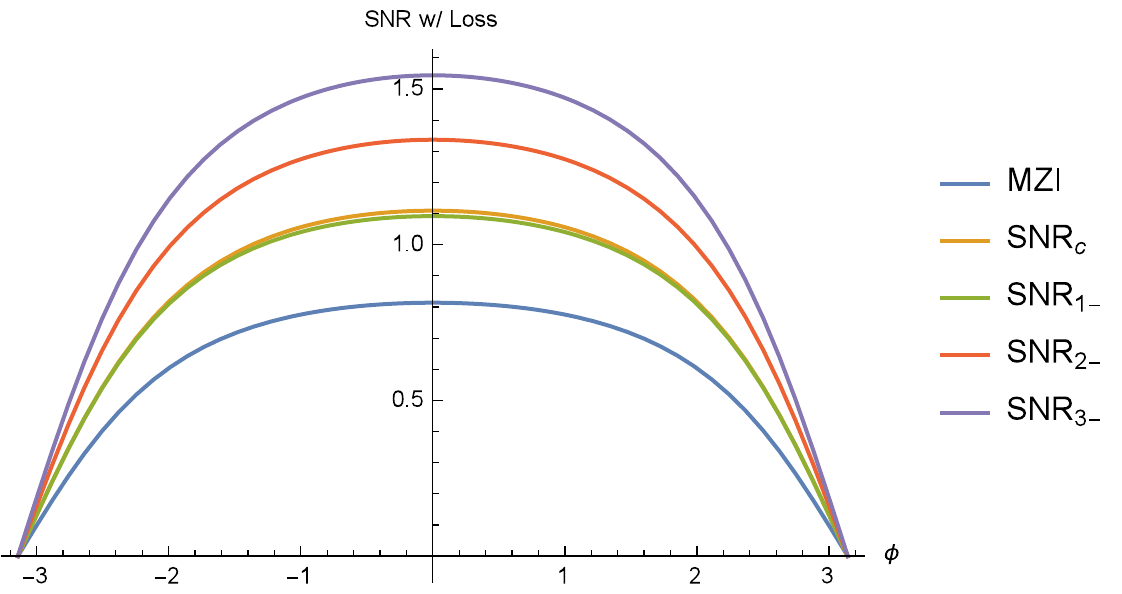}
\caption{SNR, with loss, of various photon subtracted thermal states. Average photon number is fixed at $\bar{n}=4$, before losses, and variable beam splitter is fixed at $T=0.9$. In this low photon regime, click detection and single photon subtraction achieve nearly the same SNR, while higher photon subtraction show improved benefit. Photon loss inside the MZI, equal in both arms, is set to $L=30\%$ and all detector efficiency is $D=70\%$.} \label{fig:SNRloss}
\end{figure}

 Thus far all considerations have considered ideal, lossless interferometers with perfect detectors. An immediate question one may consider is, do these benefits with photon subtraction only hold in this case? In fact, the behavior of photon subtraction is somewhat robust  when loss and detector efficiency are considered. For simplicity, we assume that photon loss occurs equally in both arms inside the MZI and that all detectors share the shame efficiency. For completeness, Figure~\ref{fig:SNRloss} shows the SNR with loss parameters of $L=30 \%$ and $D=70\%$ for all displayed curves. It is easily seen that the photon subtraction schemes described earlier maintain their advantage in lossy conditions.

\begin{figure}[!htb]
\centering
\includegraphics[width=\columnwidth]{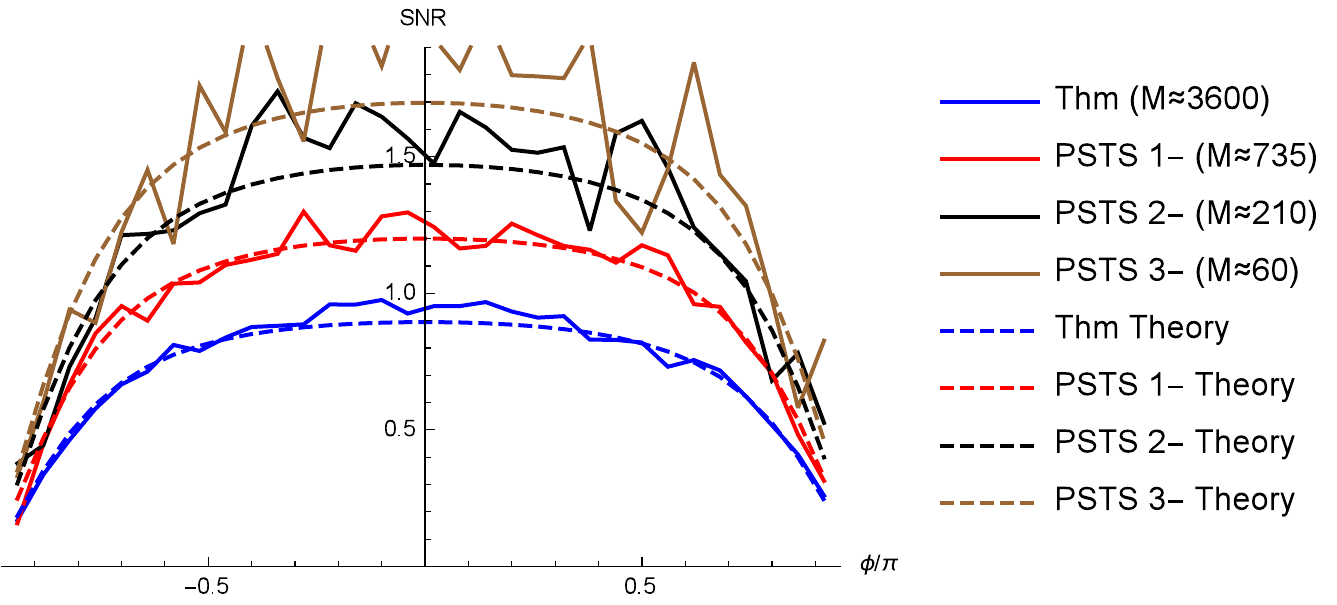}
\caption{Simulated data of the SNR for various subtraction of photons from the thermal state. Number of kept measurements, $M$, is shown for each case, which is a result of the probability of subtraction. We note that the trade for an increased SNR, is for the accuracy of the SNR measurement, but, in general can be improved with more measurements.} \label{fig:SNRprobthm}
\end{figure}

In the case of a phase measurement it is clear how the nondeterministic nature of photon subtraction affects our statistics. In the case of an SNR measurement however, the probability of generating a photon subtracted state are not directly integrated into our measurements. In order to fairly quantify the effects of this probabilistic process, we simulate the effect of this process by considering the fact that the efficiency of generating the photon subtracted state directly affects the number of our kept measurements. This has the effect of limiting to accuracy of our overall SNR measurement and is shown in Figure~\ref{fig:SNRprobthm}. Here we see that the SNR, in each case of photon subtraction, has significant scatter around the theoretical value, dependent on the efficiency. In this way, one can interpret this as the condition that, in the case of photon subtracted thermal states, if one requires a particularly accurate measurement, then use of this scheme comes at the price of longer experiment times, but can, in principle, achieve their theoretical predictions. Note that in the case of three photon subtractions, if we wish to achieve the same accuracy as compared to a standard thermal state, we would require sixty times more measurements. This condition may or may not be reasonable, depending on the application, but at the very least serves as an option, where deterministic methods provide no improvement to the SNR, regardless of the number of measurements performed.

\pagebreak
\singlespacing
\chapter{Mathematica MZI Toolbox}
\doublespacing
\section{Introduction \& Motivation}
The previous chapters have discussed the mathematical background of using Wigner functions in phase space to model the properties of quantum sensing in the form of a Mach-Zehnder Interferometer. We have used this background to construct a Mathematica notebook with the goal of requiring nearly no modification but remaining versatile enough to handle \textit{many} combinations of input states, modifications, and detection schemes. Here we will discuss the use of this toolbox, but focus more on the function of this notebook as this is a physics discussion, not one of computer science and therefore we will not discuss the deep inner workings of every command used. A few technical notes, this notebook was made in Mathematica 10.3.1, and requires no special packages or files to function. Access to this notebook is intended for any wishing to familiarize themselves with this modeling of quantum light but was originally made as a culmination of the research progress made during graduate work and serves as a parting gift to the Quantum Science and Technologies (QST) group at Louisiana State University.

\section{Function \& Usage}
We begin by listing the capabilities of this notebook. In terms of input states, one can select, using a dynamic choice of, a coherent, vacuum, thermal or single photon Fock state, in either input mode of a MZI. Some of these choices adhere to the Gaussian form and therefore this notebook will utilize Gaussian information calculations, when applicable, discussed in prior chapters. This greatly simplifies some of the calculations and allows for while you wait calculations, in most cases, detailed later. While we could choose to model a completely general Gaussian state, with the use of a displaced, squeezed, thermal state, but in some cases, this leads to overly complicated calculations when a simpler case of vacuum or a coherent state is sufficient. Our construction allows the same functionality as using a general Gaussian, but can be more simplified, in some cases. A note that all choices made in this notebook utilize the dynamic capabilities of Mathematica and therefore subsequent choices may change, depending on previous choices. 

Once the input states are selected, one has a choice of modifications to these states. These modifications include, squeezing, displacing, adding or subtracting a photon. A word of caution that while the notebook is capable of handling any combination of state and modification, the more complicated combinations can lead to lengthy calculation times and this should be considered fully. An important note is also that some combinations of states and modifications are invalid due to the fact that some combinations result in redundant states. For example, if vacuum is selected, then a modification of displacement is not available (one will not see this option even listed) due to the fact that this would result in a coherent state, which is already an option to begin with. For each state chosen as an input state, one can select the number of modifications desired, which then allows the selection of which specific modifications one desires. The order in which one selects these modifications, of course, also matters, as generally quantum operators do not commute. Therefore, squeezing and then displacing a state, can lead to different results when compared to displacing and squeezing the same state. One can also notice that choosing a first modification of, say, photon addition to a coherent state, re-opens the ability to choose further modifications such as squeezing or displacing.

Based on the choices made previously, one then calculates the propagation of the resulting state through the MZI following the mathematical description shown in previous chapters. If the state retains Gaussian form, the notebook calculates everything using Gaussian information, if the state is non-Gaussian, it constructs the corresponding non-Gaussian Wigner function and performs all calculations in this form. From here, one can now choose a detection scheme $\avg{\hat{O}}$ for either output. These choices include, homodyne, parity, intensity, click, and intensity difference. Each of these detection schemes, in general, leads to a different signal and phase variance, which can be calculated following previous chapters. A word of caution that some detection choices are significantly harder to calculate than others. In order of simplest to most complicated, in terms of computation time, we can rank these detection schemes according, homodyne, parity, click, intensity, and intensity difference. This ranking can be seen from the fact that using Gaussian information allows us to calculate homodyne detection, directly from the covariance and mean, while the other detection schemes require construction of the full Wigner function and utilization of our ``speedy" integral trick discussed in previous chapters. In the case of parity, the calculation remains fairly simple due to the usage of $\avg{\hat{\Pi}^2}=1$. Both choices of intensity and intensity difference, remain fairly complicated as the full calculation for phase variance requires a second moment calculation, which proves fairly time consuming, \textit{unless} some assumptions are made, which take the form of assuming all initial state phases are taken to be equal. Specifically, this is selected in the Optimize section and this sets all phase angles from the initial state, coherent phase, displacement angle, squeezing angle to be zero. This assumption is found to be best to minimize the phase variance and also serves to greatly speed calculations. Based on the choice of detection scheme, the phase variance is also calculated using $\Delta \phi^2=\Delta \hat{O}^2/\partial \avg{\hat{O}}/\partial \phi$. The notebook also takes advantage of a simplification possible when the mean of the output state is zero. This includes thermal, vacuum states and squeezed version of these states. In these cases, calculation of an intensity measurement and its phase variance are greatly simplified as the higher moments required for this calculation can be calculated with various identities, in terms of its covariance, since if the mean is zero, then $\Delta \hat{O}^2=\avg{\hat{O}^2}$.

An also interesting property we show in this notebook is the construction of the photon number distribution in each mode of the output of the MZI. Since this state is dependent on the unknown phase, $\phi$, this variable serves as a control (or ``steering") of the state. For a value of $\phi=0$, input 1 exits output 2 and input 2 exits output 1. For $\phi=\pi/2$, both inputs equally split into output 1 and 2 and so their distributions are identical. For $\phi=\pi$, then, as expected, the outputs swap from the case of $\phi=0$. Any other value of $\phi$ then allow one to see the arbitrary mixing of the two inputs, as a function of the relevant parameters, which depend on the choices selected previously. It is worth noting that, unless the assumption of equal phases has been made, then all possible parameters for the various combinations of input states and modifications is left completely analytical and therefore utilization of Mathematica's manipulate function is used to allow modification of these parameters, inside their plots to instantly see their effects, rather than having to turn to a numerical method with many more calculation runs.

The next section of the notebook calculates the phase variance attainable, with the selected detection scheme. Note that some of these may calculate in a few seconds, to a few minutes, depending on ones computational power and the level of complexity chosen in previous steps. Again we stress the importance of setting initial phases, prior to this stage, as they prove to severely complicate this stage, while the phase variance always achieves its minimum when these parameters are set to be equal. Therefore, unless for a specific requirement, we suggest choosing ``All initial state phases" in the ``Optimize" section of the notebook. In this section we can note that some detection schemes allow for a phase variance below the SNL, with the choice of a quantum input state, while others, do not and this effect should be considered carefully as it directly shows that some detection schemes, while perhaps complicated to implement physically (such as a Parity measurement), in some cases can allow for enhanced phase measurement, while a simpler detection scheme (such as Click detection) exhibit fairly poor statistics.

When we consider the large number of combinations possible in each stage of this notebook, we can see that this notebook is capable of reproducing results of many papers, including, in 1981, when Caves \cite{Caves1981} first suggested using a coherent state and squeezed vacuum into an MZI, up to some of the results of more recent papers, considering more exotic states into an MZI, such as Nori \cite{Nori}. We also note that likely contained in this notebook is the possibility of choices that have not been fully investigated, mostly concerning the proper use of photon addition and subtraction, which seem to have little benefit in phase estimation problems, when their probabilistic nature is properly taking into account. Therefore we suggest that this notebook, confirmed by reproducing many of various papers previous results and capable of generating new results, serves as a very useful tool in the investigation of general quantum sensors and is also possible to be adapted to specific schemes, with some minor modifications.

\pagebreak
\singlespacing
\chapter{Conclusion}
\doublespacing
In this dissertation, we have discussed many topics in the field of quantum metrology, including a basic introduction to the use of Wigner functions in phase space, a simplified model of LIGO, a full model of photon addition or subtraction, and also a description of noise sources such as photon loss, inefficient detectors, phase drift and thermal noise. 

We have shown the merits of describing quantum states of light, in terms of continuous phase space variables and discussed some of their challenges. We also briefly described the connection between a quantum Gaussian information treatment and Wigner functions, showcasing that these methods can be combined, when the Gaussian form is maintained. Using these methods for quantum metrology, we also showed the propagation of light through a typical interferometer setup used for phase measurement as well the use of SNR as a metric. In the case of phase measurement, we showed bounds given by the SNL and also discussed calculation of the QCRB through quantum Gaussian information. We contrasted these two metrics and showed that some schemes, as is the case for our example of a photon subtracted thermal state, are able to improve the SNR but not enhance a phase measurement, when the post-selection requirement of photon subtraction is taken into account. The nondeterministic nature of photon addition and subtraction was also argued in terms of a Alice and Bob gedanken discussion. This argument indicates that while our chosen physical model of photon addition and subtraction need not be the only model, any model must be described by a nondeterministic process, likely accompanied by post-selection, and this requirement directly affects claims of improved phase variance measurements. We also argue that due to this requirement, that a mathematical model of photon addition and subtraction with creation and annihilation operators is insufficient to account for the effects of this nondeterministic process. One can accommodate a mathematical model, along with inefficiencies to try to more closely model the realistic process, but we suggest it is more advantageous to model the actual physical process, rather than rely on purely numerical methods, which may or may not model the physical process.

We have also included a manual of sorts which accompanies the use of the MZI Toolbox Mathematica notebook, which serves as a useful tool to easily show the results of sending various Gaussian states of light through a MZI, with many different combinations of modifications and detection schemes. This notebook requires a minimal amount of changes, but allows the calculation of hundreds of combinations of input state, modifications, and detection schemes.

\pagebreak
\singlespacing
\addtocontents{toc}{\vspace{12pt}}
\addcontentsline{toc}{chapter}{\hspace{-1.6em} References}
\bibliographystyle{plain}
\bibliography{bib}
\pagebreak
\singlespacing
\addcontentsline{toc}{chapter}{\hspace{-1.6em} Appendix: Author Publications}
\appendix
\chapter*{Appendix: Authors Publications}
\doublespacing
\section*{Refereed Journal Articles}
\subsection*{Published}
\begin{enumerate}
\item{
Bryan T. Gard, Robert M. Cross, Petr M. Anisimov, Hwang Lee, and Jonathan P. Dowling, Quantum random walks with multiphoton interference and high order correlation functions, Journal of the Optical Society of America B, \textbf{30}, 1538, (2013)}
\item{Bryan T. Gard, Jonathan P. Olson, Robert M. Cross, Moochan B. Kim, Hwang Lee, Jonathan P. Dowling, Inefficiency of classically simulating linear optical quantum computing with Fock-state inputs, Physical Review A, \textbf{89}, 022328, (2014)}
\end{enumerate}
\subsection*{To be Published}
\begin{enumerate}
\item{
S. M. Hashemi Rafsanjani, M. Mirhosseini, O. S. Magana-Loaiza, B. Gard, B. E.
Koltenbah, C. G. Parazzoli, B. C. Capron, C. G. Gerry, J. P. Dowling, and R. W. Boyd, Enhanced thermal interferometry via photon subtraction, To be Submitted to: Nature, (2016)}
\item{Claudio G Parazzoli, Barbara A Capron, Benjamin E. Koltenbah, David R. Gerwe, Paul S. Idell, Bryan T. Gard, Richard Birrittella,  S. M. Hashemi Rafsanjani, M. Mirhosseini, O. S. Magana-Loaiza, Jonathan P. Dowling, Christopher G. Gerry, Robert W. Boyd, Enhanced Thermal Images of Faint Objects
Via Photon Addition / Subtraction, To be Submitted to: Physical Review Letters, (2016)
}
\item{Bryan T. Gard, Dong Li, Chenglong You, Kaushik P. Seshadreesan, Richard Birrittella, Jerome Luine, 
S. M. Hashemi Rafsanjani, M. Mirhosseini, O. S. Magana-Loaiza, Benjamin E. Koltenbah, Claudio G.
Parazzoli, Barbara A. Capron, R. W. Boyd, Christopher C. Gerry, Hwang Lee, and Jonathan P. Dowling, Photon Added Coherent States: Nondeterministic, Noiseless Amplification in Quantum Metrology, To be Submitted to: Physical Review Letters, (2016)}
\item{Bryan T. Gard, Chenglong You, Devendra K. Mishra, Robinjeet Singh, Hwang Lee, Thomas R. Corbitt, and Jonathan P. Dowling, Optimal measurement scheme for Advanced LIGO, To be Sumbitted to: Physical Review Letters, (2016)}
\item{Dong Li, Bryan T. Gard, Yang Gao, Chun-Hua Yuan, Weiping Zhang, Hwang Lee and Jonathan P. Dowling, Parity detection achieves Heisenberg limit in an SU(1,1) interferometer with coherent
mixed with squeezed vacuum input states, To be Submitted to: Physical Review A, (2016)}
\end{enumerate}
\section*{Book/Book Chapters}
Bryan T. Gard, Keith R. Motes, Jonathan P. Olson, Peter P. Rohde, Jonathan P. Dowling, An introduction to boson-sampling, \textit{From Atomic to Mesoscale: The Role of Quantum Coherence in Systems of Various Complexities}, World Scientific Publishing Co, (2015)
\section*{Presentations}
\begin{enumerate}
\item{Oral Presentation, Triple EX STEM at Louisiana State University, Nov 2010 and 2011}
\item{Presented Poster, Optical Society of America's Frontiers in Optics XXVII, Oct 2011, San Jose, CA}
\item{Oral Presentation, Division of Atomic, Molecular, and Optical Physics, June 2014, Madison, WI}
\item{Oral Presentation, Optical Society of America's Frontiers in Optics XXXI, Oct, 2015, San Jose, CA}
\end{enumerate}

\pagebreak
\singlespacing

\chapter*{Vita}
\doublespacing
\setlength{\parindent}{1.75em}
\vspace{0.2em}
\addtocontents{toc}{\vspace{1pt}}
\addcontentsline{toc}{chapter}{\hspace{-1.5em} Vita}
Bryan Gard was born in Baton Rouge, Louisiana, USA. He received his associate degree from Baton Rouge Community College in 2009 and bachelor degree from Louisiana State University in 2012. Bryan began working in quantum sciences, with Jonathan Dowling, during undergrad as part of the program, LASTEM. Bryan joined the Quantum Sciences and Technologies (QST) research group at LSU in 2010 and continued as a member throughout his undergrad and graduate career until graduating. He is a candidate to graduate in May 2016.

\end{document}